\documentclass[11pt]{article}
\pdfoutput=1
\usepackage{putex}
\usepackage[vcentermath]{youngtab}
\usepackage{subfig}
\usepackage{lscape}
\usepackage{pdflscape}        
\usepackage{hhline}

\usepackage{graphicx}
\usepackage{epstopdf}
\usepackage{enumerate}
\usepackage{cite}
\usepackage{tensor}
\usepackage{slashed}
\usepackage{amsmath}
\usepackage{amssymb}
\usepackage{mathrsfs}
\usepackage{lgrind}
\usepackage{multirow}

\usepackage{bbm}

\usepackage{hyperref}

\numberwithin{equation}{section}

\newcommand {\be} {\begin {equation}}
\newcommand {\ee} {\end {equation}}

\newcommand {\bes} {\begin {equation*}}
\newcommand {\ees} {\end {equation*}}

\newcommand{\eps}{\epsilon}

\def\CH{{\cal H}}

\def\CO{{\cal O}}

\newcommand{\beq}{\begin{equation}}
\newcommand{\eeq}{\end{equation}}

\def\be{ \begin{equation} }
\def\ee{ \end{equation} }

\def\XXint#1#2#3{{\setbox0=\hbox{$#1{#2#3}{\int}$}
\vcenter{\hbox{$#2#3$}}\kern-.5\wd0}}

\def\???th{{\textrm{\,???th}}}
\def \eps {\epsilon}

\def\XXint#1#2#3{{\setbox0=\hbox{$#1{#2#3}{\int}$}
     \vcenter{\hbox{$#2#3$}}\kern-.5\wd0}}

\newcommand{\ignore}[1]{}

\begin{document}


\institution{CMU}{Department of Physics, Carnegie Mellon University, Pittsburgh, PA 15213}
\institution{PU}{Joseph Henry Laboratories, Princeton University, Princeton, NJ 08544}
\institution{PCTS}{Princeton Center for Theoretical Science, Princeton University, Princeton, NJ 08544}

\title{Yang-Lee Quantum Criticality in Various Dimensions}

\authors{Erick Arguello Cruz,\worksat{\CMU} Igor R.~Klebanov,\worksat{\PU,\PCTS}  Grigory Tarnopolsky\worksat{\CMU} and  Yuan Xin\worksat{\CMU}}

\abstract{The Yang-Lee universality class arises when imaginary magnetic field is tuned to its critical value in the paramagnetic phase of the $d<6$ Ising model. In $d=2$, this non-unitary Conformal Field Theory (CFT) is exactly solvable via the $M(2,5)$ minimal model. As found long ago by von Gehlen using Exact Diagonalization, the corresponding real-time, quantum critical behavior arises in the periodic Ising spin chain when the imaginary longitudinal magnetic field is tuned to its critical value from below. Even though the Hamiltonian is not Hermitian, the energy levels are real due to the $\mathcal{PT}$ symmetry.
In this paper, we explore the analogous quantum critical behavior in higher dimensional non-Hermitian Hamiltonians on regularized spheres $S^{d-1}$. For $d=3$, we use the recently invented, powerful fuzzy sphere method, as well as discretization by the Platonic solids cube, icosahedron and dodecaherdron. The low-lying energy levels and structure constants we find are in agreement with expectations from the conformal symmetry. The energy levels are in good quantitative agreement with the high-temperature expansions and with Pad\` e extrapolations of the $6-\epsilon$ expansions in Fisher's $i\phi^3$ Euclidean field theory for the Yang-Lee criticality. In the course of this work, we clarify some aspects of matching between operators in this field theory and quasiprimary fields in the $M(2,5)$ minimal model. For $d=4$, we obtain new results by replacing the $S^3$ with the self-dual polytope called the $24$-cell.   
}

\date{}
\maketitle

\tableofcontents

\section{Introduction and Summary}

The Ising model is the most fundamental and widely explored statistical model, which has a broad range of applications across sciences. Its one-dimensional version was solved by Ernst Ising in 1925 and shown not to possess a finite-temperature phase transition \cite{Ising:1925em}. In all dimensions $d>1$, however, it has a line of first-order transitions $T<T_c$, which terminates at the critical point where the $\mathbb{Z}_2$ symmetry is restored. The properties of this second-order phase transition have been studied using many different methods, analytical and numerical. For $d>4$, the critical exponents of the Ising model are given exactly by Landau's mean field theory, but for $d<4$ this is no longer the case. Determining the dimension dependence of the critical exponents is a fascinating problem, which continues to attract a great deal of attention. 

In $d=2$, the exact Ising critical exponents have been known for many years thanks to the Onsager solution \cite{Onsager:1943jn}. The critical 2D Ising model corresponds to the simplest unitary Belavin-Polyakov-Zamolodchikov (BPZ) conformal minimal model $M(3,4)$ \cite{Belavin:1984vu}. More generally, the critical $d$-dimensional Ising model is described by the continuum theory of a massless scalar field $\phi$ with the interaction term $\sim \phi^4$. The problem of finding the critical exponents is then mapped to calculation of the scaling dimensions of operators $\phi$, $\phi^2$, $\phi^4$, etc. In dimensions slightly below $4$, the IR fixed point is weakly coupled, and the Wilson-Fisher $4-\epsilon$ expansion \cite{Wilson:1971dc} provides a useful method for approximating the $d$ dependence of operator dimensions. This has led to estimates of the 3D Ising critical exponents that are in very good agreement with experiments and Monte Carlo simulations. A significant improvement in accuracy was provided by the numerical conformal bootstrap \cite{Rattazzi:2008pe,El-Showk:2012cjh}, which is based on the conformal symmetry of critical fluctuations \cite{Polyakov:1970xd}
(for reviews, see \cite{Poland:2018epd,Simmons-Duffin:2016gjk, Rychkov:2016iqz, Chester:2019wfx}). Furthermore, the numerical bootstrap method has been applied to the operator scaling dimensions in the entire range $2\leq d < 4$ \cite{Cappelli:2018vir, Henriksson:2022gpa}, producing excellent agreement with resummed $4-\epsilon$ expansions for the $\phi^4$ theory. 

More recently, new methods for studying the 3D critical phenomena have emerged \cite{Zhu:2022gjc, Hu:2023xak, Hu:2024pen, Lao:2023zis, Lauchli:2025fii}. They rely on the critical behavior of the quantum theory in $2+1$ dimensions, which is analogous to the well-known periodic quantum spin chains in $1+1$ dimensions. In the continuum limit, the latter give the energy levels on a circle which, via the 2D Conformal Field Theory state-operator map determine the scaling dimensions of the primary operators and their descendants. To study a 3D CFT in a similar way, one must define the quantum Hamiltonian system on an appropriately regularized sphere $S^2$. In an important development \cite{Zhu:2022gjc,Hu:2023xak,Han:2023yyb}, the fuzzy sphere regularization \cite{Haldane:1983xm,Madore:1991bw} was shown to work very nicely for the critical 3D Ising model, giving the low-energy states that organize into the expected conformal multiplets (in fact, replacing the $S^2$ by an icosahedron works quite well too \cite{Lao:2023zis}, even though this method does not readily allow for extrapolation to the continuum limit).
The fuzzy sphere measurements of various CFT data approximately satisfy the conformal symmetry; this provides non-trivial checks and also can be used to correct the finite-size errors in this approach \cite{Lauchli:2025fii}. The conformal symmetry generators can be directly constructed from the fuzzy sphere observables \cite{Fardelli:2024qla,Fan:2024vcz}.
The fuzzy sphere approach has also been used to study various other 3D CFTs and defect CFTs\cite{Przetakiewicz:2025gzi,Zhou:2024zud,Dedushenko:2024nwi,Chen:2024jxe,Chen:2023xjc,Zhou:2023qfi,Han:2023lky}. 

Besides the scaling dimensions and OPE coefficients in a 3D CFT, the fuzzy sphere approach provides access to the sphere free energy $F$, which characterizes the number of degrees of freedom \cite{Jafferis:2011zi,Klebanov:2011gs,Casini:2012ei}. This quantity can be extracted from the entanglement entropy across the equator of $S^2$, and the numerical result for $F$  in the Ising model \cite{Hu:2024pen} is in excellent agreement with the extrapolation of the $4-\epsilon$ expansion of $F$ \cite{Fei:2015oha}. 

A fascinating feature of the $d$-dimensional Ising model is the appearance of zeros of the partition function $Z$ at imaginary values of the magnetic field 
\cite{Yang:1952be, Lee:1952ig}. In the paramagnetic phase $T>T_c$, these Yang-Lee (YL) zeros become dense near the edge value $h_{\rm crit}(T)$. 
As shown by Kortman and Griffiths \cite{Kortman:1971zz}, the density of zeros has the scaling form $\rho(h)\sim (h-  h_{\rm crit}(T))^\sigma$, and $\sigma$ is negative in $d=2$ but positive in $d=3$. In an important further development, Michael Fisher \cite{Fisher:1978pf} realized that this critical behavior is described by the Euclidean field theory with the imaginary potential $\sim i \phi^3$. In this continuum approach, the critical exponent is
$\sigma= \frac{\Delta_\phi}{d- \Delta_\phi}$, where $\Delta_\phi$ is the scaling dimension of the real scalar field $\phi$. 
The field theory description of the YL universality class implies that its upper critical dimension is $6$. 
The lower critical dimension is $1$, and the exact solution there gives $\sigma=-\frac 1 2$ \cite{Yang:1952be, Lee:1952ig}. 
Therefore, an interesting problem is to determine the $d$ dependence of the critical exponents in the range $1<d<6$.
One approach, analogous to the Wilson-Fisher $4-\epsilon$ expansion for the Ising universality class, is to develop the $6-\epsilon$ expansion for the massless $i\phi^3$ field theory.
This approach, in conjunction with the exact result in $d=1$, was used by Fisher \cite{Fisher:1978pf} to estimate the $d$-dependence of $\sigma$ for the YL universality class. Soon after the advent of the BPZ minimal models \cite{Belavin:1984vu}, Cardy \cite{Cardy:1985yy} realized that the critical YL model is also exactly solvable in $d=2$, where it corresponds to the simplest non-unitary minimal model $M(2,5)$ (for a recent review, see \cite{Cardy:2023lha}). The field $\phi$ corresponds to the Virasoro primary $\Phi_{1,2}$ of holomorphic dimension $h_{1,2}=-\frac 1 5$. This means that $\Delta_\phi=-\frac 2 5$ and $\sigma=-\frac 1 6$ in $d=2$.  

In the range $2<d<6$, no similar exact results are known. In this paper, we will use a combination of $6-\epsilon$ expansions and numerical diagonalizations to approximately determine the scaling dimensions of various operators. The $6-\epsilon$ expansions are available to $O(\epsilon^5)$ \cite{Borinsky:2021jdb,Kompaniets:2021hwg}.\footnote{After the original version of this paper appeared, the $6$-loop beta function and anomalous dimension became available \cite{Schnetz:2025wtu}; they determine the $O(\epsilon^6)$ terms in 
$\Delta_\phi$ and $\Delta_{\phi^3}$.}
 Their resummations have been extrapolated to $d=2$, yielding good agreement with $M(2,5)$. However, for the higher dimension operators this method can exhibit significant uncertainties.
The numerical boostrap approach is complicated by the fact that the 
YL model is non-unitary, so that the standard methods based on positivity cannot be used. A different bootstrap approach to the YL model was used in 
\cite{Gliozzi:2013ysa,Gliozzi:2014jsa,Hikami:2017hwv}, but it does not appear to be as precise as the numerical bootstrap based on positivity. We will instead follow the successes of the fuzzy sphere \cite{Zhu:2022gjc} and the icosahedron\cite{Lao:2023zis}\footnote{The icosahedron has also been used to discretize the 2D classical Ising model on a sphere \cite{Brower:2024otr,Brower:2025oti}} approaches to the 3D Ising model, adapting them to the YL case where a non-Hermitian Hamiltonian has to be tuned to quantum criticality. 
We will proceed analogously to the quantum spin chain approach which has proven to be successful for describing the 2D YL model. Here an imaginary longitudinal magnetic field $h_{z}$ was introduced and tuned to a critical value from below 
\cite{vonGehlen:1991zlm,vonGehlen:1994rp, Castro-Alvaredo:2009xex}.\footnote{More recent studies of non-Hermitian spin chain Hamiltonians were carried out in the context of $Q$-state Potts models with $Q=5$ \cite{TangY_2024, JesperJ_2024}. Let us also note that the YL universality class is related to the $Q\rightarrow \infty$ limit of the Potts model \cite{Wiese:2023vgq}.} 
Even though the spin chain Hamiltonian is complex, the energy levels are real for $h_{z} < h_{z}^{\rm crit}$ due to the $\mathcal{PT}$ symmetry \cite{Castro-Alvaredo:2009xex}. 

This paper is organized as follows. In section \ref{minimal} we review the minimal model $M(2,5)$ and its field theory description, discussing the correspondence between the two.  
In section \ref{YLchain} we review the quantum spin chain with an imaginary longitudinal magnetic field \cite{Uzelac1981, vonGehlen:1991zlm, vonGehlen:1994rp, Castro-Alvaredo:2009xex}. We carry out its numerical diagonalizations and show that the appropriate criteria for quantum criticality, analogous to those used recently \cite{ArguelloCruz:2024xzi} for a precise study of the Ising transition in the Schwinger model, 
 lead to excellent agreement with the exact results from the $M(2,5)$ minimal model. In section \ref{platonicsection} we study a quantum spin model with an imaginary longitudinal magnetic field, putting it on the Platonic solids cube, icosahedron and dodecahedron, which are viewed as regularized versions of $S^2$. We obtain good agreement with the expected YL criticality.
 In section \ref{fuzzysection}, we carry out analogous calculations on the fuzzy sphere and extrapolate them to the thermodynamic limit. Our fuzzy sphere result for the scaling dimension of $\phi$, $\Delta_\phi=0.214(2)$ is in excellent agreement with the high-temperature expansion result $0.214(6)$ from
\cite{Butera:2012tq} and is also very close to our two-sided Pad\'e estimate $\approx 0.218$. 
In section \ref{Octaplex}, we obtain first results for the quantum criticality of the $d=4$ YL model by replacing the $S^3$ by the $24$-cell, which is a selfdual regular $4$-polytope. 

{\bf Note Added:} Some of the preliminary results from this project were presented by one of us (G.T.) at the PCTS workshop ``Spheres of Influence'' \cite{Grishatalk}. 
As we were finishing this paper, we learned of two papers in preparation on related subjects \cite{Fan:2025bhc, Miro:2025jnz}.     
 
 \section{Minimal model $M(2,5)$ and its field theory description}
\label{minimal}

The minimal conformal model $M(2,5)$, which describes the YL critical behavior in two dimensions \cite{Cardy:1985yy}, has only one non-trivial Virasoro primary field, $\Phi_{1,2} \equiv \Phi_{1,3}$. It has holomorphic dimension $h_{1,2}=-\frac{1}{5}$.
The Kac table of this model, where we denote this field by $\phi$, is:
\begin{table}[h!]
  \centering
  \begin{tabular}{@{} |c|c|c|c| @{}}
  \hline
 $\Phi_{1,1}$ &     $\Phi_{1,2}$ &     $\Phi_{1,3}$ &     $\Phi_{1,4}$ \\ 
    \hline
  \end{tabular}, \qquad 
  \begin{tabular}{@{} |c|c|c|c| @{}}
  \hline
 $\textrm{I}$ &     $\phi$ &     $\phi$ &     $\textrm{I}$ \\ 
    \hline
  \end{tabular} ,
  \caption{$M(2,5)$ model with the central charge $c=-22/5$.}
  \label{tab1}
\end{table}

\begin{figure}[h!]
\center
\includegraphics[width=0.5\textwidth]{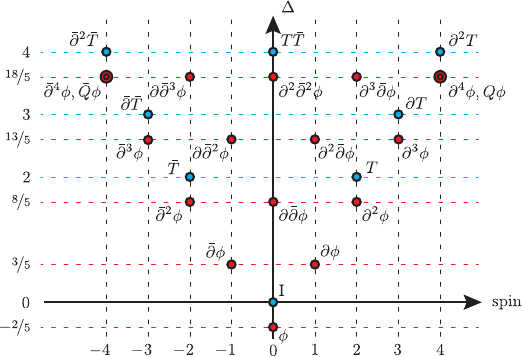}
\caption{
The lowest dimension operators of $M(2,5)$.}
\label{2DYL_states}
\end{figure}

The Operator Product Expansion (OPE) reads  
\begin{align}
\phi(z,\bar{z}) \times \phi (0,0) = |z|^{4/5}\textrm{I} + C_{\phi \phi \phi}  |z|^{2/5} \phi\, .
\end{align} 
The structure constant is purely imaginary and has the value
\begin{align}
C_{\phi \phi \phi} =  i  \left(\frac{\Gamma(\frac{6}{5})^{2}\Gamma(\frac{1}{5})\Gamma(\frac{2}{5})}{\Gamma(\frac{3}{5})\Gamma(\frac{4}{5})^{3}}\right)^{1/2}\approx 1.911312699 i\,.
\end{align} 
The lowest level null vectors in the $M(2,5)$ model are \cite{Yurov:1989yu,Xu:2022mmw}
 \begin{align}
L_{-1}|\textrm{I} \rangle, \quad \Big(L_{-2}^2+\frac{3}{5}L_{-4}\Big)| \textrm{I} \rangle, \quad \Big(L_{-2}-\frac{5}{2}L_{-1}^{2}\Big)|\phi\rangle, \quad \Big(L_{-3}-\frac{10}{3}L_{-1}L_{-2}+\frac{25}{36}L_{-1}^{3}\Big)|\phi\rangle\,.
\end{align} 
All the operators (primary and descendants) up to level $4$ are exhibited in Figure \ref{2DYL_states}, where $T=L_{-2}|\textrm{I} \rangle$ is the holomorphic part of the stress-energy tensor and  
\be 
Q \phi =(L_{-4}- \frac{625}{624} L_{-1}^{4})|\phi\rangle\ , \qquad  \bar Q \phi =(\bar L_{-4}-\frac{625}{624} \bar L_{-1}^{4})|\phi\rangle\ .
\ee 
denote the spin $4$ quasi-primary state \cite{Yurov:1989yu}. A related spin $0$ quasiprimary is
$\Xi= Q\bar Q |\phi\rangle$ \cite{Xu:2022mmw}.

Using the operator $Q_6=L_{-6}- \frac{35}{43} L_{-5} L_{-1} - \frac{105}{43} L_{-4} L_{-2}$, we can also construct a spin $6$ quasiprimary state 
\cite{Leitner:2018iyf,Katsevich:2024sov}: 
\be 
Q_{6} |\phi\rangle\ , \qquad  \bar Q_{6} |\phi\rangle\ \ .
\ee  
The simplest non-conserved spin $2$ quasiprimary is \cite{Xu:2022mmw}
\be
\label{newspintwo} 
T'= Q_{6} \bar Q|\phi\rangle\ , \qquad  \bar T'= Q\bar Q_{6} |\phi\rangle\ .
\ee  
It has dimension $\Delta_{T'}= 9.6$ and lies on the first subleading Regge trajectory.

\begin{figure}[h!]
\center
\includegraphics[width=1.03\textwidth]{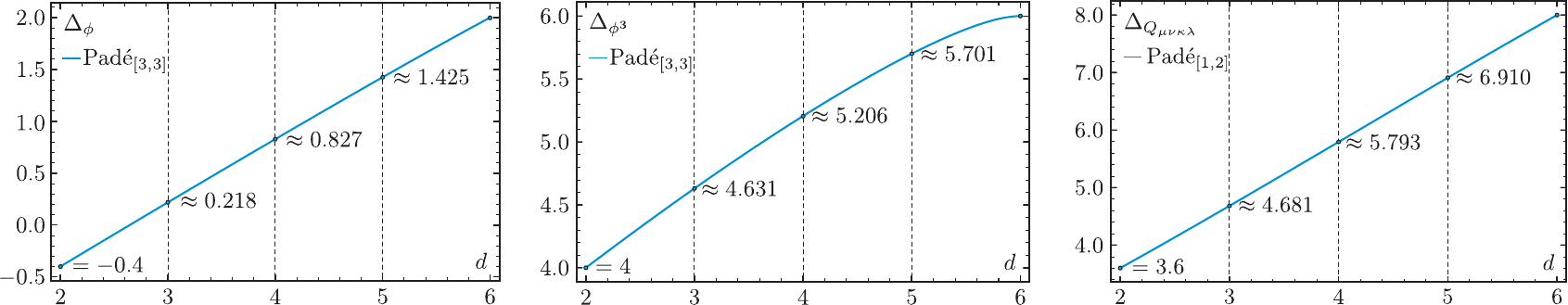}
\caption{
Two-sided Pad\'e extrapolations of the scaling dimensions of $\phi, \phi^3$ and $Q_{\mu\nu\kappa\lambda}$ operators.}
\label{2SPadeYLfig}
\end{figure}

The field theory description of the YL universality class is provided by 
\be
\label{FishGL}
S=\int d^d x \bigg ( \frac{1}{2}\left(\partial_{\mu}\phi\right)^2 + {g\over 6} \phi^{3} 
\bigg )\ ,
\ee    
where $g$ is purely imaginary \cite{Fisher:1978pf}. This action has a so-called $\mathcal{PT}$ symmetry \cite{Bender:2018pbv}, which acts by $\phi\rightarrow -\phi$, $i\rightarrow -i$. In $d=6-\epsilon$, the beta function is known to 5-loop order \cite{Borinsky:2021jdb}, but for brevity we state the results up to $3$ loops: 
\cite{deAlcantaraBonfim:1980pe,deAlcantaraBonfim:1981sy,Fei:2014xta,Borinsky:2021jdb}
\be
\beta (g)= -\frac{\epsilon g}{2} - \frac{3 g^3}{4 (4\pi)^3}- \frac{125 g^{5}}{144 (4\pi)^6}-\Big(\frac{33085}{20736} + \frac{5 \zeta(3)}{8}\Big)\frac{g^{7}}{(4\pi)^{9}}+  O(g^9)\ .
\ee
Therefore, there is an IR stable fixed point at
\be
g_*= i\sqrt{ \frac{2 (4\pi)^3 \epsilon} {3}} \Big(1+\frac{125}{324}\epsilon +\frac{5(253-972 \zeta(3))}{26244}\epsilon^2 + O(\epsilon^3)\Big)\ .
\ee 
In $d=2$, the field $\phi$ is identified with the Virasoro primary operator $\Phi_{1,2}$ of the $M(2,5)$ model. The $6-\epsilon$ expansion of $\Delta_\phi$ is 
\cite{deAlcantaraBonfim:1980pe,deAlcantaraBonfim:1981sy,Fei:2014xta,Borinsky:2021jdb}
\be 
\Delta_{\phi} = 2- \frac{\epsilon}{2}+ \gamma_\phi =2 - \frac{5}{9}\epsilon - \frac{43}{1458}\epsilon^2 
 + \left ( -\frac{8375}{472392}+ \frac{8 \zeta(3)}{243}\right ) \epsilon^3 + O(\epsilon^4)\ .
\ee
Performing Pad\' e extrapolations of this series, which is known to order $O(\epsilon^5)$ \cite{Borinsky:2021jdb}, with the condition that $\Delta_{\phi}=-\frac{2}{5}$ in $d=2$, gives the estimates of its scaling dimension that are plotted in Fig. \ref{2SPadeYLfig}. The $(3,3)$ and $(4,2)$ Pad\' e approximants produce results that are almost indistinguishable.
The dependence of $\Delta_\phi$ on $d$ is very smooth and close to linear; see also
Table \ref{2SPadeYLest}. 

The operator $\phi^2$ is a descendant by the equation of motion. 
Therefore, the next spin-zero conformal primary field is $\phi^3$. Its scaling dimension is \cite{deAlcantaraBonfim:1980pe,deAlcantaraBonfim:1981sy,Fei:2014xta,Borinsky:2021jdb}
\be
\Delta_{\phi^3} = d+ \beta' (g_*)= 6 - \frac{125}{162}\epsilon^2 + \left ( \frac{36755}{52488}+ \frac{20 \zeta(3)}{27}\right ) \epsilon^3 + O(\epsilon^4)\ .
\ee
Pad\' e extrapolations of this series, which is known to order $O(\epsilon^5)$ \cite{Borinsky:2021jdb}, indicate that this operator is irrelevant for $d<6$. In $d=2$, we identify 
the $\mathcal{PT}$ invariant operator $i\phi^3$
 with the leading $s=0$ quasiprimary irrelevant operator $T\bar T$ of scaling dimension $4$. This is analogous to the well-known identification of irrelevant operator $\phi^4$ at the IR fixed point of the massless $\phi^4$ theory with operator $T\bar T$ in the 2D critical Ising model. 
Performing Pad\' e extrapolations with the condition that $\Delta_{\phi^3}=4$ in $d=2$, gives the estimates of its scaling dimension 
that are plotted in Fig. \ref{2SPadeYLfig} and listed for integer $d$
in Table \ref{2SPadeYLest}.

The next spin-zero quasiprimary operator in $M(2,5)$ is $\Xi=Q\bar Q\phi$ \cite{Xu:2022mmw}; it has a rather high 
scaling dimension, $\frac{38}{5}$. In \cite{Gliozzi:2014jsa}, this operator was identified with $\phi^3$ in the field theory, but our Pad\' e extrapolations of $\Delta_{\phi^3}$ to $d=2$ are clearly inconsistent with this. We instead propose to identify $Q\bar Q\phi$ with field theory operator $i\phi^4$.\footnote{This is analogous to the identification of
$L_{-3} \bar L_{-3} \phi$ in the 2D Ising model $M(3,4)$ with the operator $\phi^5$ in the massless $\phi^4$ field theory \cite{Gliozzi:2014jsa}.}
Its one-loop scaling dimension can be extracted from the results in \cite{Klebanov:2022syt}, and we find
\be
\Delta_{\phi^4} = 8 + \frac{10}{9}\epsilon + O(\epsilon^2)\ ,
\ee   
Since in $d=2$ it equals $7.6$, its 
dimension dependence cannot be monotonic. This makes the extrapolation challenging, and we need at least the next term in the series
to make quantitative estimates in $d=3,4,5$. We leave this for future work.

The field theory also contains primary operators with an even number of derivatives, which have schematic form
$ \phi \partial_{\mu_1} \partial_{\mu_2} \ldots \partial_{\mu_{2n}} \phi + \ldots$. Projecting onto the traceless part of such an operator, we find the leading twist operators of spin $2n$ (they form the leading Regge trajectory).
For $n=1$ this operator is the conserved stress-energy tensor, but for $n>1$ they are not conserved. The first such operator is the traceless rank $4$ tensor
\be
Q_{\mu\nu\kappa\lambda}= \phi \partial_\mu \partial_\nu \partial_\kappa \partial_\lambda \phi + \ldots
\ee
 In $d=2$, we identify it with 
the spin $4$ quasiprimaries $iQ \phi$ and $i\bar Q \phi$ of $M(2,5)$, which have scaling dimension $\frac{18}{5}$.
The $6-\epsilon$ expansion of the dimension of $Q_{\mu\nu\kappa\lambda}$ is \cite{Giombi:2016hkj, Gopakumar:2016wkt, Dey:2017oim, Goncalves:2018nlv}
\be 
\Delta_{Q} = 8 - \frac{16}{15}\epsilon - \frac{871}{30375}\epsilon^2 
 + O(\epsilon^3)\ .
\ee    
Together with the condition that $\Delta_Q= \frac{18}{5}$ in $d=2$, this leads to the two-sided Pad\' e estimates shown in Fig. \ref{2SPadeYLfig} and Table \ref{2SPadeYLest}.
The higher spin operators on the leading Regge trajectory can be studied similarly. 

The CFT also contains operators of subleading twist. For example, the spin $s$ operators on the first subleading Regge trajectory have schematic form 
$ \phi \partial_{\mu_1} \partial_{\mu_2} \ldots \partial_{\mu_s} \partial^2 \phi + \ldots$. For $s=2$ this operator is
\be
T'_{\mu \nu} = \phi \partial_{\mu} \partial_{\nu} \partial^2 \phi + \ldots \label{TprYL}
\ee 
Its scaling dimension $\Delta_{T'} (\epsilon)= 8 + O(\epsilon)$ in $d=6-\eps$.
This operator likely extrapolates to the 2D operator (\ref{newspintwo}), so that $\Delta_{T'} (4)=9.6$. 
It would be interesting to study this function, and we leave this for future work.\footnote{In the Ising model, the dimension of $T'_{\mu \nu}$ as a function of $d$ was calculated using the numerical
bootstrap methods \cite{Cappelli:2018vir}. It equals $6$ in 4D and 2D, while in 3D it is $\approx 5.5$.}

\begin{table}[h!]
\centering
\begin{tabular}{c |c|c|c|c|c}
\hline
\hline
Operators  & Exact $\Delta$  & \multicolumn{3}{c|}{Two-sided Pad\'e for $\Delta$} & GL   \\
\cline{3-5}
$d=2$ & $d=2$ & $d=3$ & $d=4$ & $d=5$ &  description \\ 
\hline
\noalign{\vskip 2pt} 
$\phi$ &  $-2/5$ & $0.218_{[3,3]},\; 0.218_{[4,2]}$ &  $0.827_{[3,3]},\; 0.827_{[4,2]}$  & $1.425_{[3,3]},\; 1.425_{[4,2]}$  & $\phi$ \\
\noalign{\vskip 2pt} 
\hline
\noalign{\vskip 2pt} 
$T\bar{T}$ & $4$ & $4.631_{[3,3]},\; 4.639_{[4,2]}$ & $5.206_{[3,3]},\; 5.212_{[4,2]}$  & $5.701_{[3,3]},\; 5.702_{[4,2]}$  & $i\phi^3$ \\
\noalign{\vskip 2pt} 
\hline
\noalign{\vskip 2pt} 
$i Q \phi, i\bar{Q} \phi$ & $18/5$ &  $4.681_{[1,2]}, \; 4.709_{[2,1]}$ &  $5.793_{[1,2]}, \; 5.815_{[2,1]}$ & $6.910_{[1,2]}, \; 6.916_{[2,1]}$   & 
$Q_{\mu\nu\kappa\lambda}$ \\
\noalign{\vskip 2pt} 
\hline
\hline
\end{tabular}
\caption{Two-sided Pad\'e approximations of the scaling dimensions of lowest primary operators 
in $d=3, 4, 5$. All of them are below the unitarity bounds.}
\label{2SPadeYLest}
\end{table}

Our Pad\'e extrapolations can be compared with the bootstrap results of \cite{Gliozzi:2013ysa,Gliozzi:2014jsa,Hikami:2017hwv} and the high-temperature expansions \cite{Butera:2012tq}. Some of them are in good agreement with the estimates in this section.
For example, in $d=3$ the estimate $\Delta_\phi^{\rm boot}\approx 0.213$ from the original paper \cite{Gliozzi:2013ysa} is close to our value $\Delta_\phi\approx 0.218$ and to the high-temperature expansion result $0.214(6)$. However, in the later papers \cite{Gliozzi:2014jsa,Hikami:2017hwv}
the bootstrap estimates of $\Delta_\phi$ were revised to 
$0.235$ and $0.174$ respectively. These different estimates point to the significant uncertainties in the bootstrap procedure, which are related to the truncation \cite{Poland:2018epd}.
Also, some results for the higher operators are further apart. In \cite{Gliozzi:2014jsa} the estimate of $\Delta_{\phi^3}$ is $\approx 5.0$ in $d=3$ vs. our $\approx 4.63$. In $d=4$, 
\cite{Gliozzi:2014jsa}
 obtained $\approx 6.8$ vs. our $\approx 5.2$. We will find better agreement with the Pad\'e extrapolations from our numerical results on the regularized $S^2$ and $S^3$.

\section{Spin chain for 2D Yang-Lee model}
\label{YLchain}
The Hamiltonian of the 2D Yang-Lee quantum model on a chain with $N$ sites and periodic boundary conditions (PBC) is \cite{Uzelac1981, vonGehlen:1991zlm, vonGehlen:1994rp}
\begin{align}
H_{\textrm{YL}} = - J\sum_{j=1}^{N}Z_{j}Z_{j+1} - h_{x}\sum_{j=1}^{N}X_{j} - ih_{z} \sum_{j=1}^{N}Z_{j}\,, \label{2DYL_ham}
\end{align}
where $X$ and $Z$ are the Pauli matrices and  we assume that $Z_{N+1}\equiv Z_{1}$. 
The Hamiltonian has the $\mathcal{PT}$ symmetry generated by the operator \cite{Castro-Alvaredo:2009xex}
\begin{align}
\mathcal{PT} =   \prod_{j=1}^{N}X_{j} \,\mathcal{T} \,, 
\end{align}
where $\mathcal{T}$ is complex conjugation. The $\mathcal{PT}$ symmetry acts by $Z_j\rightarrow - Z_j$, $i\rightarrow - i$.

We set $J = 1$ and
assume that  the system is in paramagnetic (disordered) phase, so the transverse magnetic field $h_{x}> h_{x}^{\textrm{crit}}$, where 
$h_{x}^{\textrm{crit}}/J =1$  is the critical value for the 2D Ising quantum model.  It corresponds to the $M(3,4)$ model, whose Kac table is well known to be
\begin{table}[h!]
  \centering
  \begin{tabular}{@{} |c|c|c| @{}}
  \hline
    $\Phi_{2,1}$ &     $\Phi_{2,2}$ &     $\Phi_{2,3}$  \\ 
    \hline
        $\Phi_{1,1}$ &     $\Phi_{1,2}$ &     $\Phi_{1,3}$  \\ 
    \hline
  \end{tabular}, \qquad 
  \begin{tabular}{@{} |c|c|c| @{}}
  \hline
    $\varepsilon$ & $\sigma $ & I   \\ 
      \hline
    I & $\sigma $ & $\varepsilon $  \\ 
    \hline
  \end{tabular}
  \caption{$M(3,4)$ model with the central charge $c=1/2$.}
  \label{tab1}
\end{table}

\begin{figure}[h!]
\center
\includegraphics[width=0.5\textwidth]{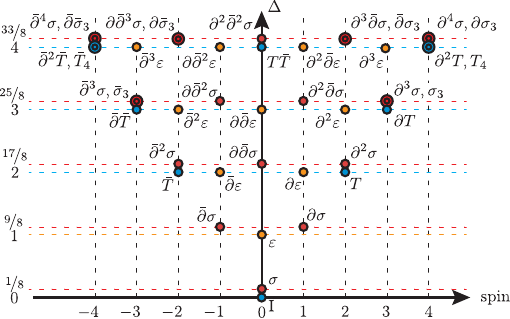}
\caption{
The lowest dimension operators of $M(3,4)$.}
\label{2DIsing_states}
\end{figure}

The Yang-Lee critical point is reached by tuning the imaginary magnetic field $ih_{z}$ to the critical point $ih_{z}^{\textrm{crit}}$. At this critical point, some pairs of energy levels merge, and beyond it they become complex conjugates of each other. Since the Hamiltonian (\ref{2DYL_ham}) is symmetric, 
$H_{\textrm{YL}}^T= H_{\textrm{YL}}$, it is pseudo-Hermitian: $H_{\textrm{YL}} = \mathcal{P}H_{\textrm{YL}}^{\dag}\mathcal{P}$. In this case, 
each eigenstate can be assigned an appropriately defined eigenvalue of $\mathcal{PT}$, which is either $1$ or $-1$;
it is also known as the Krein signature \cite{Nixon2016, Chernyavsky2018}.
 For $h_{z}=0$, the eigenvalue of $\mathcal{PT}$ coincides with the $\mathbbm{Z}_{2}$ eigenvalue of the $\mathcal{P}$ operator.
For two states to merge, they must carry opposite eigenvalues of $\mathcal{PT}$ \cite{Nixon2016, Chernyavsky2018, Starkov:2023hox}.

\begin{figure}[h!]
\center
\includegraphics[width=0.6\textwidth]{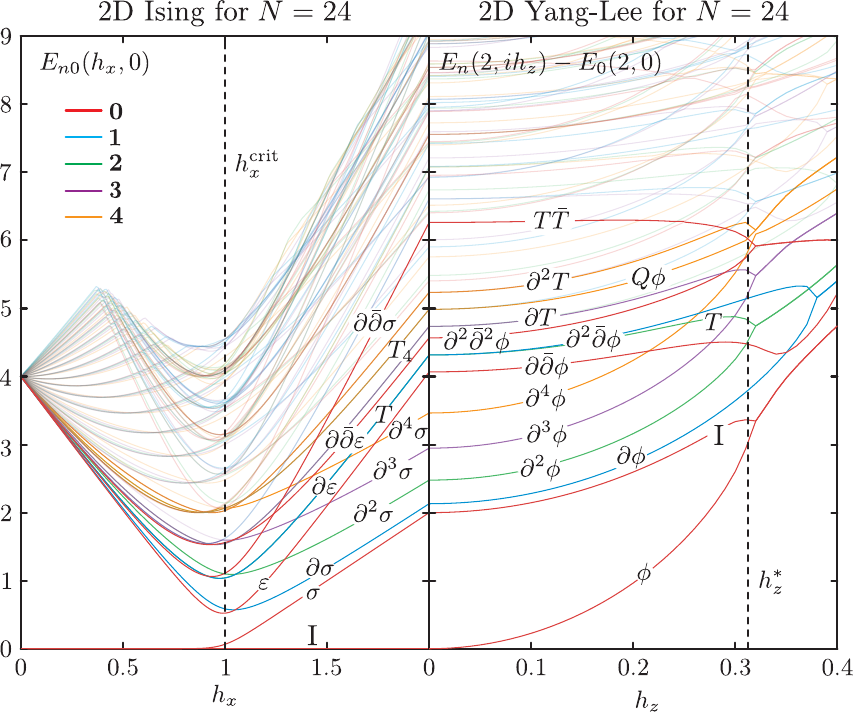}
\caption{Energy 
levels of the Hamiltonian (\ref{2DYL_ham}) for $N=24$ along the path $(h_{x},ih_{z})=(0,0)\to(2,0)\to(2,i0.4)$, with $J=1$. Beyond a merger point, two energy levels become complex conjugate, and we plot their real part (sometimes another transition is visible where the two energies become real again). We identify the lower energy levels with their corresponding operators in the 2D Ising and Yang-Lee CFTs. The first dashed lines is the critical point $h_{x}^{\textrm{crit}}=1$ for the 2D Ising model and the second dashed line is the pseudo-critical point $ih_{z}^{*}$ for the 2D Yang-Lee model, obtained fixing $r_{T}=2$.}
\label{2DIsingYL_spectrum}
\end{figure}

\begin{figure}[h!]
  \centering
  \includegraphics[width=0.5\linewidth]{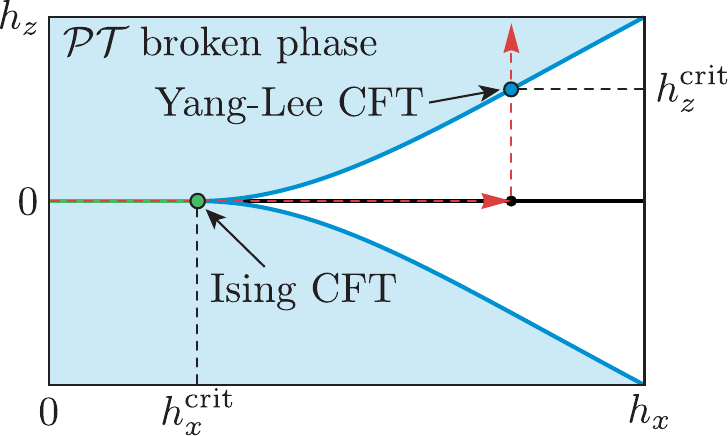}
  \caption{\label{fig:3dHypoPhaseDigm}
    The qualitative phase diagram of the quantum Ising model with transverse magnetic field $h_x$ and imaginary longitudinal magnetic field $h_z$. The $h_z = 0$ horizontal line is the $\mathbb Z_2$ symmetric line. The Ising critical point, which is located at this line, is shown by the green dot. To the right of this point, the Ising model is in the paramagnetic phase, also known as the high temperature phase, shown by the black section of the $\mathbb Z_2$ symmetric line. The paramagnetic phase with a small $h_z$ deformation still preserves $\mathcal{PT}$ symmetry, shown by the white region. When $h_z$ passes its critical value, the $\mathcal{PT}$ symmetry is spontaneously broken, shown by the blue region. The critical coupling $h_z$ depends on $h_x$ and therefore we have a line of critical points corresponding to the YL criticality, shown as the blue solid line. We study the Ising to YL criticality flow by moving along the red dashed  path.}
\end{figure}

The Hamiltonian (\ref{2DYL_ham})  was studied numerically long ago  \cite{Uzelac1981, vonGehlen:1991zlm, vonGehlen:1994rp}  and also more recently in \cite{Wei:2023ygd, Yamada:2022dka}. Our numerical results, shown in Figure \ref{2DIsingYL_spectrum}, are in agreement with these calculations and extend them. 
For a finite $N$, mergers of different pairs of levels, which are called ``exceptional points" \cite{Heiss:2012dx},  occur at somewhat different values of $h_z$. However, they are expected to approach the same critical value as $N\rightarrow \infty$.
The qualitative phase structure in the thermodynamic limit is shown in Fig. \ref{fig:3dHypoPhaseDigm}; we will see that it applies also in higher dimensions.

Figure \ref{2DIsingYL_spectrum} also shows that, as $h_z$ is increased beyond the merger point, there may be another transition, after which the energy levels become real again. Two energy levels becoming complex conjugates in a small region of parameter space is a common phenomenon in
$\mathcal{PT}$-symmetric systems; it occurs due to a small imaginary off-diagonal element in a non-Hermitian Hamiltonian \cite{Heiss:2012dx}. If the off-diagonal element is very small, the evolution of the two energy levels resembles their crossing. For the pseudo-Hermitian Hamiltonians, such as (\ref{2DYL_ham}), a pair of energy levels can ``cross" only if they have opposite eigenvalues of $\mathcal{PT}$ (Krein signatures) \cite{Nixon2016, Chernyavsky2018, Starkov:2023hox}.
For a Hermitian Hamiltonian, a small off-diagonal element instead produces an avoided crossing of two energy levels.

In search of the YL critical point $ih_{z}^{\textrm{crit}}$, we may use different criticality criteria that help us define pseudo-critical points $ih_{z}^{*}(N)$ for given $N$ and $h_{x}$, such that when taking the thermodynamic limit of any of them, we must find
\begin{align}
\label{eq_thermo_crit}
    \lim_{N \to \infty} ih_{z}^{*}(N,h_{x}) = ih_{z}^{\textrm{crit}}(h_{x}).
\end{align}
There are various examples of these criteria: Finite-Size-Scaling (FSS) \cite{CJHamer_1980, CJHamer_1981, CJHamer_1981_2}, extrapolation of merging points \cite{von1991critical, gehlen1994non}, CFT ratios of energy gaps \cite{ArguelloCruz:2024xzi}, etc. We define the pseudo-critical point as the point at which the following  criterion is satisfied:
\begin{align}
\label{eq_ratio_2d}
	r_{T}=\frac{E_{T}-E_{\textrm{I}}}{E_{\partial\phi} - E_{\phi}}=2\,.
\end{align}
In the numerical spectrum, this ratio can be written using the energy levels of distinct spin sectors, namely, $E^{(\textbf{s})}_{n}$ with $n=0,1,2,\dots$ represents the $n$th energy level in spin sector $\textbf{s}=0,1,2,\dots$, so
\begin{align}
	r_{T}=\frac{E^{(\textbf{2})}_{1}-E^{(\textbf{0})}_{1}}{E^{(\textbf{1})}_{0} - E^{(\textbf{0})}_{0}}.
\end{align}
Using the values of the pseudo-critical points, at fixed $N$, we can calculate the scaling dimensions of operators using 
\begin{align}
\label{eq_op_2d}
	\Delta_{\mathcal{O}}= 2 \frac{E^{(\textbf{s})}_{\mathcal{O}}-E^{(\textbf{0})}_{1}}{E^{(\textbf{2})}_{1}-E^{(\textbf{0})}_{1}}.
\end{align}
First, in Table \ref{Table_2d_crit_points}, we list the position of the merging points of the first two states using $N=24$ sites. 

\begin{table}[h!]
\centering
\begin{tabular}{ l  c c c c c }
\hline
\hline
$h_{x}$  & $2$ & $3$ & $4$  & $5$ & $10$ \\
\hline
\noalign{\vskip 2pt} 
$h_{z}^{\textrm{merg}}$  & 0.314574 & 0.885528 &  1.560752 & 2.294549 & 6.368703 \\
\hline
\end{tabular}
\caption{Results for the merging point $ih_{z}^{\textrm{merg}}$ of the Yang-Lee model on a circular chain for $N=24$.}
\label{Table_2d_crit_points}
\end{table}

Using \eqref{eq_op_2d}, we list the values of the anomalous dimensions for $\mathcal{O}=\phi,\; T\Bar{T},\; iQ\phi$ as well as the position of the pseudo-critical points, for the case of $N=24$, in Table \ref{Table_2d_scaling_dim}. We see that the pseudo-critical points and the merging points are very close to each other.

\begin{table}[h!]
\centering
\begin{tabular}{l c c c c c c c }
\hline
\hline
$h_{x}$  & $2$ & $3$ & $4$  & $5$ & $10$ & $\textrm{Exact}$ \\
\hline
$h_{z}^{*}$  & 0.312706 & 0.883574 & 1.558738 & 2.292472 & 6.366335 & $ - $ \\

$\Delta_{\phi}$  & $-0.375487$ & $-0.3948$ &  $-0.400085$ & $-0.402508$ & $-0.406277$ & $-0.4$ \\

$\Delta_{T\Bar{T}}$ & 3.615627 & 3.829043 & 3.890879 & 3.919301 & 3.962861 & 4 \\

$\Delta_{Q\phi}$  & 3.612873 & 3.552427 & 3.533772 & 3.524567 & 3.50883 & 3.6 \\
\hline
\hline
\end{tabular}
\caption{Results for the pseudo-critical points $ih_{z}^{*}$, defined where $r_{T}=2$, and the anomalous dimensions of the operators $\phi$, $T\Bar{T}$, and $iQ\phi$ for various values of transverse field $h_{x}$, using $N=24$.}
\label{Table_2d_scaling_dim}
\end{table}

\section{3D model on Platonic solids}
\label{platonicsection}

\begin{figure}[h!]
\center
\includegraphics[width=0.6\textwidth]{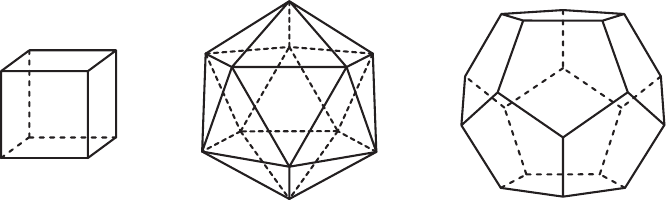}
\caption{3D Platonic solids $(N, E)$: Cube$(8, 12)$,  Icosahedron$(12, 30)$,  Dodecahedron$(20, 30)$. }
\label{3Dsolids_pic}
\end{figure}

In this section we consider the Ising and Yang-Lee models on three 3D Platonic solids \cite{Lao:2023zis}: Cube, Icosahedron and Dodecahedron, which are depicted in Figure \ref{3Dsolids_pic}. They have correspondingly $N=8$, $N=12$ and $N=20$ sites. Apart from them there are also another two 3D Platonic solids: Tetrahedron and Octahedron, which have only $N=4$ and $N=6$ sites and we don't consider them here.

We have for the  schematic Yang-Lee Hamiltonian on Cube, Icosahedron and Dodecahedron:
\begin{align}
H_{\textrm{YL}} = - J \sum_{\langle i j\rangle \in e}Z_{i}Z_{j} - h_{x}\sum_{i \in v}X_{i} - ih_{z}\sum_{i \in v}Z_{i}\,, \label{YL3Dplat}
\end{align}
where the first sum is taken over all the edges of a 3D solid and the last two sums go over the vertices.

These three Platonic solids have different ratio of number of vertices $N$ and edges $E$, therefore if we would like to compare  their properties between each other we have to adjust couplings $J$ or $h_{x}$ accordingly. We keep $h_{x}$ the same for all these Platonic solids and adjust only the coupling $J$. We set $J_{\textrm{dodecahedron}} = 1$ and adjust other solids' $J$ couplings assuming that 
$J E/N = \textrm{const.}$.
Therefore we find  $J_{\textrm{cube}} =  1$ and  $J_{\textrm{icosahedron}} =  3/5$. In the Table \ref{Merg3Dsolids} we list values for the merger point $ i h_{z}^{\textrm{merg}}$ of two lowest states: $|\textrm{I}\rangle$ and $|\phi\rangle$, which correspond to the lowest energy levels $E_{0}^{(\textbf{1})}$ and $E_{1}^{(\textbf{1})}$ in the singlet representation $\textbf{1}$ of  3D symmetry point groups. Below we use notation $E_{n}^{(\textbf{r})}$ for the energy level $n$ in irrep $\textbf{r}$ of the point group.
\begin{table}[h]
\centering
\begin{tabular}{l c c c c c c c c c c c }
\hline
\hline
$h_{x}$  & $5$ & $10$ & $15$ & $20$  & $25$ & $30$ & $35$ & $40$ & $45$ & $50$\\
\hline
\noalign{\vskip 2pt} 
$\textrm{C:}\;   h_{z}^{\textrm{merg}}$  &1.513 & 5.173& 9.283 &  13.591& 18.016  &22.517  & 27.074  &31.674 & 36.307 &40.968\\
\noalign{\vskip 2pt} 
$\textrm{I:}\;\;\,  h_{z}^{\textrm{merg}}$ & 1.339 & 4.926& 8.991 & 13.265 & 17.661 &  22.138 & 26.673 &31.253& 35.868  &40.513 \\
\noalign{\vskip 2pt} 
$\textrm{D:}\;  h_{z}^{\textrm{merg}}$ & 1.387 & 5.020 & 9.112 &13.406 & 17.817 &  22.308& 26.855  &31.438 & 36.071 &40.723 \\
\hline
\hline
\end{tabular}
\caption{Results for the merger point $ih_{z}^{\textrm{merg}}$ of the Yang-Lee model on Cube, Icosahedron and Dodecahedron.}
\label{Merg3Dsolids}
\end{table}

The other criticality criteria for pseudo-critical point $i h_{z}^{*}$ corresponds to fixing the ratio
\begin{align}
r_{T} \equiv \frac{E_{T}-E_{\textrm{I}}}{E_{\partial \phi} - E_{\phi}}
\end{align}
to be exactly $r_{T} = 3$. Numerically this ratio can be  
obtained using the following energy levels in different irreps of $O$ and $I$:
\begin{align}
\textrm{Cube}: \quad r_{T} = \frac{E_{1}^{(\textbf{3}')}-E_{0}^{(\textbf{1})}}{E_{0}^{(\textbf{3})}-E_{1}^{(\textbf{1})}}, \qquad \textrm{Icosahedron/Dodecahedron}:\quad r_{T} = \frac{E_{1}^{(\textbf{5})}-E_{0}^{(\textbf{1})}}{E_{0}^{(\textbf{3})}-E_{1}^{(\textbf{1})}}\,,
\end{align}
where we used that $\textbf{5}_{O(3)}$ splits as $\textbf{2}+\textbf{3}'$ under the cubic  symmetry group $O$, whereas it does not split for Icosahedral group $I$, and $\textbf{1}_{O(3)}$ and $\textbf{3}_{O(3)}$ do not split for $O$ and $I$ and corresponds to irrep $\textbf{1}$ and $\textbf{3}$.
Here we list values $ih_{z}^{*}$, where $r_{T} = 3$ in the Table \ref{Table_rThz}. They are slightly away from the merger points listed in the previous Table \ref{Merg3Dsolids}. 
\begin{table}[h]
\centering
\begin{tabular}{ l c c c c c c c c c c }
\hline
\hline
$h_{x}$  & $5$ & $10$ & $15$ & $20$  & $25$ & $30$ & $35$ & $40$ & $45$ & $50$\\
\hline
$\textrm{C:}\;   h_{z}^{*}$ & 1.467 & 5.140& 9.252 &  13.560& 17.984  &22.485 & 27.042  &31.641 & 36.274  &40.934\\

$\textrm{I:}\;\;\,   h_{z}^{*}$ & 1.312 & 4.907& 8.973 & 13.247 & 17.643 &  22.120 & 26.655 &31.234& 35.849  &40.493 \\

$\textrm{D:}\;   h_{z}^{*}$ &1.381 & 5.015 & 9.107 & 13.41 & 17.813 &  22.303 & 26.850  &31.441 & 36.065 & 40.718 \\
\hline
\hline
\end{tabular}
\caption{Results for the Yang-Lee pseudo-critical point $ih_{z}^{*}$  defined by the criteria $r_{T}=3$.}
\label{Table_rThz}
\end{table}
We can compute anomalous dimension $\Delta_{\phi}$  as 
 \begin{align}
\textrm{Cube}: \quad \Delta_{\phi} = 3\frac{E_{1}^{(\textbf{1})}-E_{0}^{(\textbf{1})}}{E_{1}^{(\textbf{3}')}-E_{0}^{(\textbf{1})}}\,, \qquad \textrm{Icosahedron/Dodecahedron}: \quad \Delta_{\phi} = 3\frac{E_{1}^{(\textbf{1})}-E_{0}^{(\textbf{1})}}{E_{1}^{(\textbf{5})}-E_{0}^{(\textbf{1})}}\,, 
\end{align}

We list values for $\Delta_{\phi}$ computed at the  pseudo-critical point $ih_{z}^{*}$  where $r_{T}=3$ in the Table \ref{Table_rTDphi}.
\begin{table}[h]
\centering
\begin{tabular}{l c c c c c c c c c c c}
\hline
\hline
$h_{x}$ & $5$ & $10$ & $15$ & $20$  & $25$ & $30$ & $35$ & $40$ & $45$ & $50$\\
\hline
$\textrm{C:}\; \Delta_{\phi}$  & 0.424 & 0.346& 0.323 &  0.312& 0.305  &0.300  & 0.297  &0.294 & 0.292  &0.290\\

$\textrm{I:}\;\;  \Delta_{\phi}$ & 0.385 & 0.306& 0.283 & 0.272 & 0.265 &  0.260 & 0.257 &0.254& 0.252  &0.250 \\

$\textrm{D:}\;  \Delta_{\phi}$ & 0.289 & 0.244 & 0.233 &0.228 & 0.224 &  0.222& 0.221  &0.220 & 0.219 & 0.218 \\
\hline
\hline
\end{tabular}
\caption{Results for $\Delta_{\phi}$ at the pseudo-critcal points $ih_{z}^{*}$ where $r_{T}=3$.}
\label{Table_rTDphi}
\end{table}

\begin{figure}[h!]
\center
\includegraphics[width=0.496\textwidth]{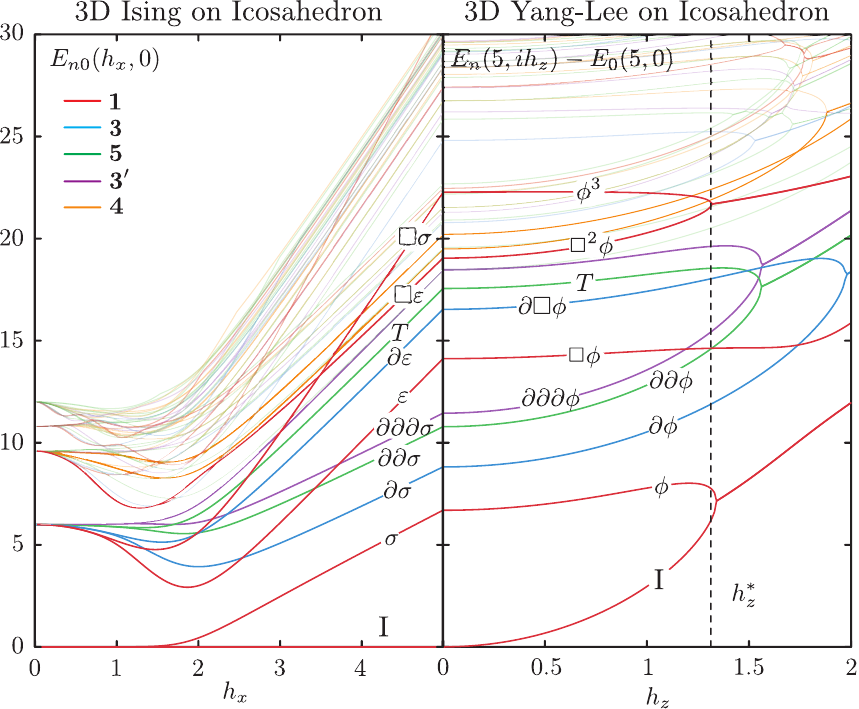}
\includegraphics[width=0.496\textwidth]{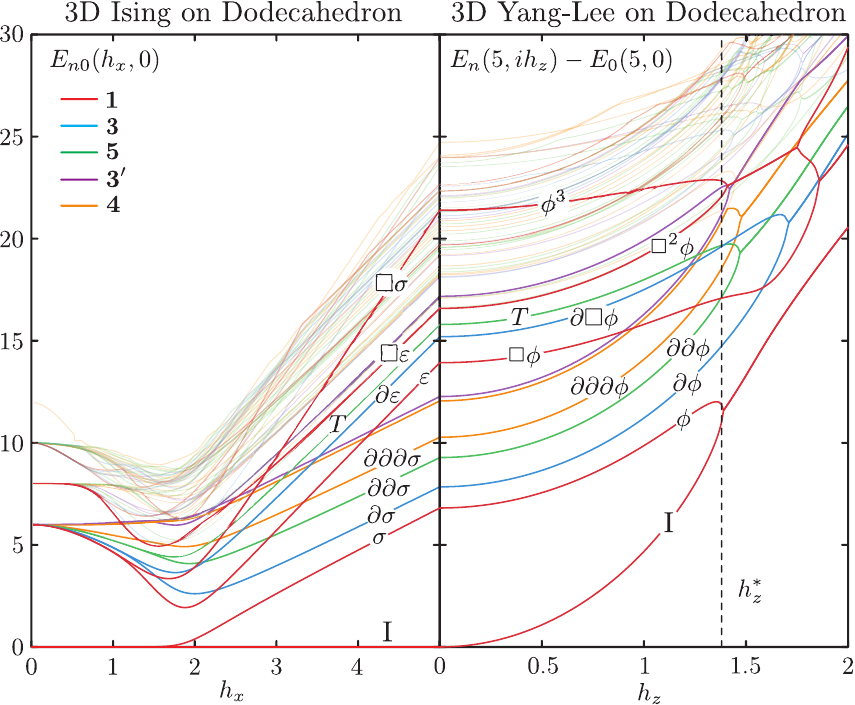}
\caption{Energy gaps of the Hamiltonian (\ref{YL3Dplat}) for Icosahedron and Dodecahedron along the path $(h_{x},ih_{z})=(0,0)\to(5,0)\to(5,i2)$, where $J=3/5$ for Icosahedron and $J=1$ for Dodecahedron. We identified the lower energy levels with their corresponding operators in the Ising and Yang-Lee CFTs. The dashed line is the pseudo-critical point $ih_{z}^{*}$ obtained fixing $r_{T}=3$.}
\label{IcoDode_spectrum}
\end{figure}

The results of this section show that the Platonic solid regulators of the sphere give a good indication of the YL quantum criticality, in spite of the fact that there are only 3 possible numbers of lattice sites. However, the determination of scaling dimensions is not as precise as in the fuzzy sphere method that we will use in the next section. 
In order to improve the Platonic solid approach to YL criticality, one can use the conformal perturbation theory, which was shown to work well for the Platonic solid approach to the 3D Ising model \cite{Lao:2023zis, Lauchli:2025fii}. We leave this for future work.

\section{3D model on fuzzy sphere}
\label{fuzzysection}

\subsection{Fuzzy sphere approach to Yang-Lee quantum criticality}

In this section we use the fuzzy sphere regulator for $S^2$.
This is a novel setup using the Quantum Hall system. It is analogous to Quantum Spin systems but preserves the SO(3) symmetry. 
The model describes non-relativistic fermions projected to the lowest Landau Level (LLL) with quadratic and quartic couplings\cite{Zhu:2022gjc,Fardelli:2024qla}
\begin{equation}\label{Hamiltonian-LLL}
\begin{aligned}
    H &= H_4 + H_2 \\
    H_4 &= R^2\int d^2\Omega  \left[\lambda_n (\psi^\dagger\psi) U (\psi^\dagger\psi)
              - \lambda_{n,z} (\psi^\dagger \sigma_z \psi) U (\psi^\dagger \sigma_z \psi) \right]\\
    H_2 &= - R^2\int d^2\Omega  \left[ h_x (\psi^\dagger \sigma_x \psi) +i h_z (\psi^\dagger \sigma_z \psi) \right]
\end{aligned}
\end{equation}
where $U = \sum_n a_n \nabla^{2n}$ is a contact potential that can be written as a series of derivative operators. 
We follow the same setup as \cite{Zhu:2022gjc,Fardelli:2024qla} and set $\lambda_n = \lambda_{n,z}$. 
We can expand the above Hamiltonian using the fermion creation and annihilation operators. The current-current 4-body interaction term is \cite{Zhu:2022gjc}
\begin{equation}\label{Hamiltonian-quartic-terms}
\begin{aligned}
  H_4 = \frac{1}{2} \sum_{m_{1,2,3,4}=-s}^{s} &V_{m_1,m_2,m_3,m_4} \delta_{m_1+m_2,m_3+m_4} \\ 
  &\times\bigg[ 
      (\mathbf{c}_{m_1}^{\dagger} \mathbf{c}_{m_4}) (\mathbf{c}_{m_2}^{\dagger} \mathbf{c}_{m_3})
      - (\mathbf{c}_{m_1}^{\dagger} \sigma^z \mathbf{c}_{m_4}) (\mathbf{c}_{m_2}^{\dagger} \sigma^z \mathbf{c}_{m_3})
  \bigg]
\end{aligned}
\end{equation}
where $\mathbf{c}_{m_1} = \begin{pmatrix} c_{m,\uparrow} \\ c_{m,\downarrow} \end{pmatrix}$. The coefficient tensor
\begin{equation}
  V_{m_1,m_2,m_3,m_4} = \sum_l V_l (4s-2l+1) \begin{pmatrix} s & s & 2s-l \\ m_1 & m_2 & -m_1-m_2 \end{pmatrix} \begin{pmatrix} s & s & 2s-l \\ m_4 & m_3 & -m_3-m_4 \end{pmatrix}
\end{equation}
is parametrized by the Haldane pseudo-potentials $V_{l}$. In our work we truncate the potential to $V_0$ and $V_1$, which is equivalent to the derivative expansion truncated to order $\nabla^2$. The overall normalization is fixed with $V_1 = 1$ and $V_0$ is a tunable parameter of the model.
The LLL-projected Hamiltonian has $(4s+2)$ levels. We fill the total number of $N = 2s+1$ fermions in these levels. Under half-filling the charge degrees of freedom are gapped, and also the particle-hole symmetry is preserved. The total number of spins $N$ can be treated as a UV cutoff, which can be most clearly seen from the fact that the commutator $\{\psi(\Omega),\psi(\Omega')\}$ approaches a delta function of size $\sqrt{1/N}$. This allows us to identify the radius of the sphere as $R \sim \sqrt{N}$.

The density terms $H_2$ can be written as 
\begin{equation}\label{Hamiltonian-quadratic-terms}
\begin{aligned}
  H_2 &= - h_x H_X - i h_z H_Z \\ 
  H_X &= \sum_{m=-s}^{s} \mathbf{c}_m^{\dagger} \sigma^x \mathbf{c}_m\\
  H_Z &= \sum_{m=-s}^{s} \mathbf{c}_m^{\dagger} \sigma^z \mathbf{c}_m
\end{aligned}
\end{equation}
The 2-body terms are chosen so that the Hamiltonian mimics the quantum spin model setup in previous sections with the identification
\begin{equation}\label{YL3D-Hamiltonian}
\boxed{
  H_{\textrm{YL}}(V_0,h_x, i h_z) = H_{\rm Ising}(V_0, h_x) - ih_z H_{Z} 
} 
\end{equation}
where the Ising part of the Hamiltonian
\begin{equation}
  H_{\rm Ising} = H_4(V_0) - h_x H_X
\end{equation}
follows the original setup of \cite{Zhu:2022gjc}. The fuzzy sphere Ising Hamiltonian is analogous to the lattice quantum Ising model that has a $ZZ$ nearest coupling term and a transverse field $X$. Apart from the SO(3) symmetry, the Ising Hamiltonian has a spin-flip $\mathbb{Z}_2$ symmetry that takes $\mathbf{c}_{m} \mapsto \sigma_x \mathbf{c}_{m}$, but this symmetry is explicitly broken by the $H_Z$ horizontal field term. This term also breaks Hermiticity. The Yang-Lee Hamiltonian is still invariant under a combination of spin-flip and complex conjugation, which serves as the $\mathcal{PT}$ symmetry. The eigenvalues of the $\mathcal{PT}$ symmetric Hamiltonian are real numbers or pairs of conjugate complex numbers.
Under half-filling the system also has a particle-hole symmetry $\mathbf{c}_{m} \mapsto i \sigma_y \mathbf{c}_{m}, i\rightarrow -i$, which can be interpreted as the spacial parity symmetry in the thermodynamic limit.

\begin{figure}[h!]
  \centering
  \includegraphics[width=0.6\linewidth]{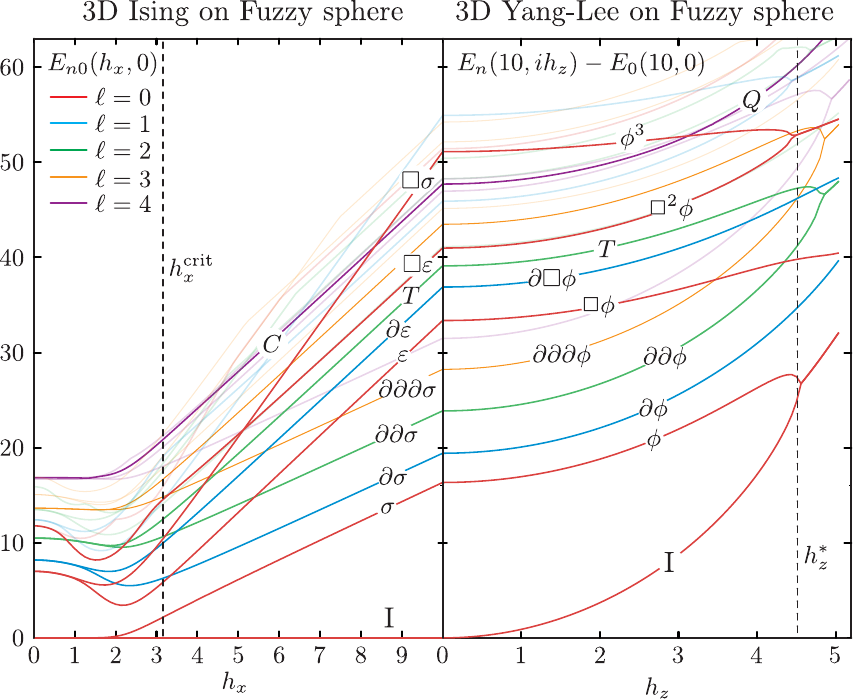}
  \caption{\label{fig:3dIsing_to_YL}
  Energy gaps of the fuzzy sphere Hamiltonian (\ref{YL3D-Hamiltonian}) along the path $(h_{x},ih_{z})=(0,0)\to(10,0)\to(10,i5.2)$, where  $N=12$, $V_{0}=4.75$. We identified the lower energy levels with their corresponding operators in the Ising and Yang-Lee CFTs. The dashed lines are the Ising critical point $h_{x}^{\textrm{crit}}\approx 3.16$ and the Yang-Lee pseudo-critical point  $ih^{*}_{z}\approx 4.5$ obtained fixing $r_{T}=3$.
We notice that operators with the same spin that merge near  $ih_{z}^{*}$ always come from different $\mathbb{Z}_2$ Ising sectors, hence they have opposite
eigenvalues of $\mathcal{PT}$.
}
\end{figure}

We expect the phase diagram of the Hamiltonian (\ref{YL3D-Hamiltonian}) to also be Fig. \ref{fig:3dHypoPhaseDigm}. Similar to the 2D case, there is a phase transition when we take the Ising model in the paramagnetic phase and turn on an imaginary longitudinal magnetic field $h_z$. To show the qualitative behavior of the phase transition, it is helpful to identify the correspondence of low-lying eigenstates between Ising criticality and YL criticality. We scan coupling constants $h_x$ and $ih_z$ along the red-dashed  trajectory in Fig. \ref{fig:3dHypoPhaseDigm} and keep track of a number of eigenvalues. The Ising criticality is reported in \cite{Zhu:2022gjc} to occur at couplings $V_0 = 4.75$ and $h_x = 3.16$. We hold $V_0$ fixed throughout the trajectory. We gradually increase $h_x$ from $0$ to $10$, then increase $h_z$ from $0$ to $h_z \approx 5.2$. The spectrum along this trajectory is shown in Fig. \ref{fig:3dIsing_to_YL}, we can identify the tower of operators that belong to the 3D YL universality class.

A summary of the operator flow is presented in Table \ref{Table_Op_flow} as well as the first low-lying merging Yang-Lee states. We highlighted in blue our conjecture about flows between the higher primary operators. We note that all flows between these low-lying operators in both CFTs admit a clear interpretation in the GL description. For example, both $\sigma$ and $\phi$ correspond to the $\phi$ field in the $\phi^4$ and $i\phi^3$ models. The operator $\varepsilon$ corresponds to $\phi^2$ in the $\phi^4$ model, and $\phi^2$ coincides with $\Box\phi$ in the $i\phi^3$ model due to the equations of motion. Likewise, the operator $\Box\sigma$ coincides with $\phi^3$ in the $\phi^4$ model for the same reason, and appears as the $\phi^3$ operator in the $i\phi^3$ model. Based on this, we conjecture that $\varepsilon'$, which corresponds to $\phi^4$ in the $\phi^4$ model, also represents $\phi^4$ in the $i\phi^3$ model. A similar approach applies to the other primary operators and their descendants.

 Finally we notice that operators from different sectors of the Ising-$\mathbb{Z}_2$ symmetry and equal spin on the left ($ih_z = 0$) merge on the Yang-Lee side on the right ($ih_z \neq 0$), where such symmetry is explicitly broken, e.g. $\textrm{I}$  merges with $\sigma$ and $\Box \varepsilon$ merges with $\Box \sigma$ at some $ih_{z}$ value.
This is consistent with the general result, reviewed in section \ref{YLchain}, that a pair of levels can merge only if they have opposite eigenvalues of $\mathcal{PT}$.

\begin{table}[h!]
  \centering
  \begin{tabular}[t]{c||c}
      Operator flow from 3D Ising CFT to 3D YL CFT & Merging states in 3D YL \\
      \hline\hline
  \begin{tabular}[t]{c|c}
  \hline
  \hline
    Ising ($\ell=0$)  & YL ($\ell=0$) \\ 
    \hline
    $\textrm{I}$ & $\textrm{I}$ \\
    \hline
    $\sigma$ & $\phi$ \\
    \hline
    $\varepsilon$ & $\Box\phi$ \\
    \hline
    \noalign{\vskip 1pt} 
    $\Box\varepsilon$ & $\Box^2 \phi$ \\
    \hline
     \noalign{\vskip 1pt} 
    $\Box\sigma$ & $\phi^3$ \\
    \hline
    \noalign{\vskip 1pt} 
    \textcolor{blue}{$\varepsilon'$} & \textcolor{blue}{$\phi^4$} \\
    \hline
    \noalign{\vskip 10pt} 
    \hline\hline
    Ising ($\ell=  1$)  & YL ($\ell= 1$) \\ 
    \hline
    $\partial_{\mu} \sigma$ & $\partial_{\mu}\phi$ \\
    \hline
    $\partial_{\mu} \varepsilon $ & $ \partial_{\mu}\Box \phi$ \\
    \hline
    \end{tabular}

   \begin{tabular}[t]{c|c}
   \hline
   \hline
    Ising ($\ell= 2$)  & YL ($\ell= 2$) \\ 
    \hline
    $\partial_{\mu_{1}}\partial_{\mu_{2}} \sigma$ & $\partial_{\mu_{1}}\partial_{\mu_{2}}\phi$ \\
    \hline
    $T_{\mu\nu}$ & $T_{\mu\nu}$ \\
    \hline
    \noalign{\vskip 1pt} 
    \textcolor{blue}{$T'_{\mu\nu}$} &  \textcolor{blue}{ $T'_{\mu\nu}$} (\ref{TprYL}) \\
    \hline
    \noalign{\vskip 10pt} 
    \hline\hline
    Ising ($\ell= 3$)  & YL ($\ell= 3$) \\ 
    \hline
    $\partial_{\mu_{1}}\partial_{\mu_{2}}\partial_{\mu_{3}} \sigma$ & $\partial_{\mu_{1}}\partial_{\mu_{2}}\partial_{\mu_{3}}\phi$ \\
    \hline
    $\partial_{\alpha}T_{\mu\nu}$ & $\partial_{\alpha}T_{\mu\nu}$ \\
    \hline
  \noalign{\vskip 10pt} 
      
    \hline\hline
    Ising ($\ell= 4$)  & YL ($\ell= 4$) \\ 
    \hline
    $\partial_{\mu_{1}}\partial_{\mu_{2}}\partial_{\mu_{3}} \partial_{\mu_{4}} \sigma$ & $\partial_{\mu_{1}}\partial_{\mu_{2}}\partial_{\mu_{3}} \partial_{\mu_{4}} \phi$ \\
    \hline
    $\partial_{\alpha_{1}}\partial_{\alpha_{2}}T_{\mu\nu}$ & $\partial_{\alpha_{1}} \partial_{\alpha_{2}}T_{\mu\nu}$ \\
    \hline
    $C_{\mu_{1}\mu_{2}\mu_{3}\mu_{4}}$ & $Q_{\mu_{1}\mu_{2}\mu_{3}\mu_{4}}$ \\
    \hline
    \end{tabular} &
  
  \begin{tabular}[t]{c|c}
  \hline
  \hline
     Merger 1 & Merger 2  \\ 
    \hline
    $\textrm{I}$ & $\phi$ \\
    \hline
    $\partial_{\mu} \phi$ & $\partial_{\mu}\Box \phi$ \\
    \hline
    \noalign{\vskip 1pt} 
    $\partial_{\mu}\partial_{\nu}  \phi$ & $T_{\mu\nu}$ \\
    \noalign{\vskip 1pt} 
    \hline
    \noalign{\vskip 1pt} 
    $\partial_{\alpha}\partial_{\mu} \partial_{\nu}  \phi$ & $\partial_{\alpha}T_{\mu\nu}$\\
    \noalign{\vskip 1pt} 
    \hline
    \noalign{\vskip 1pt} 
    $\Box^2 \phi$ & $\phi^3$ \\
    \noalign{\vskip 1pt} 
    \hline
    \noalign{\vskip 1pt} 
    $\partial_{\mu_{1}}\partial_{\mu_{2}}\partial_{\mu_{3}} \partial_{\mu_{4}} \phi$ & $ \partial_{\mu_{1}} \partial_{\mu_{2}}T_{\mu_{3}\mu_{4}} $ \\
    \noalign{\vskip 1pt} 
    \hline
  \end{tabular}
  
  \end{tabular}
\caption{\textbf{Left:} Summary of first CFT operators of 3D Ising model and their corresponding trajectories to 3D YL ones in each spin sector, the operator $C_{\mu_{1}\mu_{2}\mu_{3}\mu_{4}}$ is the first $\mathbb{Z}_2$-even, parity-even, spin 4 primary in 3D Ising CFT. We highlighted in blue color our conjecture for flows between the higher primary operators. \textbf{Right:} Low-lying spectrum merging states in 3D Yang-Lee model.}
\label{Table_Op_flow}
\end{table}

\subsection{Approaching the 3D YL critical point}

Among the list of possibilities to define a pseudo-critical point, we follow both the 2D and platonic solids sections, and found that one of the most suitable criterion is through the use of the stress tensor, whose scaling dimension at the critical point is $\Delta_{T_{\mu\nu}}=3$. So, as before, we consider the expression

\begin{align}
\label{eq_CFT_criterion}
    r_{T}=\frac{E_{T}(N)-E_{\textrm{I}}(N)}{E_{\partial \phi}(N) - E_{\phi}(N)} = 3
\end{align}
as our criticality criterion. In Fuzzy Sphere regularization, this corresponds to the ratio of the following levels (same as in the Icosahedron case):

\begin{align}
\label{eq_CFT_criterion_2}
	r_{T} = \frac{E_{1}^{(\textbf{5})}-E_{0}^{(\textbf{1})}}{E_{0}^{(\textbf{3})}-E_{1}^{(\textbf{1})}}=3\,.
\end{align}

\begin{figure}[h!]
  \centering
  \includegraphics[width=0.6\linewidth]{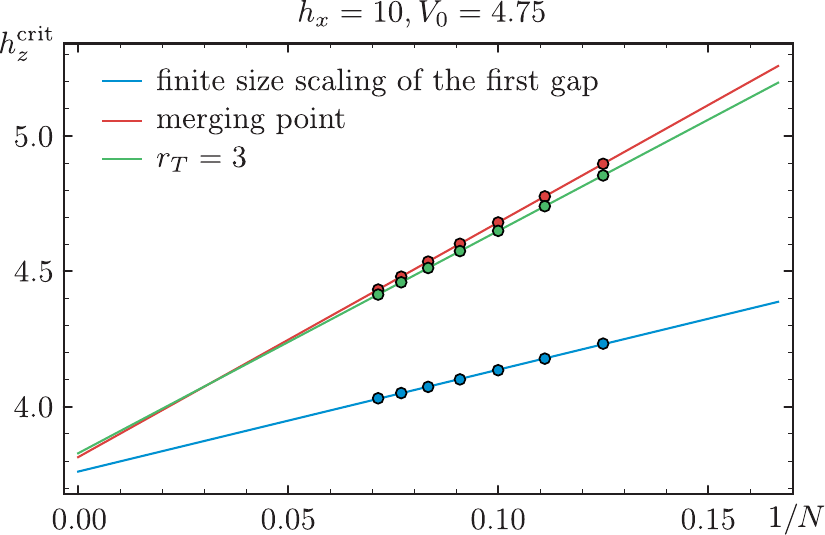}
  \caption{\label{fig:Crit_point_extrapolation} Yang-Lee thermodynamic critical point $h_{z}^{\textrm{crit}}$ extrapolation for $h_{x}=10$ and $V_{0}=4.75$ with $N\in [8,14]$ using different criticality criteria; the extrapolation is done by a linear fit in $1/N$. The results agree within a $10^{-2}$ error.}
\end{figure}

The identification of each state is taken from the operator flow from Ising to Yang-Lee, shown in Fig. \ref{fig:3dIsing_to_YL} and Table \ref{Table_Op_flow}. 
Note that the expression \eqref{eq_CFT_criterion} does not depend on the scaling dimension of $\phi$, which we will find later. From this, we obtain $ih_{z}^{*}(N)$ for $N=8,9,\dots, 14$ and the IR critical point is extrapolated via \eqref{eq_thermo_crit} using $1/N$ linear fit\footnote{ Although  Conformal Perturbation Theory (CPT) predicts the convergence rate  of $i h_{z}^{*}(N)$ to $i h_{z}^{\textrm{crit}}$ at large $N$, we observe that for small values of $N$ the points are fitted well by $1/N$ linear fit. }. In Fig. \ref{fig:Crit_point_extrapolation}, we show an example of thermodynamic critical point extrapolation as a function of $N$ for distinct criticality criteria\footnote{For the FSS criterion, we followed \cite{CJHamer_1980} and defined $h_{z}^{*}(N)$ as the point where 
$$
\sqrt{N+1}E_{10}^{(\textbf{1})}(N+1) = \sqrt{N}E_{10}^{(\textbf{1})}(N),
$$
where we used that $R\propto \sqrt{N}$ \cite{zhou2025fuzzified}.
}, we can see they all converge to the same value within $\sim 1\%$, and we notice that $ih_{z}^{\textrm{crit}}$ is far from its pseudo-critical couterpart for small $N$, in contrast to the 2D case, where $ih_{z}^{\textrm{crit}}$ and $ih_{z}^{*}(N)$ lie within a small perturbative regime \cite{von1991critical}. We present some of the extrapolated critical points for various $h_{x}$ in Table \ref{Table_Crit_points} and we observe that the deeper we go into the paramagnetic phase ($h_{x} > 3.16$), the better the convergence; however, if we take a very big $h_{x}$ then we start facing numerical convergence problems. Therefore, we chose $h_{x}\in [4,100]$.

\begin{table}[h!]
  \centering
  \begin{tabular}{ l c c c c c }
\hline
\hline
     $h_{x}$ & $4$ & $6$ & $8$ & $10$ & $11$\\
      \hline
      \noalign{\vskip 2pt} 
    $h_{z}^{*}$  & $0.465609$ & $1.60018$ & $2.9475$ & $4.41454$ & $5.17874$ \\
    \noalign{\vskip 2pt} 
      $h_{z}^{\textrm{crit}}$  & $0.20(2)$ & $1.16(3)$ & $2.41(4)$ & $3.80(4)$ & $4.53(4)$ \\
     \hline
 \end{tabular}
\caption{Results of pseudo-critical points $h_{z}^{*}$ using $r_{T}=3$ for $N=14$ and thermodynamic critical points using various criticality criteria for distinct values of $h_{x}$ and fixed $V_{0}=4.75$.}
\label{Table_Crit_points}
\end{table}

\subsection{Extracting the CFT data}

The most remarkable feature of the fuzzy sphere model is that it realizes CFTs in radial quantization, where conformal invariant data such as the scaling dimensions and OPE coefficients can be directly extracted from simple observables in the fuzzy sphere setup. In this subsection we discuss the extraction of these data numerically. Throughout this subsection we pick the critical point of the system at fixed $V_0 = 4.75$, $h_x = 20$ and tuning $i h_z $ to the pseudo-critical coupling $i h_{z}^{*}(N)$. The criterion we use for the pseudo-critical point is the CFT ratio of the stress tensor (\ref{eq_CFT_criterion}), in other words, we fix $\Delta_{T^{\mu\nu}}=3$. The rest of the CFT data will have finite size deviation from the CFT values, and we study the extrapolation to thermodynamic limit $N\rightarrow \infty$ numerically. We use a publicly available code package (\textit{FuzzifiED})\cite{zhou2025fuzzified} and compute the eigensystem through exact diagonalization up to $N=18$. Extrapolation to the thermodynamic limit is performed by fitting the data from $N=12$ to $N=18$ as a function of $N$ using models motivated by the finite size scaling properties of CFTs, also known as conformal perturbation theory \cite{CARDY1986186, Zamolodchikov:1987ti, Lao:2023zis, Lauchli:2025fii}. 

CFT systems have many symmetries. In the fuzzy sphere realization, this implies that any conformally invariant quantity can be extracted in many ways. We see remarkable agreement between many distinct ways to evaluate the same quantity, which provides nontrivial validation of our numerical procedure. In this section we will walk through these in details.

\begin{table}[h!]
\centering
   \begin{tabular}{ l c c c | c c c  }
      \hline 
       \hline 
      \multirow{2}{*}{Observable} & \multirow{2}{*}{Fuzzy sphere}  & \multirow{2}{*}{Pad\'e} & Two-sided & 5-loop &  Truncated & High-temperature  \\ 
       &  &  & Pad\'e &  All \cite{Borinsky:2021jdb} & Bootstrap\cite{Gliozzi:2014jsa} & expansion\cite{Butera:2012tq}  \\
      \hline 
      \multirow{3}{*}{$\Delta_{\phi}$} & $0.214(2)_{[E]}$ & \multirow{2}{*}{$0.222_{[3,2]}$} & $0.218_{[3,3]}$ & \multirow{2}{*}{$0.215(10)$}  & $0.235(3)$ & $0.214(6)$ \\ 
       & $0.2155(16)_{[Z]}$ &  & $0.218_{[4,2]}$ &      &  $0.174$\cite{Hikami:2017hwv} &   \\ 
       & $0.2151(8)_{[X]}$ &                   &   &                              &  \\ 
      \hline
      \multirow{2}{*}{$\Delta_{\phi^3}$} & \multirow{2}{*}{$4.613(6)_{[E]}$} & \multirow{2}{*}{$4.766_{[3,2]}$} & $4.631_{[3,3]}$ &  \multirow{2}{*}{$4.5(2)$} &\multirow{2}{*}{$5.0(1)$} & \\  
       & & & $4.639_{[4,2]}$ &  & & \\ 
      \hline 
      \multirow{2}{*}{$\Delta_{Q_{\mu\nu\kappa\lambda}}$} & \multirow{2}{*}{$4.9(1)_{[E]}$} & \multirow{2}{*}{$4.519_{[1,1]}$} & $4.681_{[1,2]}$ &   & \multirow{2}{*}{$4.75(1) $} &  \\ 
       & & & $4.709_{[2,1]}$ &     &           &   \\ 
      \hline
      \multirow{2}{*}{$|C_{\phi\phi\phi}|$} & $1.9696(31)_{[Z]}$ & & &   & \multirow{2}{*}{$1.9697(25)$}  \\ 
       & $1.969(5)_{[X]}$ & & & &  \\ 
      \hline
      \multirow{2}{*}{$|C_{\phi\phi\phi^3}|$} & $0.026(7)_{[Z]}$ & &  &   \\ 
       & $0.0238(15)_{[X]}$ & & & & \\ 
      \hline
      \multirow{2}{*}{$|C_{\phi^3\phi\phi^3}|$} & $1.3774(9)_{[Z]}$ & &  &  & \\ 
      & $1.364(7)_{[X]}$ & & & & &  \\ 
      \hline
      \multirow{2}{*}{$|C_{T\phi T}^{(0)}|$} & $1.2841(8)_{[Z]}$ & &  &   & \\ 
      & $1.277(3)_{[X]}$ & & & & &  \\ 
      \hline
      \hline 
   \end{tabular}
   \caption{\label{tab:CFT-extraction-3D}
   The CFT data from the fuzzy sphere results. We also list the values from other approaches for comparison. The subscript $[E]$ represent data extracted from eigenvalues. Subscripts $[Z]$ and $[X]$ represent data extracted from the matrix element of the fuzzy sphere $Z_\ell$ and $X_\ell$ observables, respectively. Other subscripts represent the Pad\'e level used. The error bars of the Fuzzy Sphere estimations are drawn from the largest differences of the extrapolations shown in the corresponding plots. They show the scales of possible fitting errors and should not be understood as bounds on the true values.
   }
\end{table}
In Table \ref{tab:CFT-extraction-3D} we summarize our key numerical results for the CFT data of the 3D YL criticality. Some of the quantities have been computed in the literature \cite{Borinsky:2021jdb, Gliozzi:2014jsa, Hikami:2017hwv, Butera:2012tq} and we list them in the table for comparison. The data  comes from \eqref{eq_phi_FS}, \eqref{eq_Q_FS}, \eqref{eq_OPEs}, \eqref{delta-phi-alt}, and \eqref{eq_CpTT}.

\subsubsection{Scaling dimensions}

In radial quantization, the CFT states and local operators have one-to-one correspondence. The CFT Hamiltonian is the dilatation operator, so CFT local operators correspond to eigenstates and the associated eigenvalues are the scaling dimension. Each eigenstate also diagonalize global symmetry quantum numbers, which are total spin $\ell$, spin-$z$ projection $\ell_z$, and the spacetime parity $p$. The ground state has always spin 0.

In the fuzzy sphere realization, the energy levels can be shifted by an arbitrary ground state energy and rescaled by a model-dependent speed of light. Therefore, we will read off the CFT scaling dimensions as ratios. We will assume the lowest spin-0 state, the 2nd lowest spin-0 state, and the lowest spin-1 state correspond to identity $\textrm{I}$, the fundamental primary operator $\phi$, and its descendant $\partial\phi$ operators, respectively. The difference between the scaling dimensions of the primary and its first descendant is $\Delta_{\partial\phi} - \Delta_{\phi} = 1$, which provides us a simple way to rescale the energy eigenvalues to match the CFT scaling dimensions, as in \eqref{eq_CFT_criterion}. For each state, the scaling dimension of the corresponding local operator is 
\begin{equation}
   \label{scaling-dimension-as-ratios}
   \Delta_\CO = 3\frac{E_\CO - E_{\textrm{I}}}{E_{T} - E_{\textrm{I}}}~,
 \end{equation} 
where $E_{\textrm{I}}$ is the ground state energy. All CFT data extracted from fuzzy sphere Hamiltonian at finite $N$ will have finite size error and therefore we need to study the finite size scaling behavior of the CFT data.   The finite size model near criticality is described by a CFT action with relevant and irrelevant deformations \cite{Cardy:1984epx, CARDY1986186, Zou:2018dec, Zou:2019dnc, Lao:2023zis, Lauchli:2025fii}:
\begin{equation}
  S = S_{\rm CFT} + \frac{g_{\phi}}{N^{(\Delta_\phi-d)/2}} \int d^d x \mathcal{\phi}(x) + \sum_{i} \frac{g_{i}}{N^{(\Delta_i-d)/2}} \int d^d x  \CO_i(x)\,,
\end{equation}
where we separate the relevant deformation $\phi$ and other operators $\CO_i$ which are all irrelevant, and $g_\phi$ and $g_i$ are dimensionless couplings. Near the criticality, the separation of scales is given by $\frac{\Lambda_{\rm UV}}{\Lambda_{\rm IR}} \sim \sqrt{N}$, which provides the volume dependence of the couplings. The energy eigenvalues of the finite Hamiltonian match to the CFT Hamiltonian as 
\begin{equation}\label{Hamiltonian-match}
  E_n = E_0 + \frac{\nu}{\sqrt{N}} \left( H_{\rm CFT} 
    + \frac{g_\phi}{N^{(\Delta_\phi-d)/2}} \int d^{d-1} \Omega \, \phi(\Omega) 
    + \sum_i \frac{g_i}{N^{(\Delta_i-d)/2}} \int d^{d-1} \Omega \, \CO_i(\Omega) 
  \right)\,,
\end{equation}
where $H_{\rm CFT} | n \rangle = \Delta_{n} | n \rangle$, and $\Delta_{n}$ are the CFT scaling dimensions. The relevant (irrelevant) couplings $g_\phi$ ($g_i$) contribute to the leading conformal perturbation corrections to these eigenvalues through diagonal matrix elements with a growing (suppressed) volume dependence.

In this work we measure the spectrum at the pseudo critical point defined when the stress tensor scaling dimension matches $\Delta_{T_{\mu\nu}} = d=3$, as described in (\ref{eq_CFT_criterion}). This means the deformation couplings cancel for the specific ratio (\ref{eq_CFT_criterion}) to give 3. Using the matching formula (\ref{Hamiltonian-match}), and expanding (\ref{eq_CFT_criterion}) to the first order in $g$, we can determine that
\begin{equation}
  g_\phi \propto N^{(\Delta_{\phi} - \Delta_{\CO_I})/2}\,,
\end{equation}
where $\CO_I$ represents the leading irrelevant deformation. This means the contribution from the relevant deformation and the leading irrelevant deformation are at the same order of magnitude, assuming the dimensionless coupling constant of the irrelevant deformation $g_I$ is of order 1. From this starting point, we see that the eigenvalues determined using the ratios (\ref{scaling-dimension-as-ratios}) will contain finite size error as 
\begin{equation}
  \Delta^{(N)} = \Delta + c N^{(3-\Delta_{\CO_I})/2} + \cdots
\end{equation}
where $c$ is some dimensionless coefficient that depends on $g_I$. The subleading couplings $g_i$ are suppressed but they also contribute. A comprehensive conformal perturbation analysis will help determine the explicit form of the correction terms and provide a more accurate determination of the spectrum by fitting multiple scaling dimensions using parameters $g_I$, $\Delta_I$ and the corresponding OPE coefficients, which is beyond the scope of this work. In this work, we will try accounting for the finite size error by fitting a single eigenvalue as a function of $N$. Our fitting model is 
\begin{equation}\label{fitModel}
  \Delta^{(N)} = P_{w,K}(N) \equiv \Delta + \sum_{k=0}^{K} a_k \frac{1}{N^{w+k/2}}
\end{equation}
where $\Delta$ is the scaling dimension prediction extrapolated to the thermodynamic limit. The model is essentially a polynomial in $\frac{1}{R} = \frac{1}{\sqrt{N}}$ except for beginning at power $w$. We choose this model because it takes into account that the leading correction comes in a specific power $N^{(3-\Delta_{\CO_I})/2}$ in conformal perturbation theory, and the series expansion after the leading term is agnostic of the precise form of subleading irrelevant deformations and higher order perturbation theory, which we do not know precisely. For the 3D Yang-Lee criticality, it is likely that the irrelevant deformation $\CO_I$ is the subleading scalar primary operator, which has a Ginzburg-Landau description $\phi^3$. From $(6-\epsilon)$ expansion and from our fuzzy sphere data we see that $\Delta_{\phi^3}\approx 4.6$, and we set $w = 0.8$. The extrapolations from the fit should not be, and are not, sensitive to the precise value of the input power $w$. 
For the number of powers we keep, we use $K=3$ which works for most energy levels except for a few. We detect that $K$ is too high if the coefficients of the powers abruptly increase. To validate the fitting scheme, we test it on the difference between two levels $\Delta_{\partial^2\phi} - \Delta_{\phi}$ which we expect to be 2. A linear extrapolation is still visibly deviating from $2$. A higer power series fit brings us significantly closer to the theoretical value. The precise value of the power does not matter very much, so the extrapolation from various powers $w$ and truncations $K$ agree to the third digit, though $w=0.8$, the more informed guess from the conformal perturbation theory, is more stable against small changes in truncation power.

\begin{figure}[h!]
  \includegraphics[width=0.5\linewidth]{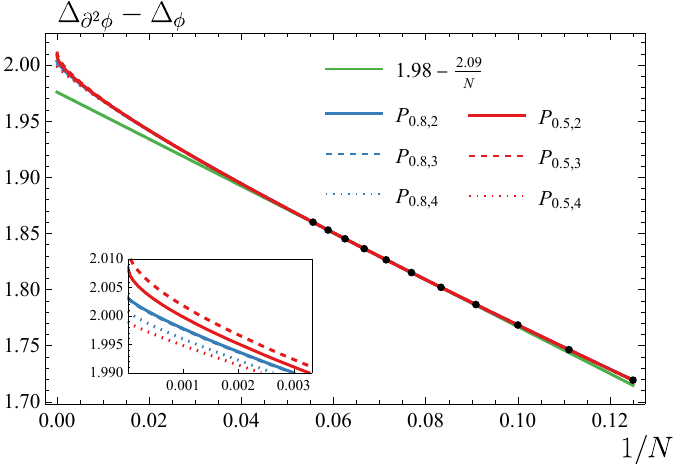}
  \centering
  \caption{\label{fig:DeltaD2PhiPlot}
    Testing the fitting scheme (\ref{fitModel}) on a known level difference $\Delta_{\partial^2\phi} - \Delta_{\phi} = 2$. The conformal perturbation theory suggests the beginning power should be $w=0.8$. We also try $w=0.5$, which amounts to a Taylor series in $\frac{1}{R}$, and a liner fit in $\frac{1}{N}$. For both non-linear models, we vary the truncation level $K$ from 2 to 4. The result shows that the non-linear model significantly improves the fit compared to the linear model, and the end result is not sensitive to the parameters. 
  }
\end{figure}

\begin{figure}[h]
  \centering
  \includegraphics[width=0.6\linewidth]{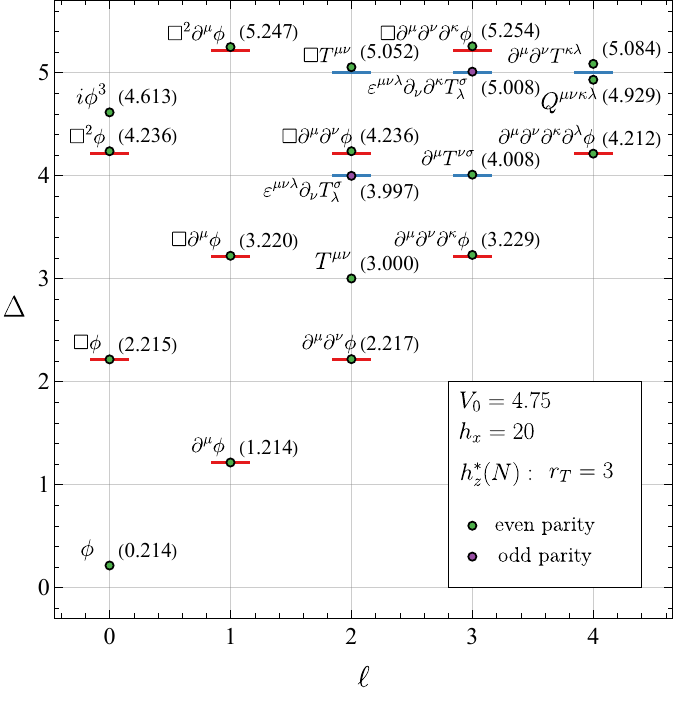}
  \caption{\label{Fig:mainSpectrumPlot}
    Estimation of the low-lying spectrum for the critical 3D YL model. We extrapolate the eigenvalues from $N=12$ to $N=18$ to the thermodynamic limit using model $P_{0.8,K}$ defined in (\ref{fitModel}). For the truncation $K$ of the fitting model, we use $K=1$ for the $Q$ state and $K=3$ for the rest. The local operator dual of each energy level is indicated on its left side, and the best fit value is indicated in the parenthesis.
    The red (blue) lines indicate the descendant operators from $\phi$ ($T$), whose position are determined by the value of $\Delta_{\phi}$ ($\Delta_T$) that appears in the plot plus integers. The four primary operators  are $\phi$, $T_{\mu\nu}$, $\phi^3$, and $Q_{\mu\nu\kappa\lambda}$.
  }
\end{figure}

In Fig. \ref{Fig:mainSpectrumPlot} we show the low spectrum of 3D YL criticality as a result of the extrapolation. In the extrapolation scheme (\ref{fitModel}) we treat each eigenvalues independently, so the conformal algebra is not taken as an input except for rescaling the gap between $\Delta_{\phi}$ and $\Delta_{\partial\phi}$ to 1. In the plot we see that relation between $\Delta_{\phi}$ and the  descendant levels are precisely integers, making it easy to distinguish primary from descendant levels. The scaling dimension of the stress tensor is fixed at $\Delta_{T_{\mu\nu}} = 3$ by the criticality criterion, and all its descendants in this plot are integers to high precision. The plot shows two other primaries, one at $(\ell = 0, \Delta \approx 4.6)$ and the other at $(\ell = 4, \Delta \approx 4.9)$. The former agrees with the ($6-\epsilon$) expansion prediction of the $\phi^3$ operator in the Ginzburg-Landau description, and the latter seems to belong to the leading double trace family $[\phi\phi]_{0,\ell}$ at spin 4, where the stress tensor also reside. We denote the spin 4 primary as $Q_{\mu\nu\kappa\lambda}$.

\begin{figure}[htbp]
  \centering
  \includegraphics[width=0.45\linewidth]{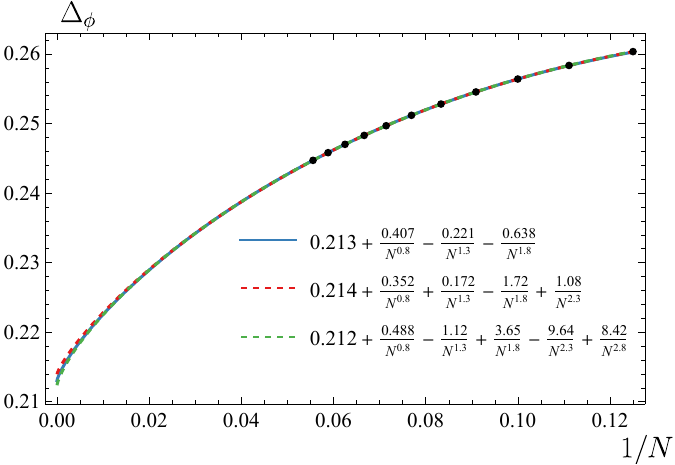}
  \includegraphics[width=0.45\linewidth]{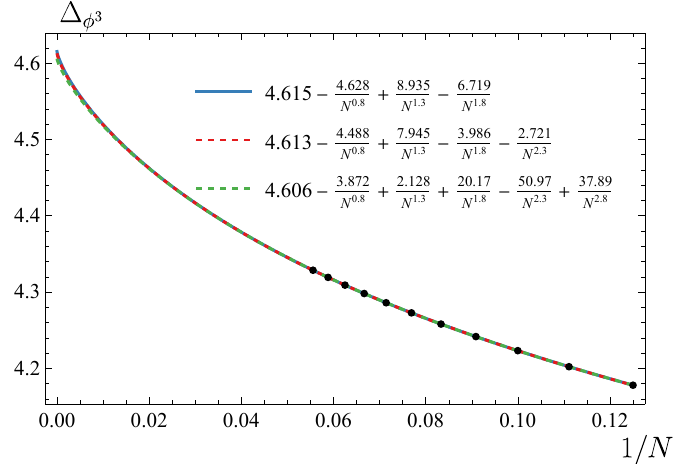}
  \caption{\label{fig:fitPowers}
    Fitting various scaling dimensions with a range of fitting models. {\bf Left panel:} The scaling dimension of the lowest primary, $\phi$ operator. Various powers converge to the same end value. Most fits agree up to within 0.002. 
    {\bf Right panel:} The scaling dimension of the next-to-leading scalar primary, the $\phi^3$ operator. Different fitting powers agree approximately up to a $0.006$ error. 
  }
\end{figure}

In Fig. \ref{fig:fitPowers} we take a closer look at the scalar primary operators $\phi$ and $\phi^3$. We see that various fit truncation levels agree to high precision. The small fluctuation of the extrapolation with respect to the fitting parameters provide us with a heuristic error bar, so we can determine
\begin{equation}
\label{eq_phi_FS}
   \Delta_{\phi} = 0.214(2)\ ,  \qquad \Delta_{\phi^3} = 4.613(6)\ .
\end{equation}
We note that the fuzzy sphere result for $\Delta_\phi$ is in excellent agreement with the high-temperature expansion result $0.214(6)$ from
\cite{Butera:2012tq} and is also very close to our two-sided Pad\'{e} estimate $\approx 0.218$.

\begin{figure}[htbp]
  \centering
  \includegraphics[width=0.45\linewidth]{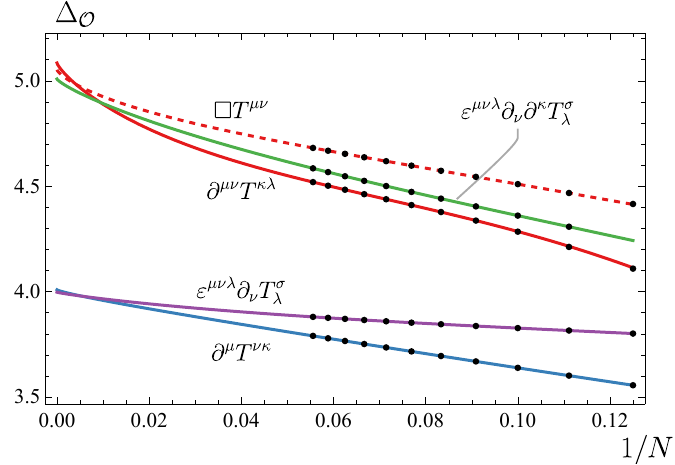}
  \includegraphics[width=0.45\linewidth]{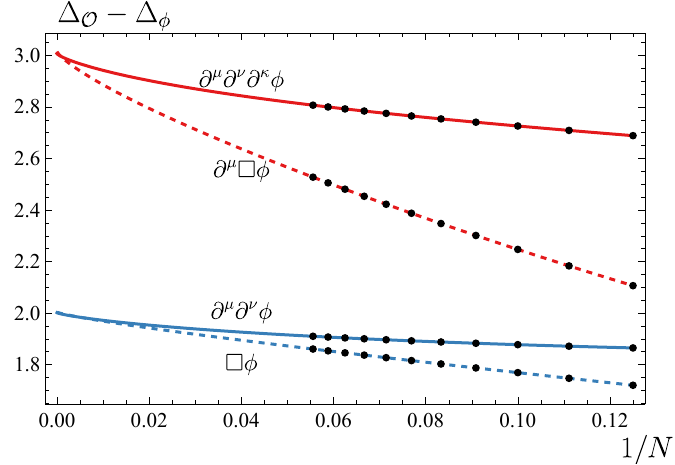}
  \caption{\label{fig:fitDescendants}
    The difference between several descendant states and their corresponding primary state. The agreement between $\square = \partial_\mu\partial^\mu $ and $\partial^2 = (\partial_\mu\partial_\nu - {\rm trace}) $ rely on conformal symmetry. At finite $N$ they receive different finite size correction. We see that the extrapolation to $N\rightarrow \infty$ bring them all to integers.  
  }
\end{figure}
\begin{figure}[htbp]
\centering
   \includegraphics[width=0.45\linewidth]{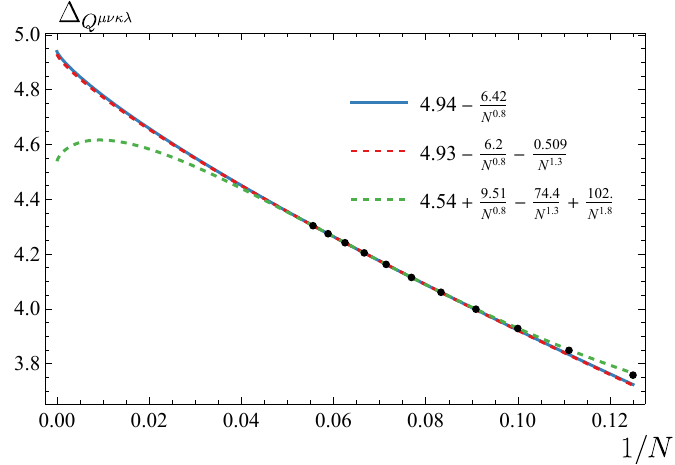}
   \includegraphics[width=0.45\linewidth]{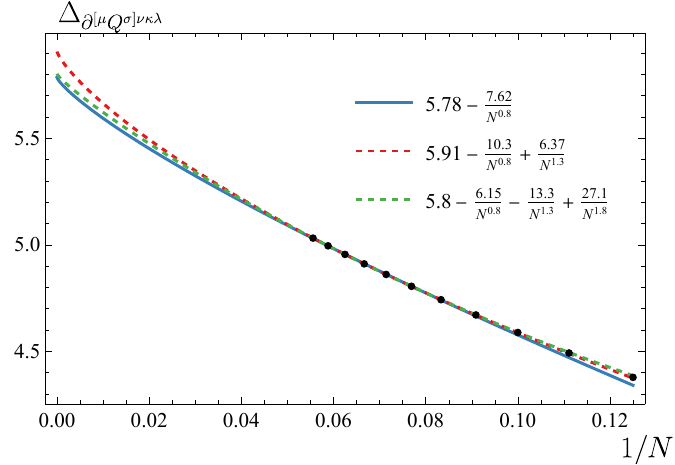}
   \caption{\label{fig:DeltaQ}
     The scaling dimension of the spin-4 primary $Q^{\mu\nu\kappa\lambda}$ and its anti-symmetric descendant $\partial^{[\mu}Q^{\sigma]\nu\kappa\lambda}$. The latter operator is in the spacetime parity odd sector and has no mixing with the descendant of $\phi$, resulting in a better-behaved extrapolation.
   }
\end{figure}
It is worth emphasizing that our confidence in the fuzzy sphere realization of the quantum criticality and the thermodynamic limit extrapolation is increased by the high precision agreement with conformal symmetry, which is presented in Fig. \ref{fig:fitDescendants}. The fact that the descendants of different spins agree with each other and are separated by integers is a result of the conformal algebra. We see that the agreement generically is broken for all states by a significant amount at finite size, but the extrapolation brings them to the corresponding integers in a non-linear manner. The agreement across many states is reached without any fine-tuning of fitting parameters. In Fig. \ref{fig:DeltaQ}, we utilize this property to determine the scaling dimension of the spin-4 primary $Q^{\mu\nu\kappa\lambda}$. As we go to the large spin regime we inevitably find a denser spectrum. 
Indeed, there are 3 operators within a small domain $4.214 \leq \Delta \leq 5$: $\partial^\mu\partial^\nu\partial^\kappa\partial^\lambda\phi$, $\partial^\mu\partial^\nu T^{\kappa\lambda}$, and $Q^{\mu\nu\kappa\lambda}$. 
As a result, the fit to $\Delta_{Q^{\mu\nu\kappa\lambda}}$ is unstable. Instead, the spacetime parity-odd sector has a cleaner spectrum: there are not anti-symmetric descendants of $\phi$, and the leading operator in the spin-4 parity odd sector is the anti-symmetric descendant of $Q$, $\partial^{[\mu}Q^{\sigma]\nu\kappa\lambda}$. This operator has a healthier extrapolation giving $\Delta_{\partial Q} = 5.9(1)$, and by conformal symmetry it is expected to be exactly 1 above the primary scaling dimension. We conclude that 
\begin{equation}
\label{eq_Q_FS}
   \Delta_{Q^{\mu\nu\kappa\lambda}} = 4.9(1)~.
\end{equation}
Still, even in the parity odd sector $\partial Q$ gets dangerously close to $\partial^3 T$, and the possible mixing between the two levels may impact the accuracy of the fit.

\begin{figure}[h!] 
  \centering
  \includegraphics[width=0.48\linewidth]{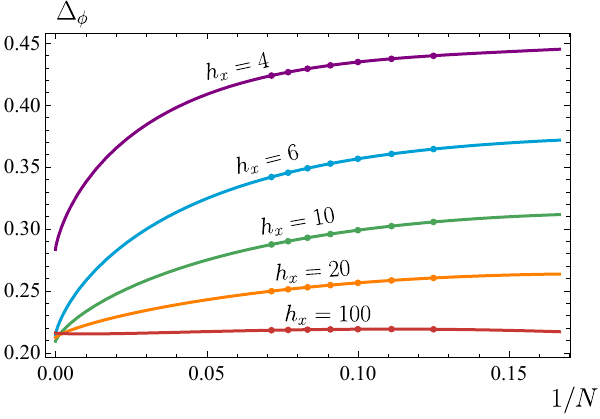}
  \includegraphics[width=0.48\linewidth]{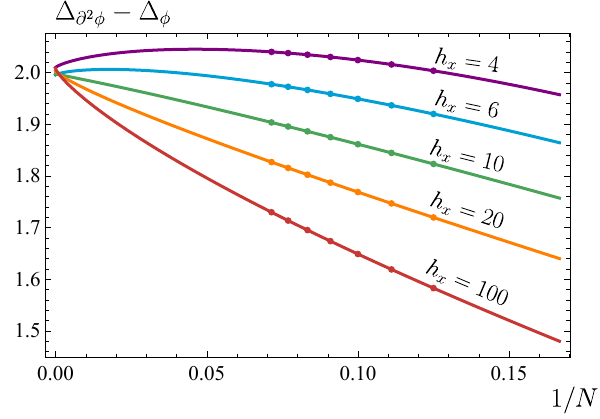}
  \caption{\label{fig:universality}
    Scaling dimension $\Delta_\phi$ (left panel) and $\Delta_{\partial^2\phi} - \Delta_\phi$ (right panel) at the critical point of systems with various $h_x$ couplings. The prediction from $N\rightarrow \infty$ extrapolation is stable for a wide range of $h_x$, indicating that they belong to the same universality class. The $\Delta_\phi$ prediction from $h_x = 4$ deviates from the rest of the result by $\sim 0.05$, indicating that small $h_x$ may be converging much slower than large $h_x$.
  }
\end{figure}

Yang-Lee criticality has 1 relevant parameter, which means in a phase diagram of more than 1 parameters it will appear as a critical line or plane. In the phase diagram proposed in Fig. \ref{fig:3dHypoPhaseDigm},
we obtain a critical line where there is a critical coupling $ih_{z,c}$ for each $h_x$. This means we should obtain the same criticality for different $h_x$ and the extracted critical data should agree. In Fig. \ref{fig:universality} we test the agreement of $\Delta_{\phi}$ and conformal algebra prediction $\Delta_{\partial^2\phi} - \Delta_{\phi} = 2$ for a range of $h_x$. The predictions from different $h_x$ all agree when extrapolated to the thermodynamic limit, realizing the critical universality. We also see that large $h_x$ is computationally favorable as it seem to receive smaller finite size error.

\subsubsection{OPE coefficients}

Apart from the operator scaling dimensions which encode the conformally invariant 2-point function data, we are also interested in the OPE coefficients which encode the conformally invariant 3-point function data. The most convenient way to extrapolate such information is through state-operator correspondence and  study the fuzzy sphere matrix elements of the form
\begin{equation}
   \langle \CO_1, \ell_1, m_1 | M_{\ell_2 m_2} | \CO_2, \ell_3, m_3 \rangle 
\end{equation}
where $| \CO, \ell, m \rangle$ is a fuzzy sphere matrix eigenstate in the spin ($\ell,m$) sector and $M_{\ell_2 m_2}$ is a fuzzy sphere observable. 
The Yang-Lee Hamiltonian (\ref{YL3D-Hamiltonian}) is a complex symmetric operator, therefore the left eigenvectors 
coincide with the right eigenvectors, $\langle \psi_{i}| = (|\psi_{i}\rangle)^T$.
With this definition we have $\langle \psi_{i}|  \psi_{j}\rangle = \delta_{ij}$.
In this work we consider the following observables
\begin{equation}\label{spherical-smeared-operators}
\begin{aligned}
   M_\ell^{a} &= \frac{1}{2s+1} \int d^2\Omega Y_{\ell 0} \psi^\dagger(\Omega)\sigma^a \psi(\Omega) ,\quad X_\ell \equiv M_\ell^x, \quad Z_\ell \equiv M_\ell^z
\end{aligned}
\end{equation}
where we suppress the quantum number $m$ as we always work in the $m=0$ representation. Unlike quantum Ising model, there is no $\mathbb Z_2$ symmetry charge distinction between the $X$ and $Z$ operators so they can generically have overlap with any operators, the $\mathcal{PT}$ symmetry will determine the phase of the matrix elements. 
As discussed previously, each fuzzy sphere eigenstate can be identified as a CFT radial quantization eigenstate which is in 1-to-1 correspondence with a CFT local operator. The fuzzy sphere observable is much less determined. $\psi^\dagger\sigma^a \psi$ is the local charge density operator in the UV, and matches to a generic linear combination of CFT local operators, and the spherical smearing in (\ref{spherical-smeared-operators}) projects to modes with specific $SO(3)$ spin, and flows to the most relevant operator in that sector in the thermodynamic limit
\begin{equation}
   M_{\ell}^a = b_{a,0} \delta_{\ell,0} \mathbbm{1} + b_{a,1} \int d^2\Omega \, Y_{\ell0}(\Omega) \phi(\Omega) + \cdots
\end{equation}
where ``$\cdots$'' contain descendants and higher dimensional primaries. We will study these matrix elements case-by-case.

\begin{figure}[htbp]
   \centering 
   \includegraphics[width=0.33\linewidth]{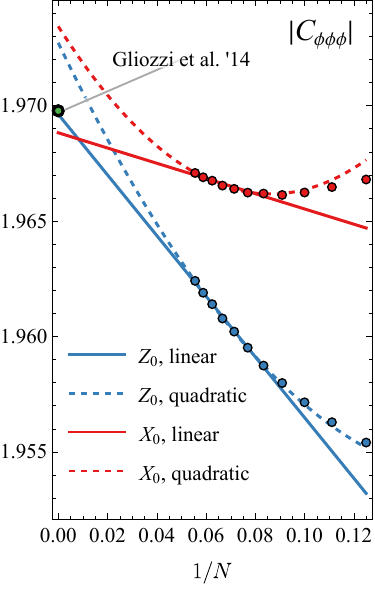} 
   \includegraphics[width=0.33\linewidth]{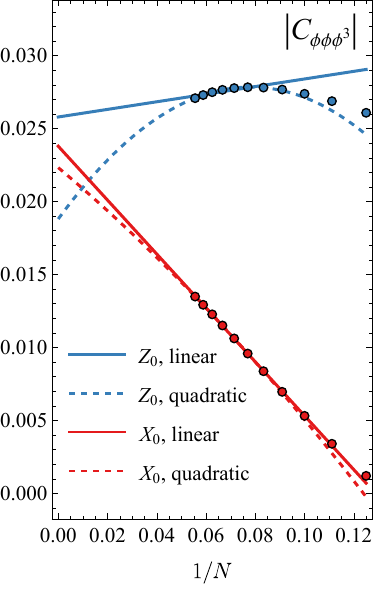} 
   \includegraphics[width=0.32\linewidth]{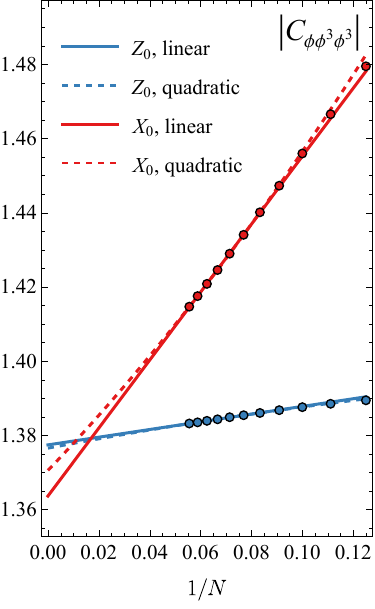}   
   \caption{\label{fig:scalar-OPE}
      The scalar OPE coefficients $C_{\phi\phi\phi}$, $C_{\phi\phi\phi^3}$, and $C_{\phi^3\phi\phi^3}$ measured from the fuzzy sphere data using the formula (\ref{phi-phi-phi}) and (\ref{phi-phi-phi3}). The blue (red) color represent the $Z_0$ ($X_0$) observable, respectively, and the solid (dashed) lines represent the fit model $a + b/N$ ($a+b/N+c/N^2$), respectively. The $C_{\phi\phi\phi}$ measurement is compared with an earlier result obtained by solving truncated crossing equations \cite{Gliozzi:2013ysa, Gliozzi:2014jsa}.
   }
\end{figure}

We begin with scalar primary 3-point functions. The scalar 2-point and 3-point functions are completely fixed by conformal symmetry 
\begin{equation}
\begin{aligned}
   \langle \CO(x_1) \CO(x_2) \rangle &= \frac{1}{
      x_{12}^{2\Delta} 
   } \\ 
   \langle \CO_1(x_1) \CO_2(x_2) \CO_3(x_3) \rangle &= \frac{c_{123}}{
      x_{12}^{\Delta_1+\Delta_2-\Delta_3} x_{23}^{\Delta_2+\Delta_3-\Delta_1} x_{31}^{\Delta_3+\Delta_1-\Delta_2}
   }
\end{aligned}
\end{equation}
The state-operator correspondence in the scalar case is
\begin{equation}
\begin{aligned}
   | \CO, 0 \rangle &= \lim_{x\rightarrow 0} \CO(x) | 0 \rangle \\
   \langle \CO, 0 | &= \lim_{x\rightarrow \infty} x^{2\Delta} \langle 0 | \CO(x) 
\end{aligned}
\end{equation}
So the fuzzy sphere matrix element matches to the scalar 3-point function as 
\begin{equation}
   \langle \CO_1, 0 | M_{0}^a | \CO_3, 0 \rangle \supset \lim_{\substack{x_1 \rightarrow \infty \\ x_3 \rightarrow 0}} x_1^{2\Delta_1} \int d^2\Omega \, Y_{\ell0}(\Omega) \langle O_1(x_1) O_2(\Omega) O_3(x_3) \rangle
\end{equation}
so the expansion is
\begin{equation}
   \langle \CO_1, 0 | M_{0}^a | \CO_3, 0 \rangle =  b_0 \delta_{\CO_1\CO_3} + b_1 C_{\CO_1 \phi \CO_3} + \cdots
\end{equation}
and we can determine the OPE coefficient $C_{\phi\phi\phi}$ as 
\begin{equation}\label{phi-phi-phi}
   C_{\phi\phi\phi} = \frac{
      \langle \phi,0 | M_{0}^a | \phi,0 \rangle - \langle 0 | M_{0}^a | 0 \rangle
   }{\langle 0 | M_{0}^a | \phi,0 \rangle}
   + O\big(N^{-1}\big)~ .
\end{equation}
The leading error term contain the contribution from descendant $\partial^k \phi$. Only even powers can contribute to preserve spacetime parity. By dimensional analysis, the leading error $\partial^2 \phi$ is suppressed by $R^{-2} \sim N^{-1}$. Similarly, $C_{\phi^3\phi\phi^3}$ can be determined by simply replacing the state $\phi$ by state $\phi^3$. An off-diagonal matrix element is given by 
\begin{equation}\label{phi-phi-phi3}
   C_{\CO\phi\CO'} = \frac{
      \langle \CO,0 | M_{0}^a | \CO',0 \rangle 
   }{\langle 0 | M_{0}^a | \phi,0 \rangle}
   + O\big(N^{-1}\big)~ ,
\end{equation}
and we can use it to determine $C_{\phi\phi\phi^3}$. The numerical results of these scalar OPE coefficients are shown in Fig. \ref{fig:scalar-OPE}. The predictions are\footnote{The error bars are determined by the difference between the linear and quadratic extrapolations. The purpose of these error bars is to show the scale of a possible fitting error, but not to assert that the true value should lie within the error bar. Note that $C_{\phi \phi^3 \phi^3}$ has non-overlapping error bars.}
\begin{equation}
\begin{aligned}
\label{eq_OPEs}
    |C_{\phi\phi\phi}| &= 1.9696(31)_{[Z]}, \quad 1.969(5)_{[X]} \\ 
    |C_{\phi\phi\phi^3}| &= 0.026(7)_{[Z]}, \quad 0.0238(15)_{[X]} \\ 
    |C_{\phi\phi^3\phi^3}| &= 1.3774(9)_{[Z]}, \quad 1.364(7)_{[X]}~.
\end{aligned}
\end{equation}

\begin{figure}[htbp]
   \centering 
   \includegraphics[width=0.45\linewidth]{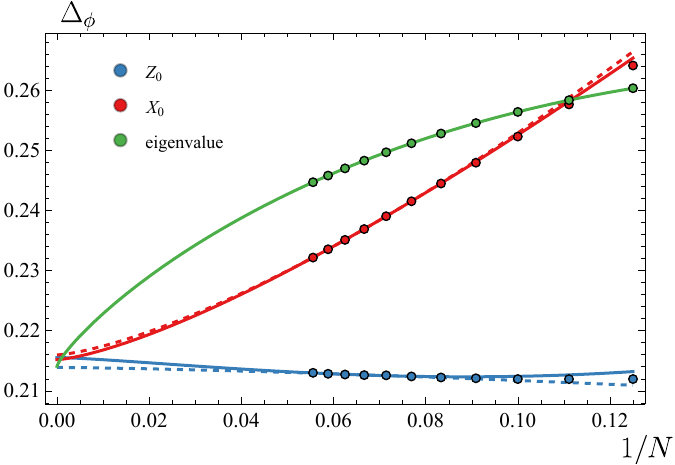}
   \hspace{0.02\linewidth}
   \raisebox{0.09\linewidth}{\includegraphics[width=0.45\linewidth]{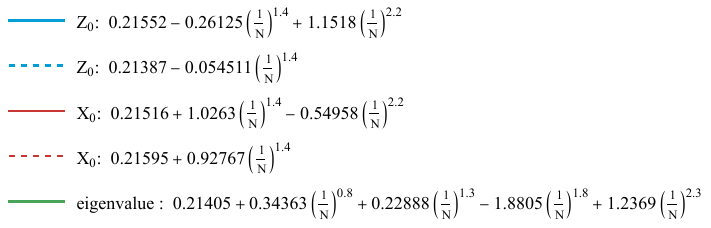}}
   \caption{\label{fig:delta-phi-from-3pt}
     The alternative determination of $\Delta_\phi$ from the ratio of 3-point functions (\ref{Delta-phi-alt}). 
   }
\end{figure}

Next, we move on to the descendants of scalar primaries. The matching between fuzzy sphere states to such descendant states is 
\begin{equation}
   | \partial^k \CO, \ell \rangle \equiv a_{\CO, k, \ell} v_{\ell 0}^{\mu_1\cdots \mu_k} P_{\mu_1} \cdots P_{\mu_k} | \CO, 0 \rangle 
\end{equation}
where $P_{\mu}$ is a conformal generator of translation, $v_{\ell 0}^{\mu_1\cdots \mu_k}$ is a vector projecting the $3^k$ tensor components to the ($\ell,0$) representation of $SO(3)$, and $a_{\CO, k, \ell}$ is a normalization factor to keep the state normalized to 1. The matrix elements involving the descendants are completely determined by the conformal algebra up to the corresponding primary data. A number of useful diagonal matrix elements have been computed in \cite{Lao:2023zis, Lauchli:2025fii} 
\begin{align}
   \label{descendant-matrix-element}
   A_{k,\ell} (\Delta,\Delta') &\equiv \frac{
      \langle \partial^k \CO', \ell | \int d^2 \Omega Y_{00}(\Omega) \CO(\Omega) | \partial^k \CO', \ell \rangle
   }{
      \langle \CO', 0 | \int d^2 \Omega Y_{00}(\Omega) \CO(\Omega) | \CO', 0 \rangle
   } \\ 
   A_{1,1} &= 1 + \frac{\mathcal{C}_\CO}{6\Delta'}, \quad \mathcal{C}_\CO \equiv \Delta(\Delta-3)~.
\end{align}
Note that $A_{1,1}$ for $\CO = \CO' = \phi$ is a simple linear function of $\Delta_\phi$, and the right hand side of (\ref{descendant-matrix-element}) can be explicitly computed from fuzzy sphere, so as an application we use (\ref{descendant-matrix-element}) to determine $\Delta_{\phi}$
\begin{equation}\label{Delta-phi-alt}
   \Delta_\phi = 6 \frac{\langle \partial\phi | M_0^a | \partial\phi \rangle - \langle 0 | M_0^a | 0 \rangle}{
   \langle \phi | M_0^a | \phi \rangle - \langle 0 | M_0^a | 0 \rangle } - 3
\end{equation}
The result is shown in Fig. \ref{fig:delta-phi-from-3pt}. We obtain a prediction of the $\Delta_\phi$ as
\begin{equation}
\label{delta-phi-alt}
   \Delta_{\phi} = 0.2155(16)_{[Z]} , ~~0.2151(8)_{[X]}
\end{equation}

It is remarkable that the $\Delta_\phi$ determined this way has significantly smaller finite size error. Indeed, the subleading terms in the $M_{0}^{a}$ have a special feature. We expand it more to include the descendants of $\phi$
\begin{equation}
   M_{0}^a = b_{a,0} \mathbbm{1} + b_{a,1} \int d^2\Omega \, Y_{00}(\Omega) \phi(\Omega) + \sum_k b_{a,1}^{(k)} \int d^2\Omega \, Y_{00}(\Omega) \partial_r^k \phi(\Omega) + {(\text{more primaries}\cdots)} ~.
\end{equation}
Note that the spacial total derivatives have to vanish due to the integral against $Y_{00}(\Omega)$ over all sphere, so the non-vanishing descendants can only be radial descendants. The action of $k$-fold radial derivatives has a simple pattern 
\begin{equation}
   \partial_r^k \frac{1}{r^{\Delta_2 + \Delta_3 - \Delta_1}} = \frac{(-1)^k(\Delta_2 + \Delta_3 - \Delta_1)_k}{r^{\Delta_2 + \Delta_3 - \Delta_1+k}}
\end{equation}
where $(a)_k = \frac{\Gamma(a+k)}{\Gamma(a)}$ is the Pochammer symbol. Thus all the descendant of $\phi$ terms can be resummed 
\begin{equation}
   M_{0}^a = b_{a,0} \mathbbm{1} + \bigg(b_{a,1} +  \sum_{k=2} (-1)^k b_{a,1}^{(k)} (\Delta_\phi)_k\bigg)  \int d^2\Omega \, Y_{00}(\Omega) \phi(\Omega) + {(\text{more primaries}\cdots)} ~.
\end{equation}
Now, note that the resummed coefficient is exactly the same when the external state is $\partial\phi$ instead of $\phi$, so the ratio (\ref{Delta-phi-alt}) gets rid of all errors up to the next primary operator $T^{\mu\nu}$. So we expect the leading finite volume error to be $O(N^{-(d-\Delta_\phi)/2}) = O(N^{-1.4})$, whose convergence is a lot faster than the eigenvalues with $N^{-0.8}$ error. This qualitatively explains why the 3-point function gives a better $\Delta_\phi$. We can also try using the fit to account for the next error which is $N^{-2.2}$, given by $\phi^3$ contribution.
 
\begin{figure}
   \center 
   \includegraphics[width=0.32\linewidth]{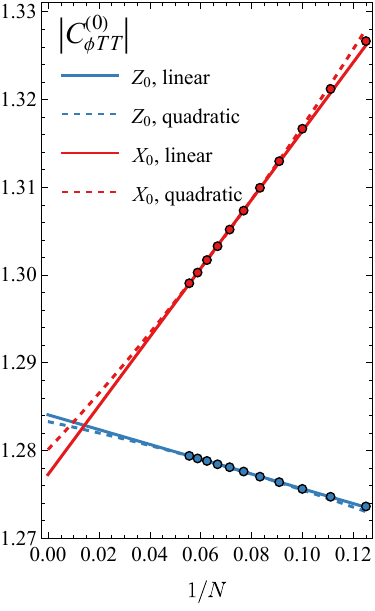}
   \includegraphics[width=0.32\linewidth]{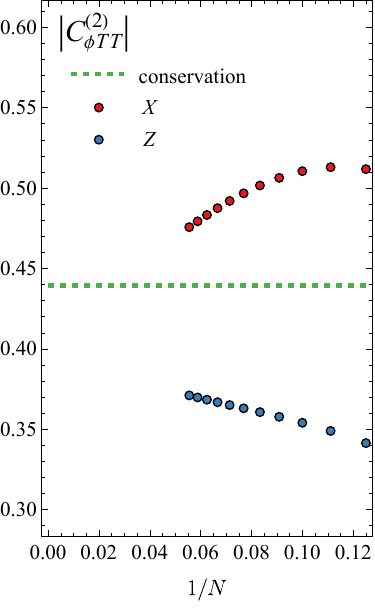}
   \includegraphics[width=0.32\linewidth]{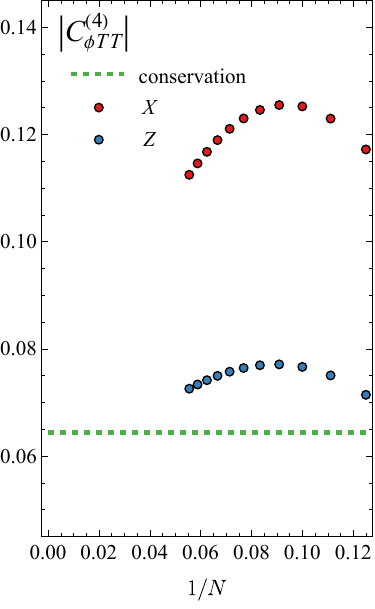}
   \caption{\label{fig:OPEPhiTT}
     The fuzzy sphere result of $C_{\phi T T}^{(s)}$ for $s=0,2,4$.
   }
\end{figure}

We conclude this section with an example of scalar-spin-spin 3-point function. In our convention, we normalize a generic spin-$\ell$ primary to have a standard form 2-point function
\begin{equation}
   \langle \CO(x_1,z_1) \CO(x_2,z_2) \rangle = \frac{\left( (z_1 \cdot z_2) - (x_{12}\cdot z_1)(x_{12}\cdot z_2)/x_{12}^2 \right)^\ell}{x_{12}^{2\Delta}}
\end{equation}
where $x_{ij} = x_i - x_j$ and $z_i^{\mu}$ is an auxiliary vector to keep index-free: $\CO(z) \equiv T^{\mu_1\cdots\mu_\ell} z_{\mu_1} \cdots z_{\mu_\ell}$.
A generic scalar-spin-spin correlator can have several polarizations carrying independent OPE coefficients. We follow the argument of \cite{Costa:2011mg,Meltzer:2018tnm} and expand the correlator as (assuming parity even)
\begin{equation}
   \langle \CO_1 (x_1,z_1) \phi (x_2) \CO_3 (x_3,z_3) \rangle = \sum_{\substack{s=|\ell_1-\ell_3|\\ s-|\ell_1-\ell_3| = 0 \mod 2}}^{\ell_1 + \ell_3} C_{\CO_1\phi\CO_3}^{(s)} \frac{H_{13}^{\frac{1}{2}(\ell_1+\ell_3-s)}V_{1,2,3}^{\frac{1}{2}(\ell_1-\ell_3+s)}V_{3,1,2}^{\frac{1}{2}(-\ell_1+\ell_3+s)}}{x_{12}^{\kappa_1+\kappa_2-\kappa_3} x_{13}^{\kappa_1+\kappa_3-\kappa_2} x_{23}^{\kappa_2+\kappa_3-\kappa_1}}
\end{equation}
where $\kappa_i = \Delta_i + \ell_i$, $x_{ij} = x_i - x_j$, and
\begin{equation}
\begin{aligned}
   H_{ij} &= x_{ij}^2 (z_i \cdot z_j) - 2 (x_{ij}\cdot z_i) (x_{ij}\cdot z_j) \\
   V_{i,j,k}  &= \frac{x_{ij}^2 (x_{ik} \cdot z_i) - x_{ik}^2(x_{ij} \cdot z_i)}{x_{jk}^2} ~.
\end{aligned}
\end{equation}
In this basis, the correspondence between the fuzzy sphere matrix elements and the OPE coefficients is 
\begin{equation}
   \langle \CO, \ell | M_{s}^a | \CO', \ell' \rangle = \delta_{\CO,\CO'} \delta_{s,0} b_0 +  b_{a,1} f_{\ell,\ell',s} C_{\CO\phi\CO'}^{(s)} + \cdots,
\end{equation}
where $f_{\ell,\ell',s}$ is an exactly computable prefactor that only depends on the spins. The detailed derivation of this correspondence is included in Appendix \ref{app:spin-corrlator}. With this convention we study the $\langle\phi T T\rangle$ correlator, which has three polarizations $s=0,2,4$. The result is shown in Fig. \ref{fig:OPEPhiTT}. For $s=0$, the convergence is good, and by extrapolating to the thermodynamic limit we obtain 
\begin{equation}
\label{eq_CpTT}
   C_{\phi T T}^{(0)} = 1.2841(8)_{[Z]} , ~~1.277(3)_{[X]}
\end{equation}
The error bar of the quantities is determined by the difference between the linear and quadratic fit. The fact that the two results differ by a gap greater than the error bar means the higher powers in $\frac{1}{N}$ still contributes significantly. 
As $s$ become larger, the convergence becomes worse. A remarkable feature of the stress tensor correlator is that the stress tensor is a conserved current, $\partial_\mu T^{\mu \nu} = 0$,
which generates additional equations of motion that relate the OPE coefficients of different polarizations. It turns out that this correlator only has one independent OPE coefficient\cite{Meltzer:2018tnm}
\begin{equation}
\begin{aligned}
   C_{\phi TT}^{(2)} &= -\frac{2\Delta_\phi(\Delta_\phi - 4)}{\Delta_\phi^2 - 6\Delta_\phi + 6}C_{\phi TT}^{(0)} \\ 
   C_{\phi TT}^{(4)} &= \frac{\Delta_\phi(\Delta_\phi + 2)}{2(\Delta_\phi^2 - 6\Delta_\phi + 6)}C_{\phi TT}^{(0)} ~.
\end{aligned}
\end{equation}
The conservation condition provides a non-trivial check of the conformal symmetry. Fig. \ref{fig:OPEPhiTT} shows that the fuzzy sphere data for $C_{\phi TT}^{(2)}$ and $C_{\phi TT}^{(4)}$ are in approximate agreement with conservation, and are converging to the theoretical values as $N$ grows. Spinning correlators converge more slowly and studying them thoroughly require us to either push to higher $N$ or implement more techniques to mitigate the finite size error. We will leave this task as a future study.

\section{4D model on the $24$-cell (Icositetrachoron)}
\label{Octaplex}
\begin{figure}[h!]
\center
\includegraphics[width=0.3\textwidth]{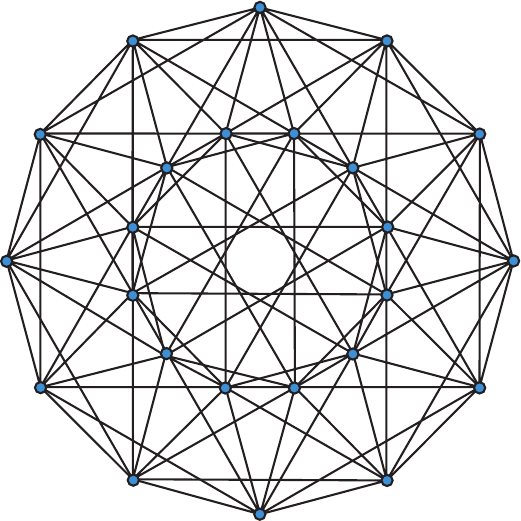}
\caption{Picture of the 24-cell as graph on the $F_{4}$ Coxeter plane.}
\label{24cellGraph}
\end{figure}

 The 24-cell  or Icositetrachoron is a self-dual regular 4D polytope (polychoron), which has $N = 24$ vertices and  $E = 96$ edges.
We can construct this polychoron by defining its vertices in a four-dimensional Cartesian coordinate system. Its $24$ vertices $v_{i}$ (for $i=1,\dots,24$) are obtained by all permutations of coordinates of the vectors $(\pm1, \pm 1,0,0)$. 
We draw the  24-cell as graph on  the $F_{4}$   Coxeter plane in Figure \ref{24cellGraph}.

\noindent The Yang-Lee Hamiltonian on the 24-cell reads:
\begin{align}
H_{\textrm{YL}} = - J \sum_{\langle i j\rangle \in e}Z_{i}Z_{j} - h_{x}\sum_{i \in v}X_{i} - ih_{z}\sum_{i \in v}Z_{i}\,, \label{HamOcta}
\end{align}
where the first sum is taken over all the $96$ edges and the last two sums go over the $24$ vertices.
The symmetry group of $24$-cell is the Weyl group of $F_{4}$, denoted as $W(F_{4})$. 
This group has $1152$ elements, $25$ conjugacy classes and $25$ irreps \cite{kondo1965characters, Carter1972, carter1993finite}:  
\begin{align}
\textbf{1}, \textbf{1}', \textbf{1}'', \textbf{1}''' ,\textbf{2}, \textbf{2}', \textbf{2}'', \textbf{2}''',\textbf{4},\textbf{4}',\textbf{4}'',\textbf{4}''', \textbf{4}'''',\textbf{6}, \textbf{6}',\textbf{8},\textbf{8}',\textbf{8}'',\textbf{8}''',\textbf{9}, \textbf{9}', \textbf{9}'', \textbf{9}''',\textbf{12},\textbf{16}\,,
\end{align}
where the number represents an irrep's dimension.
For convenience we present the character table of $W(F_{4})$ in the Appendix \ref{charWF4}. The $O(4)$ irreps  $(0,0)$, $(1/2,1/2)$ and $(1,1)$ do not split on 24-cell and correspond to irreps $\textbf{1}$, $\textbf{4}$ and $\textbf{9}$. The $O(4)$ irrep $(3/2, 3/2)$ splits into $\textbf{8}+\textbf{8}'$,   $(2, 2)$ splits into $\textbf{2}+\textbf{2}'+\textbf{9}+\textbf{12}$ and 
$(5/2, 5/2)$ splits into $\textbf{4}+\textbf{8}+\textbf{8}'+\textbf{16}$.

\begin{figure}[h!]
\center
\includegraphics[width=0.7\textwidth]{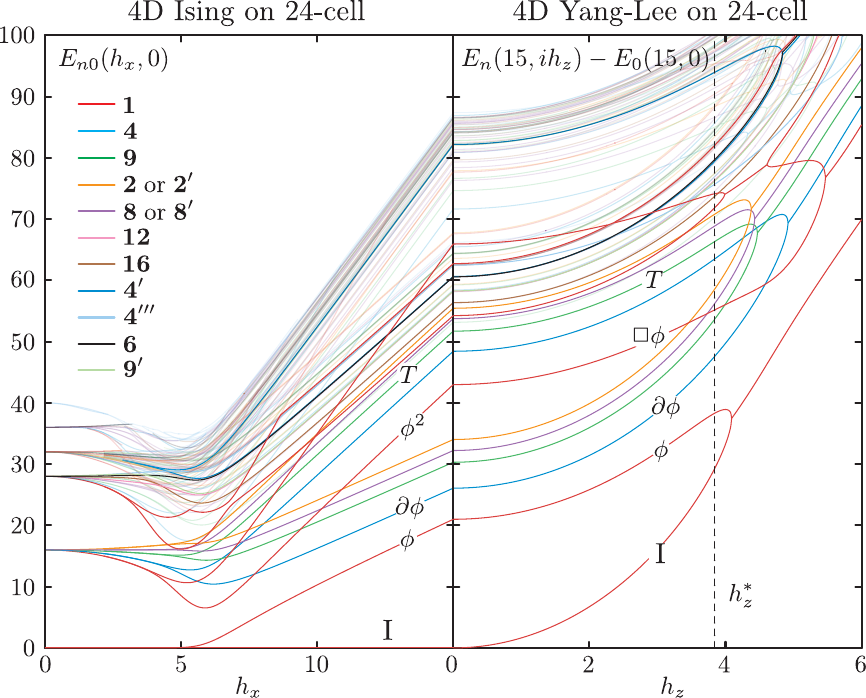}
\caption{Energy gaps of the Hamiltonian (\ref{HamOcta}) for 24-cell along the path $(h_{x},ih_{z})=(0,0)\to(15,0)\to(15,i6)$, with $J=1$. We identified the lower energy levels with their corresponding operators in the Ising (Gaussian) and Yang-Lee CFTs. The dashed line is the pseudo-critical point $h_{z}^{*}$ obtained fixing $r_{T}=4$.}
\label{Octa_spectrum}
\end{figure}

We exhibit the lower energy gaps of the Hamiltonian (\ref{HamOcta}) along the path  $(h_{x},ih_{z})=(0,0)\to(15,0)\to(15,i6)$  with $J=1$ in the Figure \ref{Octa_spectrum}, where different colors correspond to different irreps of $W(F_{4})$.  

\noindent The behavior of the energy levels is similar to the 2D and 3D cases, and we can easily identify a few low-lying operators. We note that the 4D Ising model flows to a Gaussian fixed point in  $d=4$; for example, the operators $\phi$, $\phi^2$ should have dimensions $\Delta_{\phi}=1$, $\Delta_{\phi^{2}}=2$. 
However, due to finite-size corrections for the Ising model on the 24-cell, the energy gaps are shifted and we do not observe the Gaussian fixed-point behavior for most of the levels, nevertheless one can notice that  there is a pseudo-critical point $h_{x}^{*}\approx 7.5$ where $\Delta_{\phi^2} \approx \Delta_{\partial \phi} \approx 2 \Delta_{\phi}$. It would be interesting to apply Conformal Perturbation Theory to this model to better observe the Gaussian CFT behavior.

When the imaginary magnetic field $ih_{z}$ is turned on, we observe a similar behavior for the energy gaps 
as in the 2D and 3D cases: the two lowest energy levels merge at some value $ih_{z}^{\textrm{merg}}$ . We list the values of this 
merging point in the Table \ref{Merg4Dsolids}.
\begin{table}[h!]
\centering
\begin{tabular}{l c c c c c c c c c c}
\hline
\hline
$h_{x}$ & $15$ & $20$ & $30$ & $40$  & $50$ & $60$ & $70$ & $80$ & $90$ & $100$\\
\hline
$ h_{z}^{\textrm{merg}}$  & 4.090  & 7.470&15.021& 23.135& 31.572  &40.222  &49.025  &57.945 & 66.956  &76.042\\
\hline
\end{tabular}
\caption{Results for the merger points $ih_{z}^{\textrm{merg}}$ of the Yang-Lee model on 24-cell}
\label{Merg4Dsolids}
\end{table}

To estimate the anomalous dimensions of the 4D Yang-Lee operators we use the criticality criteria for pseudo-critical point $i h_{z}^{*}$ which, corresponds to fixing the ratio
\begin{align}
r_{T} \equiv \frac{E_{T}-E_{\textrm{I}}}{E_{\partial \phi} - E_{\phi}}
\end{align}
to be exactly $r_{T} =4$. For the 24-cell this corresponds to 
$r_{T} = (E_{1}^{(\textbf{9})}-E^{(\textbf{1})}_{0})/(E_{0}^{(\textbf{4})} - E_{1}^{(\textbf{1})})$.
We list values $ih_{z}^{*}$, where $r_{T} = 4$  along with the anomalous dimensions of the scalar operators $\phi, \Box \phi, \Box^{2}\phi$ and $\phi^{3}$ in the Table \ref{Results4Dsolids}.

\begin{table}[h]
\centering
\begin{tabular}{l c c c c c c c c c c }
\hline
\hline
$h_{x}$  &$15$  & $20$ & $30$ & $40$  & $50$ & $60$ & $70$ & $80$ & $90$ & $100$\\
\hline
\noalign{\vskip 2pt}   
$ h_{z}^{*}$ & 3.839  &7.233&14.783 & 22.890& 31.317  &39.958  &48.753  &57.664 & 66.668  &75.746\\

$ \Delta_{\phi}$  & 0.976 & 0.944&0.914&  0.900& 0.891  &0.886  &0.881  &0.878 & 0.876  &0.874\\

$ \Delta_{\Box\phi }$  & 2.796 & 2.876 & 2.951 & 2.987& 3.009  &3.023  &3.034 &3.042 & 3.049  &3.054\\

$ \Delta_{\Box^{2}\phi }$& 4.593 & 4.607 & 4.638 & 4.658 & 4.672  &4.682  &4.691  &4.697 & 4.702  &4.707\\

$ \Delta_{\phi^{3}}$&       4.842 &5.076 & 5.293 & 5.401& 5.467  &5.513  &5.546  &5.575 & 5.597  &5.605\\
\noalign{\vskip 2pt}   
\hline
\hline
\end{tabular}
\caption{Results for the  anomalous dimensions  $\Delta_{\phi}, \Delta_{\Box \phi},  \Delta_{\Box^{2} \phi}$ and $\Delta_{\phi^{3}}$ at the pseudo-critical point $ih_{z}^{*}$ ($r_{T}=4$). }
\label{Results4Dsolids}
\end{table}
\noindent As in 2D and 3D cases, we expect the finite size corrections to be smaller for reasonably large $h_{x}$. We can see that at $h_{x}=50$ the 24-cell result  $\Delta_{\phi}\approx 0.891$ is within $8\%$ from the Pad\'{e} estimate $\Delta_{\phi} \approx 0.827$ (Table \ref{2SPadeYLest}); the latter is very close to the high-temperature expansion results \cite{Butera:2012tq}.
The $24$-cell result  $\Delta_{\phi^{3}}\approx 5.467 $ is within $5\%$ from the Pad\'{e} estimate $\Delta_{\phi^{3}} \approx 5.21$. We also notice that, as in 2D and 3D, the pseudo-critical points $ih_{z}^{*}$ obtained by fixing the ratio $r_{T}=4$ are slightly away from the merging points listed in Table \ref{Merg4Dsolids}.

There are five other regular 4D polytopes one can consider: the 5-cell, 16-cell, 8-cell, 120-cell, and 600-cell. The 8-cell (Hypercube) has only $N=16$ vertices, so the 4D Yang–Lee or 4D Ising (Gaussian) models can be easily analyzed numerically on this polytope via exact diagonalization. On the other hand, the 120-cell and 600-cell have $600$ and $120$ vertices, respectively, making exact diagonalization unfeasible. However, modern numerical methods such as Neural Quantum States \cite{Carleo_2017} could potentially be used to analyze the 4D Yang–Lee or 4D Ising models (the latter flows to the Gaussian one) on these large regular 4D polytopes. We leave this for future work.

\section{Future Directions}
\label{Future}

The methods we used for studying the quantum criticality of the Yang-Lee model in various dimensions can be used to calculate other observables in this non-unitary CFT. They include the entanglement entropy between the two hemispheres of $S^{d-1}$, as was done in \cite{Hu:2024pen} for the $d=3$ Ising model. The latter can lead to the determination of the sphere free energy $\tilde F$ for the YL universality class, which may be compared with the results from the $6-\epsilon$ expansion \cite{Giombi:2024zrt}. 

One of the new results in our paper is the use of the $24$-cell to approximate the $S^3$. The results for the lowest operator dimension in $d=4$ are quite good, but not as precise as in $d=3$. Obviously, it would be important to develop more precise methods for study the $d=4$ YL CFT, such as a fuzzy $S^3$ where the size of the discretization can be varied.  

We would also like to use similar methods to study other non-unitary universality classes. For example, the $D_5$ modular invariant of minimal model $M(3,8)$ has a Ginzburg-Landau description using a field theory of two scalar fields with cubic interactions containing imaginary coupling 
constants \cite{Fei:2014xta,Klebanov:2022syt,Katsevich:2024jgq}. This model may be obtained by RG flow from the $D_6$ modular invariant of $M(3,10)$
\cite{Fei:2014xta,Klebanov:2022syt,Nakayama:2024msv,Katsevich:2024jgq,Delouche:2024yuo}, and the latter is a product of two Yang-Lee models \cite{Kausch:1996vq,Quella:2006de}.
It would be interesting to calculate observables in the $M(3,8)$ universality class with the methods used in this paper and compare them with the $6-\epsilon$ expansions and the exact results in $d=2$. 

Our work also provides useful input for future conformal bootstrap studies. For non-unitary CFTs it is impossible to use the conventional convex optimization methods, but setups based on solving truncated crossing equations\cite{Gliozzi:2013ysa,Gliozzi:2014jsa,Hikami:2017hwv} can still be applied. Without positivity this method is not rigorous, so one need to start with a guess of the qualitative structure of the solution, and the CFT data we provided in this paper will be useful in this respect. Remarkably, there are recent works \cite{Poland:2023bny,Poland:2023vpn} on 5-point function bootstrap using the truncated crossing equation method, which successfully extracted spin OPE coefficients such as $C_{ \phi T T}$ for the 3D Ising model. Such quantities can also be obtained from bootstrap of the mixed $\phi$-$T$ 4-point functions system, which was recently done in \cite{Chang:2024whx}. Combining our results with these bootstrap techniques can hopefully improve our understanding of non-unitary CFTs.

\section*{Acknowledgments}
We are grateful to Richard Brower, Liam Fitzpatrick, Simone Giombi, Mina Himwich, Yin-Chen He, Luca Iliesiu, Andrei Katsevich, Yue-Zhou Li, Max Metlitski, David Poland, Zimo Sun, Matthew Walters, Zheng Zhou and Yijian Zou for useful discussions. 
We also thank the participants of the Princeton Center for Theoretical Science workshop ``Spheres of Influence" for stimulating interactions. Some of this research was carried out during the Kavli Institute for Theoretical Physics (KITP) workshop
``Lattice and Continuum Approaches to Strongly Coupled QFT" and supported in part by grant NSF PHY-2309135 to KITP. 
The work of EAC, GT and YX was supported in part by the DOE under Grant No.~DE-SC0010118 and by the 
Simons Foundation under Grant No. 994316.
The work of IRK  was supported in part by the US NSF under Grant No.~PHY-2209997 and by the Simons Foundation under Grant No.~917464.

\appendix

\section{Characters of $W(F_{4})$}
\label{charWF4}

We denote representatives of the conjugacy classes of $W(F_{4})$ by  $R_{q_{\textrm{L}}, q_{\textrm{R}}}$ for  proper rotations and by $\tilde{R}_{q_{\textrm{L}}, q_{\textrm{R}}}$ for  improper rotations  \cite{sym2031423}, where $q_{\textrm{L}}$ and $q_{\textrm{R}}$ are two unit quaternions: $qq^{\dag}=\sum_{i=0}^{3}q_{i}^{2}=1$, where $q= q_{0}+q_{1}\textbf{i}+q_{2}\textbf{j}+q_{3}\textbf{k}$, $q^{\dag}= q_{0}-q_{1}\textbf{i}-q_{2}\textbf{j}-q_{3}\textbf{k}$ and $\textbf{i}^{2}=\textbf{j}^{2}=\textbf{k}^{2}=\textbf{i}\textbf{j}\textbf{k}=-1$.  The actions of the proper rotation  $R_{q_{\textrm{L}}, q_{\textrm{R}}}$ and the improper rotation  $\tilde{R}_{q_{\textrm{L}}, q_{\textrm{R}}}$ on a vector $q=(q_{0},q_{1},q_{2},q_{3})$ in 4D space, represented as a quaternion, are 
\begin{align}
R_{q_{\textrm{L}}, q_{\textrm{R}}} (q) = q_{\textrm{L}}qq_{\textrm{R}}^{\dag}, \quad \tilde{R}_{q_{\textrm{L}}, q_{\textrm{R}}} (q) = q_{\textrm{L}}q^{\dag}q_{\textrm{R}}^{\dag}\,.
\end{align}
The explicit expressions for the unit quaternions $q_{1}, q_{2}, \dots  q_{8}$ that define the representatives 
of the conjugacy classes of $W(F_{4})$ are  \cite{sym2031423}:
\begin{align}
&q_{1}= (1, 0, 0, 0), \quad q_{2}= (1, 0, 0, 0),  \quad q_{4}= (0, 0, 1, 1)/\sqrt{2}, \quad q_{5}= (1, 0, 0, 1)/\sqrt{2}, \quad  \notag\\
&q_{6}= (1, 1, 0, 0)/\sqrt{2}, \quad q_{7}= (1, 1, 1,1)/2, \quad q_{8}= (1, -1, -1,-1)/2\,.
\end{align}
The characters of $W(F_{4})$ were computed in \cite{kondo1965characters}  (see also \cite{carter1993finite, GeckGoetz}).
We present the characters of $W(F_{4})$ for the proper rotations in the Table \ref{CharW4prop} and for the improper rotations in the Table \ref{CharW4improp}.

\begin{landscape}

  \begin{table}[h!]
  \caption{Characters of $W(F_{4})$ (proper rotations)}
  \label{CharW4prop} 
    \setlength{\tabcolsep}{4pt}   
\begin{tabular}{l r r r r r r r r r r r r r r r r   }
\hline
\hline
\noalign{\vskip 2pt}   
Conjugacy class & $1$ & $4A_{1}$ & $2A_{1}$ & $A_{2}$ & $D_{4}$ & $D_{4}(a_{1})$ & $\tilde{A}_{2}$ & $C_{3}+ A_{1}$ & $A_{2}+ \tilde{A}_{2}$ & $F_{4}(a_{1})$ & $F_{4}$ & $A_{1}+ \tilde{A}_{1}$ & $B_{2}$ & $A_{3}+ \tilde{A}_{1}$ & $B_{4}$  \\
\noalign{\vskip 2pt}    
Size &  $1$ &  $1$ &  $18$ & $32$ &  $32$ & $12$ & $32$ & $32$  & $16$  & $16$ & $96$ & $72$ & $36$ & $36$ & $144$ \\
\noalign{\vskip 2pt}   
Representative &  $R_{q_{1},q_{1}}$ &  $R_{q_{1},-q_{1}}$ &  $R_{q_{2},-q_{2}}$ & $R_{q_{7},q_{7}}$ &  $R_{q_{7},-q_{7}}$& $R_{q_{2},q_{1}}$ & $R_{q_{7},q_{8}}$ & $R_{q_{7},-q_{8}}$  & $R_{q_{7},-q_{1}}$  & $R_{q_{7},q_{1}}$  & $R_{q_{2},q_{7}}$ & $R_{q_{4},q_{4}}$ & $R_{q_{6},q_{6}}$  & $R_{q_{6},-q_{6}}$ &  $R_{q_{6},-q_{4}}$   \\
\noalign{\vskip 2pt}      
  \hhline{|----------------|} 
$\textbf{1}(\phi_{1,0})$ & $1$ & $1$ & $1$ & $1$ & $1$ & $1$ & $1$ & $1$ & $1$ & $1$ & $1$ & $1$ & $1$ & $1$ & $1$  \\  
$\textbf{1}'(\phi_{1,12}'')$ & $1$ & $1$ & $1$ & $1$ & $1$ & $1$ & $1$ & $1$ & $1$ & $1$ & $1$ & $-1$ & $-1$ & $-1$ & $-1$ \\ 
$\textbf{1}''(\phi_{1,12}')$ & $1$ & $1$ & $1$ & $1$ & $1$ & $1$ & $1$ & $1$ & $1$ & $1$ & $1$ & $-1$ & $-1$ & $-1$ & $-1$ \\ 
$\textbf{1}'''(\phi_{1,24})$ & $1$ & $1$ & $1$ & $1$ & $1$ & $1$ & $1$ & $1$ & $1$ & $1$ & $1$ & $1$ & $1$ & $1$ & $1$ \\ 
  \hhline{|----------------|} 
$\textbf{2}(\phi_{2,4}'')$ & $2$ & $2$ & $2$ & $2$ & $2$ & $2$ & $-1$ & $-1$ & $-1$ & $-1$ & $-1$ & $0$ & $0$ & $0$ & $0$  \\ 
$\textbf{2}'(\phi_{2,4}')$ & $2$ & $2$ & $2$ & $-1$ & $-1$ & $2$ & $2$ & $2$ & $-1$ & $-1$ & $-1$ & $0$ & $0$ & $0$ & $0$  \\ 
$\textbf{2}''(\phi_{2,16}')$ & $2$ & $2$ & $2$ & $2$ & $2$ & $2$ & $-1$ & $-1$ & $-1$ & $-1$ & $-1$ & $0$ & $0$ & $0$ & $0$ \\ 
$\textbf{2}'''(\phi_{2,16}'')$ & $2$ & $2$ & $2$ & $-1$ & $-1$ & $2$ & $2$ & $2$ & $-1$ & $-1$ & $-1$ & $0$ & $0$ & $0$ & $0$  \\ 
  \hhline{|----------------|} 
$\textbf{4}(\phi_{4,13})$ & $4$ & $-4$ & $0$ & $1$ & $-1$ & $0$ & $1$ & $-1$ & $-2$ & $2$ & $0$ & $0$ & $2$ & $-2$ & $0$  \\ 
$\textbf{4}'(\phi_{4,8})$ & $4$ & $4$ & $4$ & $-2$ & $-2$ & $4$ & $-2$ & $-2$ & $1$ & $1$ & $1$ & $0$ & $0$ & $0$ & $0$ \\ 
$\textbf{4}''(\phi_{4,1})$ & $4$ & $-4$ & $0$ & $1$ & $-1$ & $0$ & $1$ & $-1$ & $-2$ & $2$ & $0$ & $0$ & $2$ & $-2$ & $0$  \\ 
$\textbf{4}'''(\phi_{4,7}'')$ & $4$ & $-4$ & $0$ & $1$ & $-1$ & $0$ & $1$ & $-1$ & $-2$ & $2$ & $0$ & $0$ & $-2$ & $2$ & $0$  \\ 
$\textbf{4}''''(\phi_{4,7}')$ & $4$ & $-4$ & $0$ & $1$ & $-1$ & $0$ & $1$ & $-1$ & $-2$ & $2$ & $0$ & $0$ & $-2$ & $2$ & $0$  \\ 
  \hhline{|----------------|} 
$\textbf{6}(\phi_{6,6}')$ & $6$ & $6$ & $-2$ & $0$ & $0$ & $2$ & $0$ & $0$ & $3$ & $3$ & $-1$ & $2$ & $-2$ & $-2$ & $0$  \\ 
$\textbf{6}'(\phi_{6,6}'')$ & $6$ & $6$ & $-2$ & $0$ & $0$ & $2$ & $0$ & $0$ & $3$ & $3$ & $-1$ & $-2$ & $2$ & $2$ & $0$  \\ 
  \hhline{|----------------|} 
 $\textbf{8}(\phi_{8,9}')$ & $8$ & $-8$ & $0$ & $2$ & $-2$ & $0$ & $-1$ & $1$ & $2$ & $-2$ & $0$ & $0$ & $0$ & $0$ & $0$  \\ 
 $\textbf{8}'(\phi_{8,9}'')$ & $8$ & $-8$ & $0$ & $-1$ & $1$ & $0$ & $2$ & $-2$ & $2$ & $-2$ & $0$ & $0$ & $0$ & $0$ & $0$  \\   
$\textbf{8}''(\phi_{8,3}'')$ & $8$ & $-8$ & $0$ & $2$ & $-2$ & $0$ & $-1$ & $1$ & $2$ & $-2$ & $0$ & $0$ & $0$ & $0$ & $0$  \\ 
$\textbf{8}'''(\phi_{8,3}')$ & $8$ & $-8$ & $0$ & $-1$ & $1$ & $0$ & $2$ & $-2$ & $2$ & $-2$ & $0$ & $0$ & $0$ & $0$ & $0$ \\ 
  \hhline{|----------------|} 
$\textbf{9}(\phi_{9,2})$ & $9$ & $9$ & $1$ & $0$ & $0$ & $-3$ & $0$ & $0$ & $0$ & $0$ & $0$ & $1$ & $1$ & $1$ & $-1$   \\ 
$\textbf{9}'(\phi_{9,6}'')$ & $9$ & $9$ & $1$ & $0$ & $0$ & $-3$ & $0$ & $0$ & $0$ & $0$ & $0$ & $-1$ & $-1$ & $-1$ & $1$\\ 
$\textbf{9}''(\phi_{9,6}')$ & $9$ & $9$ & $1$ & $0$ & $0$ & $-3$ & $0$ & $0$ & $0$ & $0$ & $0$ & $-1$ & $-1$ & $-1$ & $1$  \\ 
$\textbf{9}'''(\phi_{9,10})$ & $9$ & $9$ & $1$ & $0$ & $0$ & $-3$ & $0$ & $0$ & $0$ & $0$ & $0$ & $1$ & $1$ & $1$ & $-1$  \\
  \hhline{|----------------|} 
$\textbf{12}(\phi_{12,4})$ & $12$ & $12$ & $-4$ & $0$ & $0$ & $4$ & $0$ & $0$ & $-3$ & $-3$ & $1$ & $0$ & $0$ & $0$ & $0$ \\ 
$\textbf{16}(\phi_{16,1})$ & $16$ & $-16$ & $0$ & $-2$ & $2$ & $0$ & $-2$ & $2$ & $-2$ & $2$ & $0$ & $0$ & $0$ & $0$ & $0$  \\ 
\hline
\hline
\end{tabular}
\end{table}

\newpage

\begin{table}[h!]
\center
\caption{Characters of $W(F_{4})$ (improper rotations)}
\label{CharW4improp} 
  
\begin{tabular}{l  r r r r r r r r r r r  }
\hline
\hline
\noalign{\vskip 2pt} 
Conjugacy class  &  $A_{1}$ & $3A_{1}$ & $\tilde{A}_{2}+ A_{1}$ & $C_{3}$ & $A_{3}$ & $\tilde{A}_{1}$ & $2A_{1}+ \tilde{A}_{1}$ & $A_{2}+ \tilde{A}_{1}$ & $B_{3}$ & $B_{2}+ A_{1}$ \\ 
\noalign{\vskip 2pt} 
 Size &   $12$  & $12$ & $96$ & $96$  & $72$  & $12$ & $12$  & $96$  & $96$  & $72$  \\  
\noalign{\vskip 2pt}   
Representative &   $\tilde{R}_{q_{4},-q_{4}}$ & $\tilde{R}_{q_{4},q_{4}}$ & $\tilde{R}_{q_{6},-q_{5}}$  & $\tilde{R}_{q_{6}q_{5}}$  & $\tilde{R}_{q_{6},q_{4}}$   & $\tilde{R}_{q_{1}q_{1}}$  & $\tilde{R}_{q_{2}q_{2}}$   & $\tilde{R}_{q_{2}q_{7}}$  & $\tilde{R}_{q_{7}q_{1}}$   & $\tilde{R}_{q_{2}q_{1}}$   \\
\noalign{\vskip 2pt}    
  \hhline{|------------|} 
$\textbf{1}(\phi_{1,0})$ &  $1$ & $1$ & $1$ & $1$ & $1$ & $1$ & $1$ & $1$ & $1$ & $1$ \\ 
$\textbf{1}'(\phi_{1,12}'')$ &  $1$ & $1$ & $1$ & $1$ & $1$ & $-1$ & $-1$ & $-1$ & $-1$ & $-1$ \\ 
$\textbf{1}''(\phi_{1,12}')$ &  $-1$ & $-1$ & $-1$ & $-1$ & $-1$ & $1$ & $1$ & $1$ & $1$ & $1$ \\ 
$\textbf{1}'''(\phi_{1,24})$ &  $-1$ & $-1$ & $-1$ & $-1$ & $-1$ & $-1$ & $-1$ & $-1$ & $-1$ & $-1$ \\ 
  \hhline{|------------|} 
$\textbf{2}(\phi_{2,4}'')$ & $2$ & $2$ & $-1$ & $-1$ & $2$ & $0$ & $0$ & $0$ & $0$ & $0$ \\ 
$\textbf{2}'(\phi_{2,4}')$  & $0$ & $0$ & $0$ & $0$ & $0$ & $2$ & $2$ & $-1$ & $-1$ & $2$ \\ 
$\textbf{2}''(\phi_{2,16}')$ & $-2$ & $-2$ & $1$ & $1$ & $-2$ & $0$ & $0$ & $0$ & $0$ & $0$ \\ 
$\textbf{2}'''(\phi_{2,16}'')$ & $0$ & $0$ & $0$ & $0$ & $0$ & $-2$ & $-2$ & $1$ & $1$ & $-2$ \\ 
  \hhline{|------------|} 
$\textbf{4}(\phi_{4,13})$ & $-2$ & $2$ & $1$ & $-1$ & $0$ & $-2$ & $2$ & $1$ & $-1$ & $0$ \\ 
$\textbf{4}'(\phi_{4,8})$  & $0$ & $0$ & $0$ & $0$ & $0$ & $0$ & $0$ & $0$ & $0$ & $0$ \\ 
$\textbf{4}''(\phi_{4,1})$  & $2$ & $-2$ & $-1$ & $1$ & $0$ & $2$ & $-2$ & $-1$ & $1$ & $0$ \\ 
$\textbf{4}'''(\phi_{4,7}'')$  & $2$ & $-2$ & $-1$ & $1$ & $0$ & $-2$ & $2$ & $1$ & $-1$ & $0$ \\ 
$\textbf{4}''''(\phi_{4,7}')$  & $-2$ & $2$ & $1$ & $-1$ & $0$ & $2$ & $-2$ & $-1$ & $1$ & $0$ \\ 
  \hhline{|------------|} 
$\textbf{6}(\phi_{6,6}')$ &  $0$ & $0$ & $0$ & $0$ & $0$ & $0$ & $0$ & $0$ & $0$ & $0$ \\ 
$\textbf{6}'(\phi_{6,6}'')$ &  $0$ & $0$ & $0$ & $0$ & $0$ & $0$ & $0$ & $0$ & $0$ & $0$ \\ 
  \hhline{|------------|} 
 $\textbf{8}(\phi_{8,9}')$ &  $-4$ & $4$ & $-1$ & $1$ & $0$ & $0$ & $0$ & $0$ & $0$ & $0$ \\ 
 $\textbf{8}'(\phi_{8,9}'')$  & $0$ & $0$ & $0$ & $0$ & $0$ & $-4$ & $4$ & $-1$ & $1$ & $0$ \\   
$\textbf{8}''(\phi_{8,3}'')$  & $4$ & $-4$ & $1$ & $-1$ & $0$ & $0$ & $0$ & $0$ & $0$ & $0$ \\ 
$\textbf{8}'''(\phi_{8,3}')$ & $0$ & $0$ & $0$ & $0$ & $0$ & $4$ & $-4$ & $1$ & $-1$ & $0$ \\ 
  \hhline{|------------|} 
$\textbf{9}(\phi_{9,2})$ & $3$ & $3$ & $0$ & $0$ & $-1$ & $3$ & $3$ & $0$ & $0$ & $-1$ \\ 
$\textbf{9}'(\phi_{9,6}'')$  & $3$ & $3$ & $0$ & $0$ & $-1$ & $-3$ & $-3$ & $0$ & $0$ & $1$ \\ 
$\textbf{9}''(\phi_{9,6}')$  & $-3$ & $-3$ & $0$ & $0$ & $1$ & $3$ & $3$ & $0$ & $0$ & $-1$ \\ 
$\textbf{9}'''(\phi_{9,10})$ & $-3$ & $-3$ & $0$ & $0$ & $1$ & $-3$ & $-3$ & $0$ & $0$ & $1$ \\
  \hhline{|------------|} 
$\textbf{12}(\phi_{12,4})$ & $0$ & $0$ & $0$ & $0$ & $0$ & $0$ & $0$ & $0$ & $0$ & $0$ \\ 
$\textbf{16}(\phi_{16,1})$ & $0$ & $0$ & $0$ & $0$ & $0$ & $0$ & $0$ & $0$ & $0$ & $0$ \\ 
\hline\hline
\end{tabular}
\end{table}

\end{landscape}

\section{Constructing Spinning Correlation Functions in Fuzzy Sphere}
\label{app:spin-corrlator}
In this appendix, we show an explicit construction of the spinning correlation functions using the fuzzy sphere observables. To fix our convention, we normalize the operators to have the standard form of 2-point function\cite{Poland:2018epd}
\begin{equation}
   \langle \CO(x_1,z_1) \CO(x_2,z_2) \rangle = \frac{\left( (z_1 \cdot z_2) - (x_{12}\cdot z_1)(x_{12}\cdot z_2)/x_{12}^2 \right)^\ell}{x_{12}^{2\Delta}}
\end{equation}
and 3-point function\cite{Costa:2011mg,Meltzer:2018tnm}
\begin{equation}
   \langle \CO_1 (x_1,z_1) \phi (x_2) \CO_3 (x_3,z_3) \rangle = \sum_{\substack{s=|\ell_1-\ell_3|\\ s-|\ell_1-\ell_3| = 0 \mod 2}}^{\ell_1 + \ell_3} C_{\CO_1\phi\CO_3}^{(s)} \frac{H_{13}^{\frac{1}{2}(\ell_1+\ell_3-s)}V_{1,2,3}^{\frac{1}{2}(\ell_1-\ell_3+s)}V_{3,1,2}^{\frac{1}{2}(-\ell_1+\ell_3+s)}}{x_{12}^{\kappa_1+\kappa_2-\kappa_3} x_{13}^{\kappa_1+\kappa_3-\kappa_2} x_{23}^{\kappa_2+\kappa_3-\kappa_1}}
\end{equation}
where $\kappa_i = \Delta_i + \ell_i$, $x_{ij} = x_i - x_j$. The building blocks are 
\begin{equation}
\begin{aligned}
H_{ij} &= \left(x_i-x_j\right)\cdot \left(x_i-x_j\right) z_i\cdot z_j-2 \left(x_i-x_j\right)\cdot z_i \left(x_i-x_j\right)\cdot z_j\\ 
V_{i, j,k} &= \frac{\left(x_i-x_j\right)\cdot \left(x_i-x_j\right) \left(x_i-x_k\right)\cdot z_i
- \left(x_i-x_k\right)\cdot \left(x_i-x_k\right) \left(x_i-x_j\right)\cdot z_i
}{\left(x_j-x_k\right)\cdot \left(x_j-x_k\right)}
\end{aligned}
\end{equation}
The above formulas are in the index-free notation, and the indices can be recovered using the Todorov derivative
\begin{equation}
\begin{aligned}
   \CO^{\mu_1\mu_2\cdots \mu_\ell}(x) &= \frac{1}{l!(1/2)_\ell} D_{z,\mu_1} D_{z,\mu_2} \cdots D_{z,\mu_\ell} \CO(x,z)  \\ 
   D_{z,\mu} &= \left( \frac{d}{2}-1 + z\cdot \frac{\partial}{\partial z}\right) \frac{\partial}{\partial z^\mu} - 
\frac{1}{2}z_\mu \frac{\partial^2}{\partial z \cdot \partial z}~,
\end{aligned}
\end{equation}
where $d=3$.
We match the CFT radial quantization states to the fuzzy sphere states
\begin{equation}
\begin{aligned}
   | \CO_{\ell m} \rangle &= v_{\ell,\ell,m}^{\mu_1\mu_2\cdots\mu_\ell} \lim_{x\rightarrow 0} \CO_{\mu_1\mu_2\cdots \mu_\ell}(x) | 0 \rangle  \\ 
   &\equiv \frac{\big(\mathbf{v}_{\ell,\ell,m} \cdot D_z^\ell \big)}{\ell!(1/2)_\ell} \lim_{x\rightarrow 0}\CO(x,z) | 0 \rangle
\end{aligned}
\end{equation}
where the tensor $\mathbf{v}_{n,\ell,m}$ ($v_{\ell,\ell,m}^{\mu_1\mu_2\cdots\mu_\ell}$ being the component form) projects the tensor product of $n$ indices to the $(\ell,m)$ representation. The vector is normalized to $\mathbf{v}\cdot \mathbf{v} = 1$. A construction of such vectors is shown below
\begin{equation}
   \mathbf{v}_{n,\ell,m} = \frac{1}{n} \sum_{k=0}^{n-1} \mathcal{R}^k \big( \mathbf{v}_{1,1,m'} \otimes \mathbf{v}_{n-1,|\ell-1|,m-m'} \big) 
   \big\langle |\ell-1|,m-m';1,m' \big| \ell, m \big\rangle
\end{equation}
where $\mathcal{R}: f^{\mu_1 \mu_2 \cdots \mu_n} \mapsto f^{\mu_n \mu_1 \cdots \mu_{n-1}}$ generates the cyclic permutation and $\mathcal{R}^k$ is permuting $k$ times. $\big\langle |\ell-1|,m-m';1,m' \big| \ell, m \big\rangle$ is the Clebsch-Gordan coefficient.
The ket-state is obtained from spacial inversion \cite{Simmons-Duffin:2016gjk}
\begin{equation}
\begin{aligned}
   \langle \CO_{\ell m} | &= v_{\ell,\ell,m}^{\mu_1,\mu_2\cdots\mu_\ell} \lim_{x\rightarrow 0} x^{-2\Delta_\CO} I_{\mu_1\nu_1}(x) I_{\mu_2\nu_2}(x) \cdots I_{\mu_\ell\nu_\ell}(x)   \CO^{\nu_1\nu_2\cdots \nu_\ell}\left(\frac{x}{|x|^2}\right) | 0 \rangle  \\ 
   &= \frac{\big(\mathbf{v}_{\ell,\ell,m} \cdot D_z^\ell \big)}{\ell!(1/2)_\ell} \lim_{x\rightarrow \infty} x^{2\Delta_\CO}\CO(x,z-2(x\cdot z)z) | 0 \rangle
\end{aligned}
\end{equation}
where $I_{\mu\nu} = \delta_{\mu\nu} - 2\frac{x_\mu x_\nu}{x^2}$. Strictly speaking, the ket-state is not quite the conjugation but the {\it left eigenvector} and the inversion will also take $\CO \mapsto \CO^\dagger$. In YL model the left and right eigenvector are related by complex conjugation which cancels the conjugation of the operator if it is not real. The net effect is that we do not need to worry about the conjugation of $\CO$.
By extrapolating $\CO_1$ and $\CO_3$ to $\infty$ and $0$, we get a much cleaner expression
\begin{equation}
    \langle \CO (z_1) | \CO(x_2) | \CO(z_3) \rangle \propto (z_1\cdot z_3)^{\frac{1}{2}(\ell_1+\ell_3-\ell_2)} (z_1\cdot x_2)^{\frac{1}{2}(\ell_1+\ell_2-\ell_3)} (z_3\cdot x_2)^{\frac{1}{2}(\ell_2+\ell_3-\ell_1)} 
\end{equation}
We match the leading behavior fuzzy sphere operator $M_{\ell m}^a$ to the integral of $\phi$ over the sphere weighted by spherical harmonics 
\begin{equation}
   M_{\ell m}^a = b_{a,0} \delta_{\ell,0} \mathbbm{1} + b_{a,1} \int d^2\Omega \, Y_{\ell m}(\Omega) \phi(\Omega) + \cdots
\end{equation}
and this integral can be alternatively evaluated by contracting the spinning tensor indices to $\mathbf{v}$ tensors but this time the variable carrying the indices is the spacial coordinate $x$ (which is also $\Omega$)
\begin{equation}
   \int d^2\Omega \, Y_{\ell m}(\Omega)  = \alpha_{\ell}  \frac{\big(\mathbf{v}_{\ell,\ell,m} \cdot D_x^\ell \big)}{\ell!(1/2)_\ell}, \qquad \alpha_{\ell} = \sqrt{ \frac{4\pi(1/2)_\ell}{(1+2\ell)2^{-\ell}\ell!} }
\end{equation}
This helps us to obtain the spin-dependent coefficient in the fuzzy sphere matrix element
\begin{equation}
\begin{aligned}
\langle \CO, \ell | M_{s}^a | \CO', \ell' \rangle &= \delta_{\CO,\CO'} \delta_{s,0} b_0 +  b_{a,1} f_{\ell,\ell',s} C_{\CO\phi\CO'}^{(s)} + \cdots \\
        f_{\ell,\ell',s} &= \alpha_{s} \frac{\big(\mathbf{v}_{s,s,0} \cdot D_{x_2}^s \big)}{s!(1/2)_s} 
    \frac{\big(\mathbf{v}_{\ell,\ell,0} \cdot D_{z_1}^\ell \big)}{\ell!(1/2)_\ell}
    \frac{\big(\mathbf{v}_{\ell',\ell',0} \cdot D_{z_3}^{\ell'} \big)}{\ell'!(1/2)_{\ell'}} \\ 
    &~~~~(z_1\cdot z_3)^{\frac{1}{2}(\ell_1+\ell_3-s)} (z_1\cdot x_2)^{\frac{1}{2}(\ell_1+s-\ell_3)} (z_3\cdot x_2)^{\frac{1}{2}(s+\ell_3-\ell_1)} ~.
    \end{aligned}
\end{equation}

\typeout{} 
\bibliographystyle{ssg}
\bibliography{YLrefs}

\begingroup\raggedright\begin{thebibliography}{100}

\bibitem{Ising:1925em}
E.~Ising, ``{Contribution to the Theory of Ferromagnetism},'' {\em Z. Phys.}
  {\bf 31} (1925) 253--258.

\bibitem{Onsager:1943jn}
L.~Onsager, ``{Crystal statistics. 1. A Two-dimensional model with an order
  disorder transition},'' {\em Phys. Rev.} {\bf 65} (1944) 117--149.

\bibitem{Belavin:1984vu}
A.~Belavin, A.~M. Polyakov, and A.~Zamolodchikov, ``{Infinite Conformal
  Symmetry in Two-Dimensional Quantum Field Theory},'' {\em Nucl.Phys.} {\bf
  B241} (1984) 333--380.

\bibitem{Wilson:1971dc}
K.~G. Wilson and M.~E. Fisher, ``{Critical exponents in 3.99 dimensions},''
  {\em Phys.Rev.Lett.} {\bf 28} (1972) 240--243.

\bibitem{Rattazzi:2008pe}
R.~Rattazzi, V.~S. Rychkov, E.~Tonni, and A.~Vichi, ``{Bounding scalar operator
  dimensions in 4D CFT},'' {\em JHEP} {\bf 0812} (2008) 031,
  \href{https://arxiv.org/abs/0807.0004}{{\tt 0807.0004}}.

\bibitem{El-Showk:2012cjh}
S.~El-Showk, M.~F. Paulos, D.~Poland, S.~Rychkov, D.~Simmons-Duffin, and
  A.~Vichi, ``{Solving the 3D Ising Model with the Conformal Bootstrap},'' {\em
  Phys. Rev. D} {\bf 86} (2012) 025022,
  \href{https://arxiv.org/abs/1203.6064}{{\tt 1203.6064}}.

\bibitem{Polyakov:1970xd}
A.~M. Polyakov, ``{Conformal symmetry of critical fluctuations},'' {\em JETP
  Lett.} {\bf 12} (1970) 381--383.

\bibitem{Poland:2018epd}
D.~Poland, S.~Rychkov, and A.~Vichi, ``{The Conformal Bootstrap: Theory,
  Numerical Techniques, and Applications},'' {\em Rev. Mod. Phys.} {\bf 91}
  (2019) 015002, \href{https://arxiv.org/abs/1805.04405}{{\tt 1805.04405}}.

\bibitem{Simmons-Duffin:2016gjk}
D.~Simmons-Duffin, ``{The Conformal Bootstrap},'' in {\em {Theoretical Advanced
  Study Institute in Elementary Particle Physics}: {New Frontiers in Fields and
  Strings}}, pp.~1--74, 2017.
\newblock \href{https://arxiv.org/abs/1602.07982}{{\tt 1602.07982}}.

\bibitem{Rychkov:2016iqz}
S.~Rychkov, {\em {EPFL Lectures on Conformal Field Theory in D\ensuremath{>}= 3
  Dimensions}}.
\newblock SpringerBriefs in Physics. 1, 2016.

\bibitem{Chester:2019wfx}
S.~M. Chester, ``{Weizmann lectures on the numerical conformal bootstrap},''
  {\em Phys. Rept.} {\bf 1045} (2023) 1--44,
  \href{https://arxiv.org/abs/1907.05147}{{\tt 1907.05147}}.

\bibitem{Cappelli:2018vir}
A.~Cappelli, L.~Maffi, and S.~Okuda, ``{Critical Ising Model in Varying
  Dimension by Conformal Bootstrap},'' {\em JHEP} {\bf 01} (2019) 161,
  \href{https://arxiv.org/abs/1811.07751}{{\tt 1811.07751}}.

\bibitem{Henriksson:2022gpa}
J.~Henriksson, S.~R. Kousvos, and M.~Reehorst, ``{Spectrum continuity and level
  repulsion: the Ising CFT from infinitesimal to finite
  \ensuremath{\varepsilon}},'' {\em JHEP} {\bf 02} (2023) 218,
  \href{https://arxiv.org/abs/2207.10118}{{\tt 2207.10118}}.

\bibitem{Zhu:2022gjc}
W.~Zhu, C.~Han, E.~Huffman, J.~S. Hofmann, and Y.-C. He, ``{Uncovering
  Conformal Symmetry in the 3D Ising Transition: State-Operator Correspondence
  from a Quantum Fuzzy Sphere Regularization},'' {\em Phys. Rev. X} {\bf 13}
  (2023), no.~2 021009, \href{https://arxiv.org/abs/2210.13482}{{\tt
  2210.13482}}.

\bibitem{Hu:2023xak}
L.~Hu, Y.-C. He, and W.~Zhu, ``{Operator Product Expansion Coefficients of the
  3D Ising Criticality via Quantum Fuzzy Spheres},'' {\em Phys. Rev. Lett.}
  {\bf 131} (2023), no.~3 031601, \href{https://arxiv.org/abs/2303.08844}{{\tt
  2303.08844}}.

\bibitem{Hu:2024pen}
L.~Hu, W.~Zhu, and Y.-C. He, ``{Entropic $F$-function of 3D Ising conformal
  field theory via the fuzzy sphere regularization},''
  \href{https://arxiv.org/abs/2401.17362}{{\tt 2401.17362}}.

\bibitem{Lao:2023zis}
B.-X. Lao and S.~Rychkov, ``{3D Ising CFT and exact diagonalization on
  icosahedron: The power of conformal perturbation theory},'' {\em SciPost
  Phys.} {\bf 15} (2023), no.~6 243,
  \href{https://arxiv.org/abs/2307.02540}{{\tt 2307.02540}}.

\bibitem{Lauchli:2025fii}
A.~M. L\"auchli, L.~Herviou, P.~H. Wilhelm, and S.~Rychkov, ``{Exact
  Diagonalization, Matrix Product States and Conformal Perturbation Theory
  Study of a 3D Ising Fuzzy Sphere Model},''
  \href{https://arxiv.org/abs/2504.00842}{{\tt 2504.00842}}.

\bibitem{Han:2023yyb}
C.~Han, L.~Hu, W.~Zhu, and Y.-C. He, ``{Conformal four-point correlators of the
  three-dimensional Ising transition via the quantum fuzzy sphere},'' {\em
  Phys. Rev. B} {\bf 108} (2023), no.~23 235123,
  \href{https://arxiv.org/abs/2306.04681}{{\tt 2306.04681}}.

\bibitem{Haldane:1983xm}
F.~D.~M. Haldane, ``{Fractional quantization of the Hall effect: A Hierarchy of
  incompressible quantum fluid states},'' {\em Phys. Rev. Lett.} {\bf 51}
  (1983) 605--608.

\bibitem{Madore:1991bw}
J.~Madore, ``{The Fuzzy sphere},'' {\em Class. Quant. Grav.} {\bf 9} (1992)
  69--88.

\bibitem{Fardelli:2024qla}
G.~Fardelli, A.~L. Fitzpatrick, and E.~Katz, ``{Constructing the Infrared
  Conformal Generators on the Fuzzy Sphere},''
  \href{https://arxiv.org/abs/2409.02998}{{\tt 2409.02998}}.

\bibitem{Fan:2024vcz}
R.~Fan, ``{Note on explicit construction of conformal generators on the fuzzy
  sphere},'' \href{https://arxiv.org/abs/2409.08257}{{\tt 2409.08257}}.

\bibitem{Przetakiewicz:2025gzi}
D.~Przetakiewicz, S.~Wessel, and F.~P. Toldin, ``{Boundary Operator Product
  Expansion Coefficients of the Three-dimensional Ising Universality Class},''
  \href{https://arxiv.org/abs/2502.14965}{{\tt 2502.14965}}.

\bibitem{Zhou:2024zud}
Z.~Zhou and Y.-C. He, ``{A new series of 3D CFTs with $\mathrm{Sp}(N)$ global
  symmetry on fuzzy sphere},'' \href{https://arxiv.org/abs/2410.00087}{{\tt
  2410.00087}}.

\bibitem{Dedushenko:2024nwi}
M.~Dedushenko, ``{Ising BCFT from Fuzzy Hemisphere},''
  \href{https://arxiv.org/abs/2407.15948}{{\tt 2407.15948}}.

\bibitem{Chen:2024jxe}
B.-B. Chen, X.~Zhang, and Z.~Yang~Meng, ``{Emergent conformal symmetry at the
  multicritical point of (2+1)D SO(5) model with Wess-Zumino-Witten term on a
  sphere},'' {\em Phys. Rev. B} {\bf 110} (2024), no.~12 125153,
  \href{https://arxiv.org/abs/2405.04470}{{\tt 2405.04470}}.

\bibitem{Chen:2023xjc}
B.-B. Chen, X.~Zhang, Y.~Wang, K.~Sun, and Z.~Y. Meng, ``{Phases of (2+1)D
  SO(5) Nonlinear Sigma Model with a Topological Term on a Sphere:
  Multicritical Point and Disorder Phase},'' {\em Phys. Rev. Lett.} {\bf 132}
  (2024), no.~24 246503, \href{https://arxiv.org/abs/2307.05307}{{\tt
  2307.05307}}.

\bibitem{Zhou:2023qfi}
Z.~Zhou, L.~Hu, W.~Zhu, and Y.-C. He, ``{SO(5) Deconfined Phase Transition
  under the Fuzzy-Sphere Microscope: Approximate Conformal Symmetry,
  Pseudo-Criticality, and Operator Spectrum},'' {\em Phys. Rev. X} {\bf 14}
  (2024), no.~2 021044, \href{https://arxiv.org/abs/2306.16435}{{\tt
  2306.16435}}.

\bibitem{Han:2023lky}
C.~Han, L.~Hu, and W.~Zhu, ``{Conformal operator content of the Wilson-Fisher
  transition on fuzzy sphere bilayers},'' {\em Phys. Rev. B} {\bf 110} (2024),
  no.~11 115113, \href{https://arxiv.org/abs/2312.04047}{{\tt 2312.04047}}.

\bibitem{Jafferis:2011zi}
D.~L. Jafferis, I.~R. Klebanov, S.~S. Pufu, and B.~R. Safdi, ``{Towards the
  F-Theorem: ${\mathcal{N}} = 2$ Field Theories on the Three- Sphere},'' {\em
  JHEP} {\bf 06} (2011) 102, \href{https://arxiv.org/abs/1103.1181}{{\tt
  1103.1181}}.

\bibitem{Klebanov:2011gs}
I.~R. Klebanov, S.~S. Pufu, and B.~R. Safdi, ``{F-Theorem without
  Supersymmetry},'' {\em JHEP} {\bf 1110} (2011) 038,
  \href{https://arxiv.org/abs/1105.4598}{{\tt 1105.4598}}.

\bibitem{Casini:2012ei}
H.~Casini and M.~Huerta, ``{On the RG running of the entanglement entropy of a
  circle},'' {\em Phys.Rev.} {\bf D85} (2012) 125016,
  \href{https://arxiv.org/abs/1202.5650}{{\tt 1202.5650}}.

\bibitem{Fei:2015oha}
L.~Fei, S.~Giombi, I.~R. Klebanov, and G.~Tarnopolsky, ``{Generalized
  $F$-Theorem and the $\epsilon$ Expansion},'' {\em JHEP} {\bf 12} (2015) 155,
  \href{https://arxiv.org/abs/1507.01960}{{\tt 1507.01960}}.

\bibitem{Yang:1952be}
C.-N. Yang and T.~D. Lee, ``{Statistical theory of equations of state and phase
  transitions. 1. Theory of condensation},'' {\em Phys. Rev.} {\bf 87} (1952)
  404--409.

\bibitem{Lee:1952ig}
T.~D. Lee and C.-N. Yang, ``{Statistical theory of equations of state and phase
  transitions. 2. Lattice gas and Ising model},'' {\em Phys. Rev.} {\bf 87}
  (1952) 410--419.

\bibitem{Kortman:1971zz}
P.~J. Kortman and R.~B. Griffiths, ``{Density of Zeros on the Lee-Yang Circle
  for Two Ising Ferromagnets},'' {\em Phys. Rev. Lett.} {\bf 27} (1971)
  1439--1442.

\bibitem{Fisher:1978pf}
M.~Fisher, ``{Yang-Lee Edge Singularity and $\phi^3$ Field Theory},'' {\em
  Phys.Rev.Lett.} {\bf 40} (1978) 1610--1613.

\bibitem{Cardy:1985yy}
J.~L. Cardy, ``{Conformal Invariance and the Yang-lee Edge Singularity in
  Two-dimensions},'' {\em Phys.Rev.Lett.} {\bf 54} (1985) 1354--1356.

\bibitem{Cardy:2023lha}
J.~Cardy, ``{The Yang-Lee Edge Singularity and Related Problems},'' 5, 2023.
\newblock \href{https://arxiv.org/abs/2305.13288}{{\tt 2305.13288}}.

\bibitem{Borinsky:2021jdb}
M.~Borinsky, J.~A. Gracey, M.~V. Kompaniets, and O.~Schnetz, ``{Five-loop
  renormalization of ${\phi}^3$ theory with applications to the Lee-Yang edge
  singularity and percolation theory},'' {\em Phys. Rev. D} {\bf 103} (2021),
  no.~11 116024, \href{https://arxiv.org/abs/2103.16224}{{\tt 2103.16224}}.

\bibitem{Kompaniets:2021hwg}
M.~Kompaniets and A.~Pikelner, ``{Critical exponents from five-loop scalar
  theory renormalization near six-dimensions},'' {\em Phys. Lett. B} {\bf 817}
  (2021) 136331, \href{https://arxiv.org/abs/2101.10018}{{\tt 2101.10018}}.

\bibitem{Schnetz:2025wtu}
O.~Schnetz, ``{$\phi^3$ theory at six loops},''
  \href{https://arxiv.org/abs/2505.15485}{{\tt 2505.15485}}.

\bibitem{Gliozzi:2013ysa}
F.~Gliozzi, ``{More constraining conformal bootstrap},'' {\em Phys. Rev. Lett.}
  {\bf 111} (2013) 161602, \href{https://arxiv.org/abs/1307.3111}{{\tt
  1307.3111}}.

\bibitem{Gliozzi:2014jsa}
F.~Gliozzi and A.~Rago, ``{Critical exponents of the 3d Ising and related
  models from Conformal Bootstrap},'' {\em JHEP} {\bf 10} (2014) 042,
  \href{https://arxiv.org/abs/1403.6003}{{\tt 1403.6003}}.

\bibitem{Hikami:2017hwv}
S.~Hikami, ``{Conformal bootstrap analysis for the Yang\textendash{}Lee edge
  singularity},'' {\em PTEP} {\bf 2018} (2018), no.~5 053I01,
  \href{https://arxiv.org/abs/1707.04813}{{\tt 1707.04813}}.

\bibitem{Brower:2024otr}
R.~C. Brower and E.~K. Owen, ``{The Ising Model on $\mathbb S^2$},''
  \href{https://arxiv.org/abs/2407.00459}{{\tt 2407.00459}}.

\bibitem{Brower:2025oti}
R.~C. Brower, G.~T. Fleming, N.~Matsumoto, and R.~Misra, ``{Ising on
  $\mathbb{S}^2$-- The Affine Conjecture},'' in {\em {41st International
  Symposium on Lattice Field Theory}}, 3, 2025.
\newblock \href{https://arxiv.org/abs/2503.05621}{{\tt 2503.05621}}.

\bibitem{vonGehlen:1991zlm}
G.~von Gehlen, ``{Critical and off critical conformal analysis of the Ising
  quantum chain in an imaginary field},'' {\em J. Phys. A} {\bf 24} (1991)
  5371--5400.

\bibitem{vonGehlen:1994rp}
G.~von Gehlen, ``{NonHermitian tricriticality in the Blume-Capel model with
  imaginary field},'' \href{https://arxiv.org/abs/hep-th/9402143}{{\tt
  hep-th/9402143}}.

\bibitem{Castro-Alvaredo:2009xex}
O.~A. Castro-Alvaredo and A.~Fring, ``{A Spin chain model with non-Hermitian
  interaction: The Ising quantum spin chain in an imaginary field},'' {\em J.
  Phys. A} {\bf 42} (2009) 465211, \href{https://arxiv.org/abs/0906.4070}{{\tt
  0906.4070}}.

\bibitem{TangY_2024}
Y.~Tang, H.~Ma, Q.~Tang, Y.-C. He, and W.~Zhu, ``Reclaiming the Lost
  Conformality in a Non-Hermitian Quantum 5-State Potts Model,'' {\em Phys.
  Rev. Lett.} {\bf 133} (Aug, 2024) 076504.

\bibitem{JesperJ_2024}
J.~L. Jacobsen and K.~J. Wiese, ``Lattice Realization of Complex Conformal
  Field Theories: Two-Dimensional Potts Model with $Q>4$ States,'' {\em Phys.
  Rev. Lett.} {\bf 133} (Aug, 2024) 077101.

\bibitem{Wiese:2023vgq}
K.~J. Wiese and J.~L. Jacobsen, ``{The two upper critical dimensions of the
  Ising and Potts models},'' {\em JHEP} {\bf 05} (2024) 092,
  \href{https://arxiv.org/abs/2311.01529}{{\tt 2311.01529}}.

\bibitem{Uzelac1981}
K.~Uzelac and R.~Jullien, ``The Yang-Lee edge singularity by the
  phenomenological renormalisation group,'' {\em Journal of Physics A:
  Mathematical and General} {\bf 14} (may, 1981) L151.

\bibitem{ArguelloCruz:2024xzi}
E.~Arguello~Cruz, G.~Tarnopolsky, and Y.~Xin, ``{Precision study of the massive
  Schwinger model near quantum criticality},''
  \href{https://arxiv.org/abs/2412.01902}{{\tt 2412.01902}}.

\bibitem{Butera:2012tq}
P.~Butera and M.~Pernici, ``{Yang-Lee edge singularities from extended activity
  expansions of the dimer density for bipartite lattices of dimensionality 2
  \ensuremath{<}= d \ensuremath{<}= 7},'' {\em Phys. Rev. E} {\bf 86} (2012)
  011104, \href{https://arxiv.org/abs/1206.0872}{{\tt 1206.0872}}.

\bibitem{Grishatalk}
G.~Tarnopolsky, ``{Numerical Analysis of the Yang-Lee Critical Point Across
  Different Dimensions}.''
\newblock Available from \url{https://www.youtube.com/watch?v=zfw63uGGiDE}.

\bibitem{Fan:2025bhc}
R.~Fan, J.~Dong, and A.~Vishwanath, ``{Simulating the non-unitary Yang-Lee
  conformal field theory on the fuzzy sphere},''
  \href{https://arxiv.org/abs/2505.06342}{{\tt 2505.06342}}.

\bibitem{Miro:2025jnz}
J.~E. Miro and O.~Delouche, ``{Flowing from the Ising Model on the Fuzzy Sphere
  to the 3D Lee-Yang CFT},'' \href{https://arxiv.org/abs/2505.07655}{{\tt
  2505.07655}}.

\bibitem{Yurov:1989yu}
V.~P. Yurov and A.~B. Zamolodchikov, ``{Truncated Conformal Space Approach to
  Scaling Lee-Yang Model},'' {\em Int. J. Mod. Phys. A} {\bf 5} (1990)
  3221--3246.

\bibitem{Xu:2022mmw}
H.-L. Xu and A.~Zamolodchikov, ``{2D Ising Field Theory in a magnetic field:
  the Yang-Lee singularity},'' {\em JHEP} {\bf 08} (2022) 057,
  \href{https://arxiv.org/abs/2203.11262}{{\tt 2203.11262}}.

\bibitem{Leitner:2018iyf}
M.~Leitner, ``{The $(2,5)$ minimal model on genus two surfaces},''
  \href{https://arxiv.org/abs/1801.08387}{{\tt 1801.08387}}.

\bibitem{Katsevich:2024sov}
A.~Katsevich, ``{The spectrum of perturbed (3, 10) minimal model},''
  \href{https://arxiv.org/abs/2410.18069}{{\tt 2410.18069}}.

\bibitem{Bender:2018pbv}
C.~M. Bender, N.~Hassanpour, S.~P. Klevansky, and S.~Sarkar, ``{$PT$-symmetric
  quantum field theory in $D$ dimensions},'' {\em Phys. Rev. D} {\bf 98}
  (2018), no.~12 125003, \href{https://arxiv.org/abs/1810.12479}{{\tt
  1810.12479}}.

\bibitem{deAlcantaraBonfim:1980pe}
O.~de~Alcantara~Bonfim, J.~Kirkham, and A.~McKane, ``{Critical Exponents to
  Order $\epsilon^3$ for $\phi^3$ Models of Critical Phenomena in $6-\epsilon$
  dimensions},'' {\em J.Phys.} {\bf A13} (1980) L247.

\bibitem{deAlcantaraBonfim:1981sy}
O.~de~Alcantara~Bonfim, J.~Kirkham, and A.~McKane, ``{Critical Exponents for
  the Percolation Problem and the Yang-lee Edge Singularity},'' {\em J.Phys.}
  {\bf A14} (1981) 2391.

\bibitem{Fei:2014xta}
L.~Fei, S.~Giombi, I.~R. Klebanov, and G.~Tarnopolsky, ``{Three loop analysis
  of the critical O(N) models in 6-\ensuremath{\varepsilon} dimensions},'' {\em
  Phys. Rev. D} {\bf 91} (2015), no.~4 045011,
  \href{https://arxiv.org/abs/1411.1099}{{\tt 1411.1099}}.

\bibitem{Klebanov:2022syt}
I.~R. Klebanov, V.~Narovlansky, Z.~Sun, and G.~Tarnopolsky, ``{Ginzburg-Landau
  description and emergent supersymmetry of the (3, 8) minimal model},'' {\em
  JHEP} {\bf 02} (2023) 066, \href{https://arxiv.org/abs/2211.07029}{{\tt
  2211.07029}}.

\bibitem{Giombi:2016hkj}
S.~Giombi and V.~Kirilin, ``{Anomalous dimensions in CFT with weakly broken
  higher spin symmetry},'' {\em JHEP} {\bf 11} (2016) 068,
  \href{https://arxiv.org/abs/1601.01310}{{\tt 1601.01310}}.

\bibitem{Gopakumar:2016wkt}
R.~Gopakumar, A.~Kaviraj, K.~Sen, and A.~Sinha, ``{Conformal Bootstrap in
  Mellin Space},'' {\em Phys. Rev. Lett.} {\bf 118} (2017), no.~8 081601,
  \href{https://arxiv.org/abs/1609.00572}{{\tt 1609.00572}}.

\bibitem{Dey:2017oim}
P.~Dey and A.~Kaviraj, ``{Towards a Bootstrap approach to higher orders of
  epsilon expansion},'' {\em JHEP} {\bf 02} (2018) 153,
  \href{https://arxiv.org/abs/1711.01173}{{\tt 1711.01173}}.

\bibitem{Goncalves:2018nlv}
V.~Goncalves, ``{Skeleton expansion and large spin bootstrap for $\phi^3$
  theory},'' \href{https://arxiv.org/abs/1809.09572}{{\tt 1809.09572}}.

\bibitem{Nixon2016}
S.~Nixon and J.~Yang, ``Nonlinear wave dynamics near phase transition in
  PT-symmetric localized potentials,'' {\em Physica D: Nonlinear Phenomena}
  {\bf 331} (2016) 48--57.

\bibitem{Chernyavsky2018}
A.~Chernyavsky and D.~E. Pelinovsky, ``Krein signature for instability of
  PT-symmetric states,'' {\em Physica D: Nonlinear Phenomena} {\bf 371} (2018)
  48--59.

\bibitem{Starkov:2023hox}
G.~A. Starkov, M.~V. Fistul, and I.~M. Eremin, ``{Formation of exceptional
  points in pseudo-Hermitian systems},'' {\em Phys. Rev. A} {\bf 108} (2023),
  no.~2 022206, \href{https://arxiv.org/abs/2302.14672}{{\tt 2302.14672}}.

\bibitem{Wei:2023ygd}
Y.-J. Wei and Z.-C. Gu, ``{Tensor network renormalization: application to
  dynamic correlation functions and non-hermitian systems},''
  \href{https://arxiv.org/abs/2311.18785}{{\tt 2311.18785}}.

\bibitem{Yamada:2022dka}
M.~G. Yamada, T.~Sanno, M.~O. Takahashi, Y.~Akagi, H.~Suwa, S.~Fujimoto, and
  M.~Udagawa, ``{Matrix Product Renormalization Group: Potential Universal
  Quantum Many-Body Solver},'' \href{https://arxiv.org/abs/2212.13267}{{\tt
  2212.13267}}.

\bibitem{Heiss:2012dx}
W.~D. Heiss, ``{The physics of exceptional points},'' {\em J. Phys. A} {\bf 45}
  (2012) 444016, \href{https://arxiv.org/abs/1210.7536}{{\tt 1210.7536}}.

\bibitem{CJHamer_1980}
C.~J. Hamer and M.~N. Barber, ``Finite-size scaling in Hamiltonian field
  theory,'' {\em Journal of Physics A: Mathematical and General} {\bf 13} (may,
  1980) L169.

\bibitem{CJHamer_1981}
C.~J. Hamer and M.~N. Barber, ``Finite-lattice methods in quantum Hamiltonian
  field theory. I. The Ising model,'' {\em Journal of Physics A: Mathematical
  and General} {\bf 14} (jan, 1981) 241.

\bibitem{CJHamer_1981_2}
C.~J. Hamer and M.~N. Barber, ``Finite-lattice methods in quantum Hamiltonian
  field theory. I. O(2) and O(3) Heisenberg models,'' {\em Journal of Physics
  A: Mathematical and General} {\bf 14} (jan, 1981) 259.

\bibitem{von1991critical}
G.~Von~Gehlen, ``Critical and off-critical conformal analysis of the Ising
  quantum chain in an imaginary field,'' {\em Journal of Physics A:
  Mathematical and General} {\bf 24} (1991), no.~22 5371.

\bibitem{gehlen1994non}
G.~V. Gehlen, ``Non-Hermitian tricriticality in the Blume-Capel model with
  imaginary field,'' {\em International Journal of Modern Physics B} {\bf 8}
  (1994), no.~25n26 3507--3529.

\bibitem{zhou2025fuzzified}
Z.~Zhou, ``FuzzifiED: Julia package for numerics on the fuzzy sphere,'' {\em
  arXiv preprint arXiv:2503.00100} (2025).

\bibitem{CARDY1986186}
J.~L. Cardy, ``Operator content of two-dimensional conformally invariant
  theories,'' {\em Nuclear Physics B} {\bf 270} (1986) 186--204.

\bibitem{Zamolodchikov:1987ti}
A.~B. Zamolodchikov, ``{Renormalization Group and Perturbation Theory Near
  Fixed Points in Two-Dimensional Field Theory},'' {\em Sov. J. Nucl. Phys.}
  {\bf 46} (1987) 1090.

\bibitem{Cardy:1984epx}
J.~L. Cardy, ``{Conformal invariance and universality in finite-size
  scaling},'' {\em J. Phys. A} {\bf 17} (1984), no.~7 L385.

\bibitem{Zou:2018dec}
Y.~Zou, A.~Milsted, and G.~Vidal, ``Conformal Data and Renormalization Group
  Flow in Critical Quantum Spin Chains Using Periodic Uniform Matrix Product
  States,'' {\em Phys. Rev. Lett.} {\bf 121} (Dec, 2018) 230402.

\bibitem{Zou:2019dnc}
Y.~Zou, A.~Milsted, and G.~Vidal, ``{Conformal fields and operator product
  expansion in critical quantum spin chains},'' {\em Phys. Rev. Lett.} {\bf
  124} (2020), no.~4 040604, \href{https://arxiv.org/abs/1901.06439}{{\tt
  1901.06439}}.

\bibitem{Costa:2011mg}
M.~S. Costa, J.~Penedones, D.~Poland, and S.~Rychkov, ``{Spinning Conformal
  Correlators},'' {\em JHEP} {\bf 11} (2011) 071,
  \href{https://arxiv.org/abs/1107.3554}{{\tt 1107.3554}}.

\bibitem{Meltzer:2018tnm}
D.~Meltzer, ``{Higher Spin ANEC and the Space of CFTs},'' {\em JHEP} {\bf 07}
  (2019) 001, \href{https://arxiv.org/abs/1811.01913}{{\tt 1811.01913}}.

\bibitem{kondo1965characters}
T.~Kondo, ``The characters of the Weyl group of type F4,'' {\em J. Fac. Sci.
  Univ. Tokyo Sect. I} {\bf 11} (1965), no.~145-153 1965.

\bibitem{Carter1972}
R.~W. Carter, ``Conjugacy classes in the weyl group,'' {\em Compositio
  Mathematica} {\bf 25} (1972), no.~1 1--59.

\bibitem{carter1993finite}
R.~Carter, {\em Finite Groups of Lie Type: Conjugacy Classes and Complex
  Characters}.
\newblock Wiley Classics Library. Wiley, 1993.

\bibitem{Carleo_2017}
G.~Carleo and M.~Troyer, ``Solving the quantum many-body problem with
  artificial neural networks,'' {\em Science} {\bf 355} (Feb., 2017) 602–606.

\bibitem{Giombi:2024zrt}
S.~Giombi, E.~Himwich, A.~Katsevich, I.~Klebanov, and Z.~Sun, ``{Sphere free
  energy of scalar field theories with cubic interactions},''
  \href{https://arxiv.org/abs/2412.14086}{{\tt 2412.14086}}.

\bibitem{Katsevich:2024jgq}
A.~Katsevich, I.~R. Klebanov, and Z.~Sun, ``{Ginzburg-Landau description of a
  class of non-unitary minimal models},'' {\em JHEP} {\bf 03} (2025) 170,
  \href{https://arxiv.org/abs/2410.11714}{{\tt 2410.11714}}.

\bibitem{Nakayama:2024msv}
Y.~Nakayama and T.~Tanaka, ``{Infinitely many new renormalization group flows
  between Virasoro minimal models from non-invertible symmetries},'' {\em JHEP}
  {\bf 11} (2024) 137, \href{https://arxiv.org/abs/2407.21353}{{\tt
  2407.21353}}.

\bibitem{Delouche:2024yuo}
O.~Delouche, J.~Elias~Miro, and J.~Ingoldby, ``{Testing the RG-flow $M(3, 10) +
  \phi_{1,7} \rightarrow{} M(3, 8)$ with Hamiltonian Truncation},'' {\em JHEP}
  {\bf 04} (2025) 144, \href{https://arxiv.org/abs/2412.09295}{{\tt
  2412.09295}}.

\bibitem{Kausch:1996vq}
H.~Kausch, G.~Takacs, and G.~Watts, ``{On the relation between Phi(1,2) and
  Phi(1,5) perturbed minimal models},'' {\em Nucl. Phys. B} {\bf 489} (1997)
  557--579, \href{https://arxiv.org/abs/hep-th/9605104}{{\tt hep-th/9605104}}.

\bibitem{Quella:2006de}
T.~Quella, I.~Runkel, and G.~M. Watts, ``{Reflection and transmission for
  conformal defects},'' {\em JHEP} {\bf 0704} (2007) 095,
  \href{https://arxiv.org/abs/hep-th/0611296}{{\tt hep-th/0611296}}.

\bibitem{Poland:2023bny}
D.~Poland, V.~Prilepina, and P.~Tadi\'c, ``{Improving the five-point
  bootstrap},'' {\em JHEP} {\bf 05} (2024) 299,
  \href{https://arxiv.org/abs/2312.13344}{{\tt 2312.13344}}.

\bibitem{Poland:2023vpn}
D.~Poland, V.~Prilepina, and P.~Tadi\'c, ``{The five-point bootstrap},'' {\em
  JHEP} {\bf 10} (2023) 153, \href{https://arxiv.org/abs/2305.08914}{{\tt
  2305.08914}}.

\bibitem{Chang:2024whx}
C.-H. Chang, V.~Dommes, R.~S. Erramilli, A.~Homrich, P.~Kravchuk, A.~Liu, M.~S.
  Mitchell, D.~Poland, and D.~Simmons-Duffin, ``{Bootstrapping the 3d Ising
  stress tensor},'' {\em JHEP} {\bf 03} (2025) 136,
  \href{https://arxiv.org/abs/2411.15300}{{\tt 2411.15300}}.

\bibitem{sym2031423}
S.~Mamone, G.~Pileio, and M.~H. Levitt, ``Orientational Sampling Schemes Based
  on Four Dimensional Polytopes,'' {\em Symmetry} {\bf 2} (2010), no.~3
  1423--1449.

\bibitem{GeckGoetz}
M.~Geck and G.~Pfeiffer, {\em Characters of finite Coxeter groups and
  Iwahori-Hecke algebras}.
\newblock London Math. Soc. Monographs, New Series. Oxford University Press,
  Aug., 2000.

\end{thebibliography}\endgroup

\end{document}